\documentclass[%
 reprint,
showpacs,
 amsmath,amssymb,
 aps,pre
]{revtex4-1}

\usepackage{graphicx}% Include figure files
\usepackage{dcolumn}% Align table columns on decimal point
\usepackage{bm}% bold math
\usepackage[usenames,dvipsnames]{color}
\usepackage[normalem]{ulem}
\begin{document}

\preprint{APS/123-QED}

\title{Death and rebirth of neural activity in sparse inhibitory networks}

\author{David Angulo-Garcia}
\email{david.angulo-garcia@univ-amu.fr}
\affiliation{Aix Marseille Univ, INSERM, INMED and INS, Inst Neurosci Syst, Marseille, France}
\affiliation{Aix Marseille Univ, Universit\'{e} de Toulon, CNRS, CPT, UMR 7332, 13288 Marseille, France} 

\author{Stefano Luccioli}
\email{stefano.luccioli@fi.isc.cnr.it}
\affiliation{CNR - Consiglio Nazionale delle Ricerche - Istituto dei Sistemi Complessi, 50019 Sesto Fiorentino, Italy}
\affiliation{INFN - Istituto Nazionale di Fisica Nucleare - Sezione di Firenze, 50019 Sesto Fiorentino, Italy} 

\author{Simona Olmi} 
\email{simona.olmi@isc.cnr.it}
\affiliation{Aix Marseille Univ, INSERM, INS, Inst Neurosci Syst, Marseille, France}
\affiliation{CNR - Consiglio Nazionale delle Ricerche - Istituto dei Sistemi Complessi, 50019 Sesto Fiorentino, Italy}%
\affiliation{Weierstrass Institute, Mohrenstra$\ss$e 39, 10117 Berlin, Germany}

\author{Alessandro Torcini}
\email{alessandro.torcini@univ-amu.fr}
\affiliation{Laboratoire de Physique Th\'eorique et Mod\'elisation, Universit\'e de Cergy-Pontoise, CNRS, UMR 8089,
95302 Cergy-Pontoise cedex, France}
\affiliation{Aix Marseille Univ, INSERM, INMED and INS, Inst Neurosci Syst, Marseille, France}
\affiliation{Aix Marseille Univ, Universit\'{e} de Toulon, CNRS, CPT, UMR 7332, 13288 Marseille, France} 
\affiliation{CNR - Consiglio Nazionale delle Ricerche - Istituto dei Sistemi Complessi, 50019 Sesto Fiorentino, Italy}
\affiliation{Max-Planck-Institut f\"ur Physik komplexer Systeme, N\"othnitzer Stra{\ss}e 38, 01187 Dresden, Germany}


\date{\today}% It is always \today, today,

\begin{abstract}
Inhibition is a key aspect of neural dynamics playing a fundamental role for the emergence of neural rhythms 
and the implementation of  various information coding strategies. Inhibitory populations are present in several
brain structures and the comprehension of their dynamics is strategical for the understanding of neural processing.
In this paper, we clarify the mechanisms underlying a general phenomenon present in pulse-coupled heterogeneous inhibitory networks: inhibition can induce not only  suppression of the neural activity, as expected, but it can also promote neural reactivation. In particular, for globally coupled systems, the number of firing neurons monotonically reduces upon increasing the strength of inhibition (neurons' death). However, the random pruning of the connections is able to reverse the action of inhibition, i.e. in a random sparse network a sufficiently strong synaptic strength can surprisingly promote, rather than depress, the activity of the neurons (neurons' rebirth). Thus the number of firing neurons reveals a minimum at some intermediate synaptic strength. We show that this minimum signals a transition 
from a regime dominated by the neurons with higher firing activity
to a phase where all neurons are effectively sub-threshold and  their irregular firing is driven by current fluctuations. 
We explain the origin of the transition by deriving a mean field formulation of the problem able to provide the fraction of active neurons as well as the first two moments of their firing statistics.  The introduction of a synaptic time scale does not modify the main aspects of the reported phenomenon. However, for sufficiently slow synapses the transition becomes dramatic, the system passes from a perfectly regular evolution to an irregular bursting dynamics. In this latter regime  
the model provides predictions consistent with experimental findings for a specific class of neurons, namely the medium spiny neurons in the striatum.
\end{abstract}

\maketitle

%\tableofcontents

\section{Introduction}

\begin{figure*}
\centering
\includegraphics[scale=0.35]{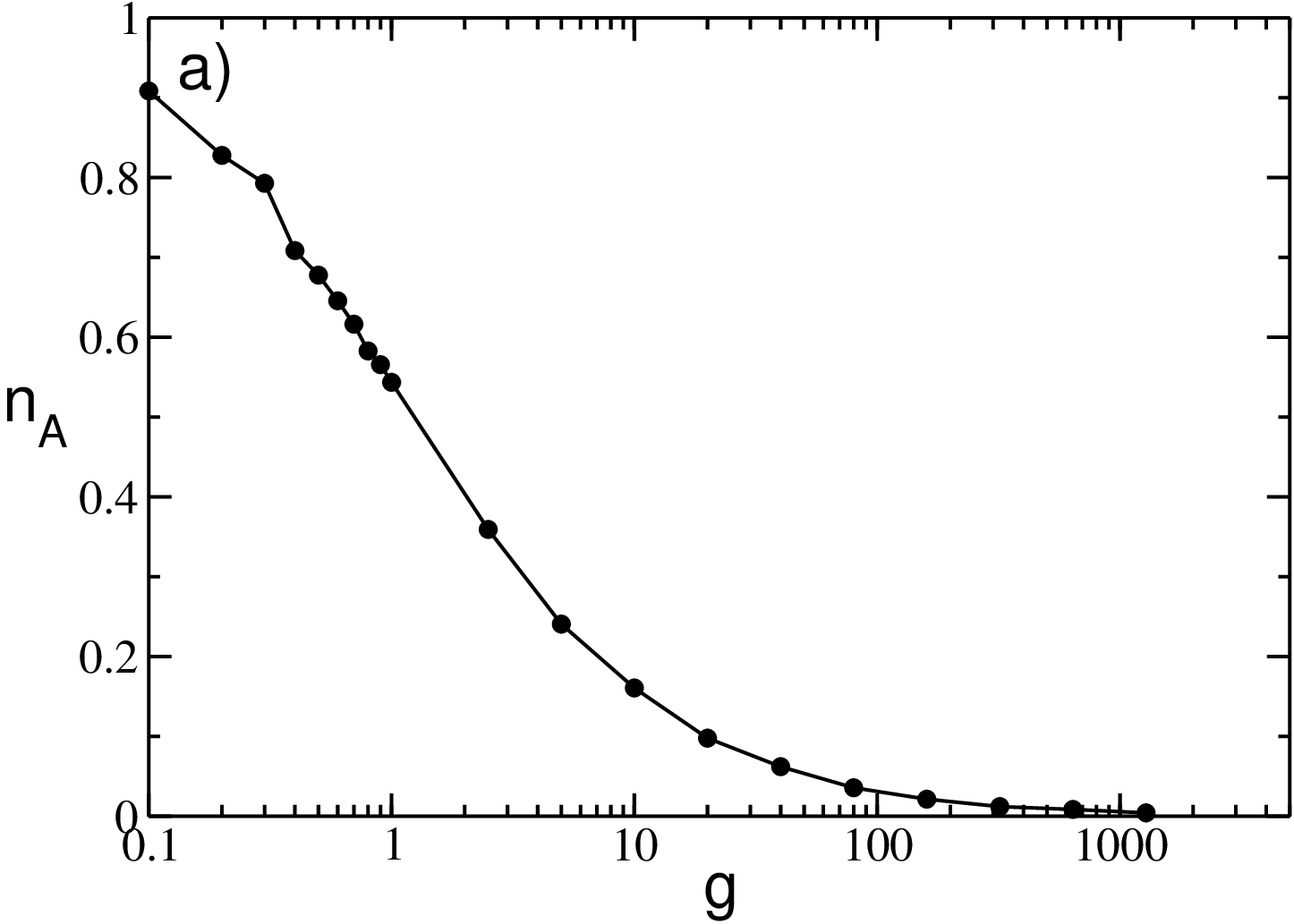}
\includegraphics[scale=0.35]{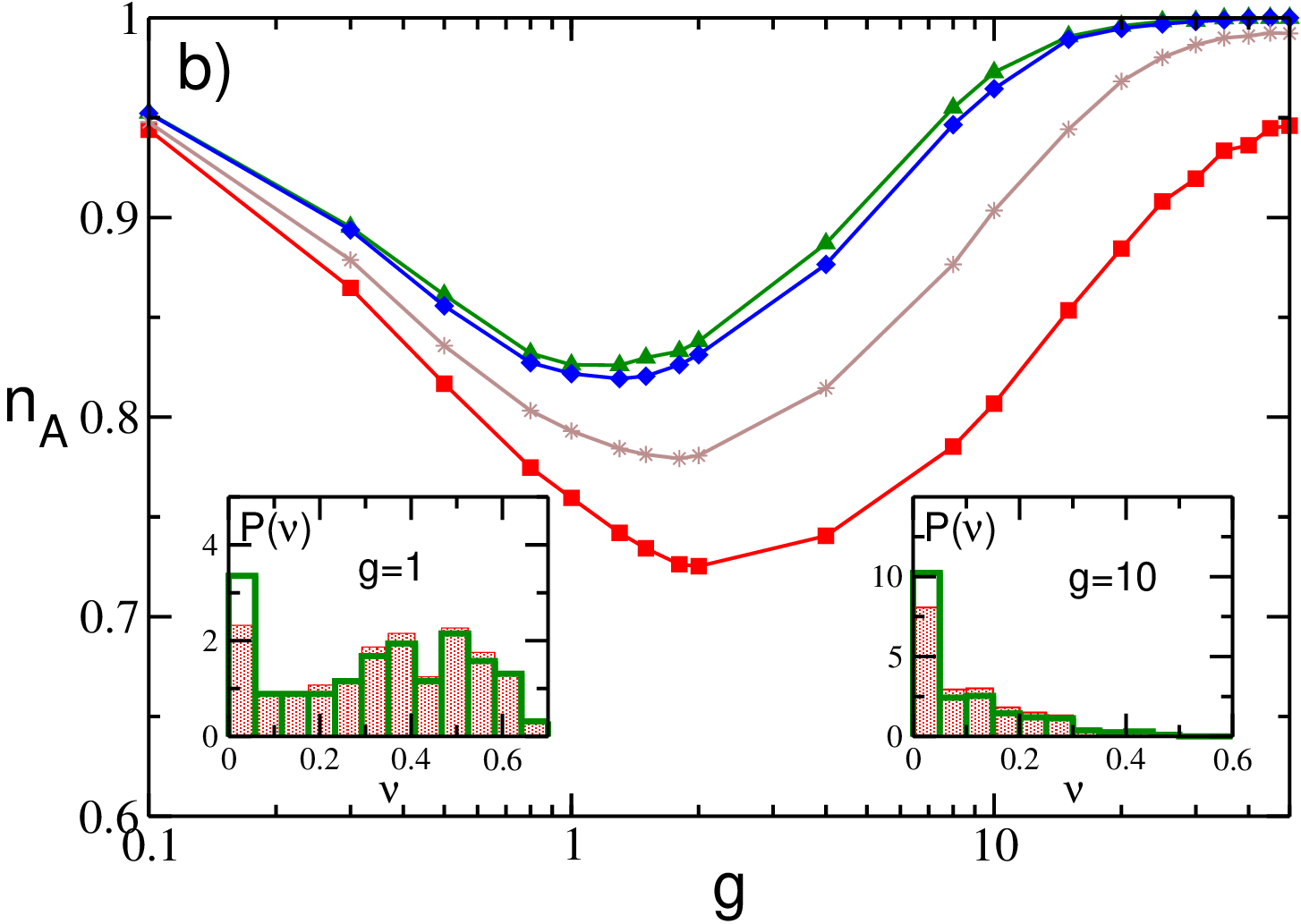}
\caption{a-b) Fraction of active neurons $n_A$ as a function of the inhibitory 
synaptic strength $g$ for a globally coupled system (a), where $K=N-1$,
and a randomly connected (sparse) network with $K=20$ (b). 
In a) is reported the asymptotic value $n_A$ calculated after
a time $t_S= 1 \times 10^6$. Conversely in b), $n_A$ is
reported at successive times: namely, $t_S = 985$ (red squares), 
$t_S = 1.1 \times 10^4$ (brown stars), $t_S = 5 \times 10^5$ (blue diamonds) and $t_S = 1 \times 10^6$ (green triangles).
An estimation of the times needed to reach $n_A = 1$  can be obtained by 
employing Eq.~\eqref{eq:kramer_eqNstar} these values range
from $t_s = 5 \times 10^9$ for $g=0.1$ to $5 \times 10^5$ for $g=50$.
Insets in b) depict the probability distributions $P(\nu)$
of the single neuron firing rate $\nu$
for the sparse network for a given $g$ indicated in the inset at two different times: 
$t_S = 985$ (red filled histograms) and $t_S = 1 \times 10^6$
(thick empty green histograms). The histograms are calculated by considering only active neurons.
The reported data refer to instantaneous synapses, to a system size $N=400$ and 
to an uniform distribution $P(I)$ with  $[l_1,l_2] = [1.0,1.5]$ and $\theta=1$, the values
reported in a) and b) have been also averaged over 10 random realizations of the network.}
\label{fig:SilecingNeuron_FC_Sparse}
\end{figure*}

The presence of inhibition in excitable systems induces a rich dynamical repertoire,
which is extremely relevant for biological~\cite{camazine2003}, 
physical~\cite{kerner1990} and chemical systems~\cite{vanag2000}. 
In particular, inhibitory coupling has been invoked to explain cell navigation~\cite{xiong2010cells}, morphogenesis in animal coat pattern formation~\cite{meinhardt1982}, and 
the rhythmic activity of central pattern generators in many biological systems~\cite{harris1992,marder2001}.
In brain circuits the role of inhibition is fundamental to balance massive recurrent 
excitation~\cite{somogyi1998salient} in order to generate 
physiologically relevant cortical rhythms~\cite{shu2003turning,buz2004}.
%In particular, $\gamma$-rhythm can emerge as collective oscillations in 
%hyppocampal and neocortical networks of mutually inhibitory interneurons when tonically 
%excited~cite{jefferys1996neuronal,buzsaki2012mechanisms}, 
%spindle oscillations during early sleep stage
%are initiated by the inhibitory action of thalamic reticular neurons~\cite{steriade2005sleep},
%and rhythmic whisking in rodents is controlled by inhibition~\cite{deschenes2016inhibition}.

Inhibitory networks are important not only for the emergence of rhythms in the brain, but also 
for the fundamental role they play in information encoding in the olfactory system~\cite{laurent2002olfactory}
as well as in controlling and regulating motor and learning activity
in the basal ganglia~\cite{berke2004oscillatory,miller2008dysregulated,carrillo2008encoding}.
Furthermore, stimulus dependent sequential activation of neurons or group of neurons,
reported for  asymmetrically connected inhibitory cells~\cite{nowotny2007dynamical,komarov2009sequentially},  
has been suggested  as a possible mechanism to explain sequential memory storage 
and feature binding~ \cite{rabinovich2011}.

These explain the long term interest for numerical and theoretical investigations
of  the dynamics of inhibitory networks. Already the study of globally coupled homogeneous systems 
revealed interesting dynamical features, ranging from full synchronization to
clustering appearance~\cite{golomb1994clustering,vanVreeswijk1996Partial,wang1996gamma}, 
from the emergence of splay states~\cite{Zillmer2006} to oscillator death~\cite{bressloff2000dynamics}.
The introduction of  disorder, e.g. random dilution, noise or other form of heterogeneity in these systems 
leads to more complex dynamics, ranging from fast global oscillations~\cite{BrunelHakim1999} in neural networks and self-sustained 
activity in excitable systems~\cite{larremore2014}, to irregular dynamics~\cite{Zillmer2006,JhankeTimme2008PRL,Timme2009ChaosBalance,MonteforteBalanced2012,
Luccioli2010Irregular,angulo2014,ostojic2014,kadmon2015,ullner2016self}.
In particular, inhibitory spiking networks, due to {\it stable chaos}~\cite{politi2010stable}, can display extremely long erratic transients even in linearly stable regimes~\cite{Zillmer2006,Zillmer2009LongTrans,JhankeTimme2008PRL,Timme2009ChaosBalance,
MonteforteBalanced2012,angulo2014,ullner2016self,Luccioli2010Irregular}.

One of the most studied inhibitory neural population is represented by medium spiny neurons (MSNs) 
in the striatum (which is the main input structure
of the basal ganglia)~\cite{parent1995functional,lanciego2012functional}.
In a series of papers, Ponzi and Wickens have shown that the main features of the 
MSN dynamics can be reproduced by considering a randomly
connected inhibitory network of conductance based neurons subject to external stochastic excitatory inputs~\cite{ponzi2010sequentially,ponzi2012input,ponzi2013optimal}.
Our study has been motivated by an interesting phenomenon reported 
for this model in~\cite{ponzi2013optimal}: namely,  upon increasing the synaptic  strength the system passes from a regularly 
firing regime, characterized by a large part of quiescent neurons, 
to a biologically relevant regime where almost all cells exhibit a bursting activity, characterized by an alternation of periods 
of silence and of high firing. The same phenomenology has been recently reproduced by employing a much simpler neural 
model~\cite{angulo2015Striatum}. Thus suggesting that this behaviour is not related to the specific
model employed, but it is indeed a quite general property of inhibitory networks.   
However, it is still unclear the origin of the phenomenon and the minimal ingredients required 
to observe the emergence of this effect.

In order to exemplify the problem addressed in this paper 
we report in Fig.~\ref{fig:SilecingNeuron_FC_Sparse}
the fraction of  active neurons $n_A$ (i.e. the ones emitting at least one spike during the simulation time) 
as a function of the strength of the synaptic inhibition $g$ in an heterogeneous network.
For a fully coupled network, $n_A$ has a monotonic decrease with $g$ (Fig.~\ref{fig:SilecingNeuron_FC_Sparse} (a)), while for a 
random sparse network $n_A$ has a non monotonic behaviour, displaying a minimum at an intermediate strength $g_m$ (Fig.~\ref{fig:SilecingNeuron_FC_Sparse} (b)). 
In fully coupled networks the effect of inhibition is simply to reduce
the number of active neurons ({\it neurons' death}). However, quite counter-intuitively, in presence 
of dilution by increasing the synaptic strength the previously silenced neurons can return to fire ({\it neurons' rebirth}).
Our aim is to clarify the physical mechanisms underlying neuron's death 
and rebirth, which are at the origin of the behaviour reported in ~\cite{ponzi2013optimal, angulo2015Striatum}.
 
In particular, we consider a deterministic network of purely inhibitory pulse-coupled 
Leaky Integrate-and-Fire (LIF) neurons with an heterogeneous distribution of excitatory DC
currents, accounting for the different level of excitability of the neurons. The evolution
of this model is studied for fully coupled and for random sparse topology, as well as for synapses with different time courses. For the fully coupled case,
it is possible to derive,  within a self-consistent mean field approach, the
analytic expressions for the fraction of active neurons and for the average firing frequency  ${\bar \nu}$ as a function of the coupling strength $g$. 
In this case the monotonic decrease of $n_A$ with $g$
can be interpreted as a  {\it Winner Takes All}  (WTA) mechanism~\cite{fukai1997simple,coultrip1992cortical,yuille1989},
where only the most excitable neurons survive to the inhibition
increase. For random sparse networks,  the neurons' rebirth 
can be interpreted as a re-activation process induced by erratic fluctuations in the
synaptic currents. Within this framework it is possible to obtain semi-analytically,
for instantaneous synapses, a closed set of equations for $n_A$ as well as for 
the average firing rate and coefficient of variation as a function of the coupling strength.
In particular, the firing statistics of the network can be obtained via a mean-field approach by extending 
the formulation derived in~\cite{richardson2010firing} to account for synaptic shot noise with constant
amplitude. The introduction of a finite synaptic time scale do not modify the overall scenario 
as far as this is shorter than the membrane time constant.
As soon as the synaptic dynamics becomes slower,  the phenomenology of
the transition is modified. At $g < g_m$ we have a {\it frozen phase} 
where $n_A$ does not evolve in time on the explored time scales, since
the current fluctuations are negligible. Above $g_m$ we have a bursting regime,
which can be related to the emergence of correlated fluctuations induced by
slow synaptic times, as discussed in the framework of the adiabatic approach in~\cite{moreno2004,moreno2010response}.
  
The paper is organized as follows: In Sect.~\ref{sec:Model} we present the models that will be 
considered in the paper as well as the methods adopted to characterize its dynamics.
In Sect.~\ref{sec:FC} we consider the globally coupled network 
where we provide analytic self-consistent expressions accounting for the fraction of active neurons and the average firing rate. 
Section \ref{sec:sparse} is devoted to the study of sparsely connected networks with instantaneous synapses and to the
derivation of the set of semi-analytic self-consistent equations 
providing $n_A$, the average firing rate and the coefficient of variation. 
In section \ref{sec:alpha} we discuss the effect of synaptic filtering
with a particular attention on slow synapses. 
Finally in Sect.~\ref{sec:discussion} we briefly discuss the obtained results with a
focus on the biological relevance of our model.

\section{Model and Methods}
\label{sec:Model}

We examine the dynamical properties of an heterogeneous 
inhibitory sparse network made of $N$ LIF neurons. 
The time evolution of the membrane potential $v_i$ of 
the $i$-th neuron is ruled by the following first order
ordinary differential equation:
\begin{equation}
\label{eq:generic_model}
\dot{v}_i(t) = I_i - v_i(t) - g E_i(t) \qquad ;
\end{equation}
where $g > 0$ is the inhibitory synaptic strength, 
$I_i$ is the neuronal excitability of the $i$-th neuron
encompassing both the intrinsic neuronal properties
and the excitatory stimuli originating from areas outside the considered neural circuit and 
$E_i(t)$ represents the synaptic current due to the recurrent interactions
within the considered network. The membrane potential $v_i$ of neuron $i$
evolves accordingly to Eq.~\eqref{eq:generic_model} until it overcomes a constant threshold $\theta = 1$, this leads to the emission of a spike (action potential) transmitted to all the connected
post-synaptic neurons, while $v_i$ is reset to its rest value $v_r = 0$. 
The model in \eqref{eq:generic_model} is expressed in adimensional units, this amounts to assume
a membrane time constant $\tau_m =1$, for the conversion to dimensional variables see
Appendix A. The heterogeneity is introduced in the model by assigning to each neuron a different value of input excitability $I_i$ drawn from a flat distribution $P(I)$, whose support
is $I \in [l_1,l_2]$ with $l_1 \ge \theta$, therefore all the neurons 
are supra-threshold.

The synaptic current $E_i(t)$ is given by the linear super-position of all the 
inhibitory post-synaptic potentials (IPSPs) $\eta(t)$ emitted 
at previous times $t_n^j < t$ by the pre-synaptic neurons 
connected to neuron $i$, namely
\begin{equation}
\label{eq:recurrent_current}
E_i(t) =  \frac{1}{K} \sum_{j \ne i} C_{ij} \sum_{n|t_n < t} \eta(t-t^{j}_n) \; ;
\end{equation}
where $K$ is the number of pre-synaptic neurons. 
$C_{ij}$ represent the elements of the $N\times N$ connectivity 
matrix associated to an undirected random network, whose entries are $1$ 
if there is a synaptic connection from neuron $j$ to neuron $i$, and $0$ otherwise. 
For the sparse network, we select randomly the matrix entries,
however to reduce the sources of variability in the network, we assume that the number of 
pre-synaptic neurons is fixed, namely $\sum_{j \ne i} C_{ij} = K  << N$ for each neuron $i$, 
where autaptic connections are not allowed. We have verified that the results
do not change if we choose randomly the links accordingly to an Erd\"os-Renyi distribution 
with a probability $K/N$. For a fully coupled network we have $K=N-1$.

The shape of the IPSP characterize the type of filtering performed by the synapses
on the received action potentials. We have considered two kind of synapses, instantaneous ones, where $\eta(t)=\delta(t)$, and synapses where the PSP is an $\alpha$-pulse, namely 
\begin{equation}
\label{eq:alpha}
\eta(t)= H(t) {\alpha}^{2} t {\rm e}^{-t{\alpha}} \qquad ;
\end{equation}
with $H$ denoting the Heaviside step function.
In this latter case the rise and decay time of the pulse are the same, namely  $\tau_\alpha = 1/\alpha$,
and therefore the pulse duration $\tau_P$ can be assumed to be twice the characteristic time $\tau_\alpha$.
The equations of the model Eqs.~\eqref{eq:generic_model} and \eqref{eq:recurrent_current}
are integrated exactly in terms of the associated {\it event driven maps} for 
different synaptic filtering, these correspond to Poincar\'e maps performed at the firing 
times (for details see Appendix A)~\cite{Zillmer2007,Olmi2014linear}.

For instantaneous synapses, we have usually considered system sizes $N=400$ and $N=1,400$
and for the sparse case in-degrees $20 \le K \le 80$ for $N=400$ and $20 \le K \le 600$ for
$N=1400$ with integration times up to $t_S = 1 \times 10^6$.
For synapses with a finite decay time we limit the analysis to $N=400$ 
and $K=20$ and to maximal integration times $t_S = 1 \times 10^5$.
Finite size dependences on $N$ are negligible with these parameter choices,
as we have verified.

In order to characterize the network dynamics we measure the fraction of active neurons $n_A (t_S)$
at time $t_S$, i.e. the fraction of neurons emitting at least one spike
in the time interval $[0,t_S]$. Therefore a neuron will be considered silent if it has a frequency 
smaller than $1/t_S$, with our choices of $t_S = 10^5 - 10^6$, this corresponds
to neurons with frequencies smaller than $10^{-3} - 10^{-4}$ Hz, by assuming as
timescale a  membrane time constant $\tau_m = 10$ ms. 
The estimation of the number of active neurons is always started 
after a sufficiently long transient time has been discarded, usually corresponding to
the time needed to deliver $10^6$ spikes in the network.  

Furthermore, for each neuron we estimate the time averaged inter-spike interval (ISI) $T_{ISI}$,
the associated firing frequency $\nu = 1/T_{ISI}$, as well as the coefficient of 
variation $CV$, which is the ratio of the standard deviation of the ISI distribution divided by $T_{ISI}$. For a regular spike train $CV=0$, for a Poissonian distributed one $CV=1$, while $CV > 1$ is an indication of bursting activity. The indicators usually reported in the following to characterize the network activity are ensemble averages over all the active neurons, which we denote as $\bar{a}$ for a generic observable $a$.

To analyze the linear stability of the dynamical evolution we measure
the maximal Lyapunov exponent $\lambda$, which is positive for chaotic evolution,
and negative (zero) for stable (marginally stable) dynamics~\citep{BenettinLyapunov1980}.
In particular, by following~\cite{Olmi2010Oscillations,angulo2015stochastic} $\lambda$ is estimated by
linearizing the corresponding event driven map.

%%%%% =============== %%%%

\section{Fully Coupled Networks: \\ Winner Takes All}
\label{sec:FC}

In the fully coupled case we observe that the number of active neurons $n_A$ saturates,
after a short transient, to a value which remains constant in time.
In this case, it is possible to derive a self-consistent mean field approach 
to obtain analytic expressions for the fraction of active neurons $n_A$ and for the average firing 
frequency $\bar \nu$ of the neurons in the network. In a fully coupled network each neuron 
receives the spikes  emitted by the other $K=N-1$ neurons, therefore each neuron is essentially 
subject to the same effective input $\mu$, apart corrections ${\cal O}(1/N)$. 

The effective input current, for a neuron with an excitability $I$,
is given by
\begin{equation}
\label{eq:eff_input_GC}
\mu = I - g \bar{\nu} n_A \; \quad ;
\end{equation}
where $n_A (N-1)$ is the number of active pre-synaptic neurons assumed
to fire with the same average frequency  $\bar \nu$.

In a mean field approach, each neuron can be seen as isolated from the network
and driven by the effective input current $\mu$. Taking into account the distribution
of the excitabilities $P(I)$ one obtains the following self-consistent
expression for the average firing frequency 
\begin{equation}
\bar{\nu} = \frac{1}{\Delta}
\int_{\{ I_A \}} dI \enskip P(I) 
\left[\ln \left( \frac{I - g \bar{\nu} n_A - v_r}{I -g\bar{\nu} n_A - \theta} \right) \right]^{-1}
\label{eq:nu}
\end{equation}
where the integral is restricted only to active neurons, i.e. to
$I \in \{I_A\}$ values for which the logarithm is defined,  while 
$\Delta = \int_{\{I_A\}} dI \enskip P(I)$ is the measure of their support.
In \eqref{eq:nu} we have used the fact that for an isolated LIF neuron with constant 
excitability $C$, the ISI is simply given by $ T_{ISI}= \ln [(C-v_r)/(C-\theta)] $~\cite{burkitt2006LIFreviewI}.

An implicit expression for $n_A$ can be obtained by estimating the neurons 
with effective input $\mu > \theta$, in particular the number of silent neurons is given by 
\begin{equation}
\label{eq:Cumulative_Eff}
1 - n_A = \int_{l_1}^{l^*} dI P(I) \; ,
\end{equation}
where $l_1$ is the lower limit of the support of the distribution,
while $l^* = g \bar{\nu} n_A + \theta$. By solving self-consistently
Eqs.\eqref{eq:nu} and \eqref{eq:Cumulative_Eff} one can obtain the 
analytic expression for $n_A$ and $\bar \nu$ for any distribution $P(I)$.

In particular, for excitabilities distributed uniformly in the interval $[l_1,l_2]$, the expression 
for the average frequency Eq. \eqref{eq:nu} becomes
\begin{eqnarray}
\label{eq:unif_nu}
\bar{\nu} & = & \frac{1}{n_A(l_2 - l_1)}\int_{\{ I_A \}} dI 
\left[\ln\left( \frac{I - g \bar{\nu} n_A - v_r}{I -g\bar{\nu}n_A - \theta} \right) \right]^{-1} \quad ;
\end{eqnarray}
while the number of active neurons is given by the following expression
\begin{equation}
\label{eq:nStar_FC}
n_A = \frac{l_2 - \theta}{l_2 - l_1 + g \bar{\nu}} \quad ;
\end{equation}
with the constraint that $n_A$ cannot be larger than one.

\begin{figure}
\centering
\includegraphics[width=0.95\linewidth]{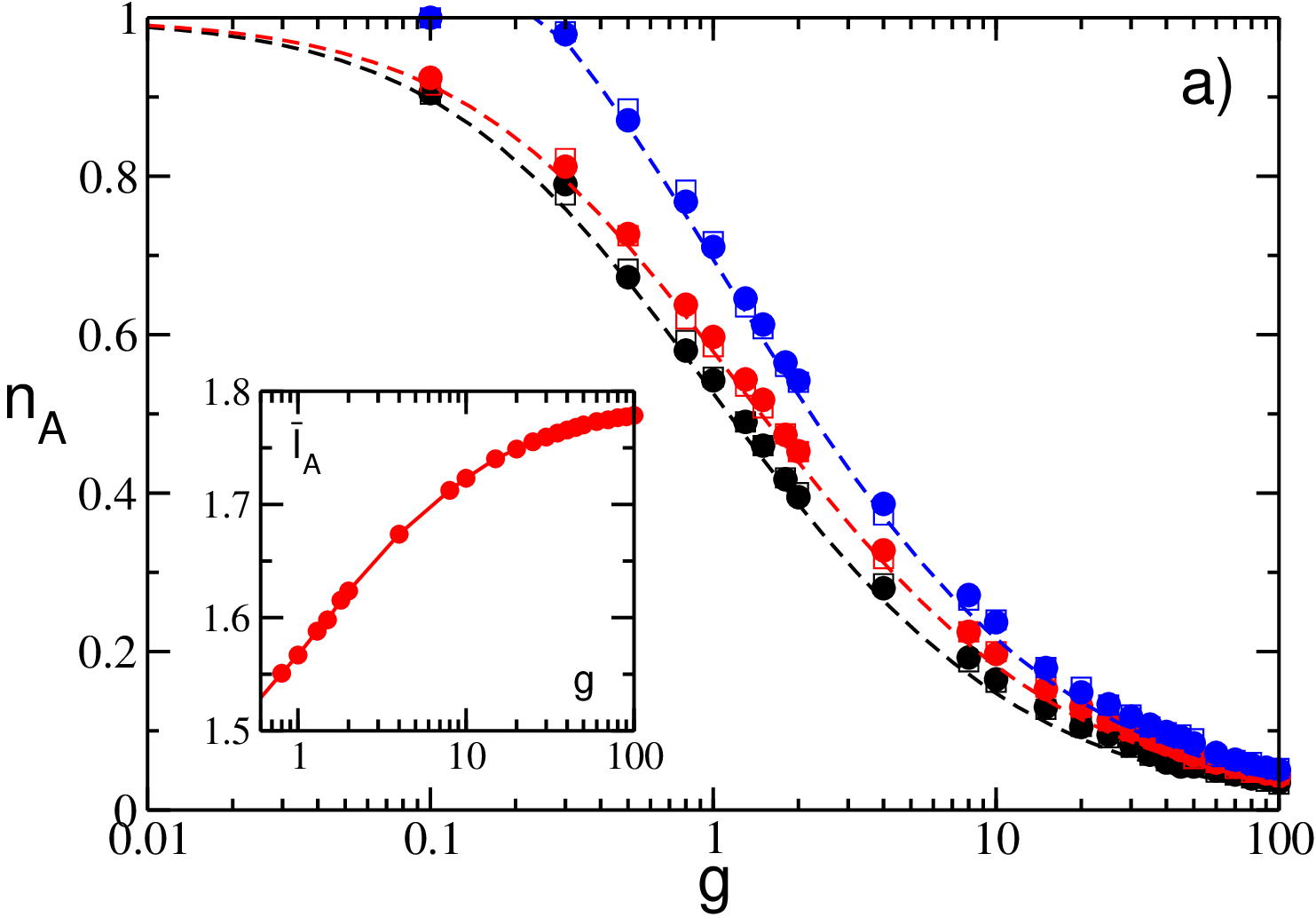}
\includegraphics[width=0.95\linewidth]{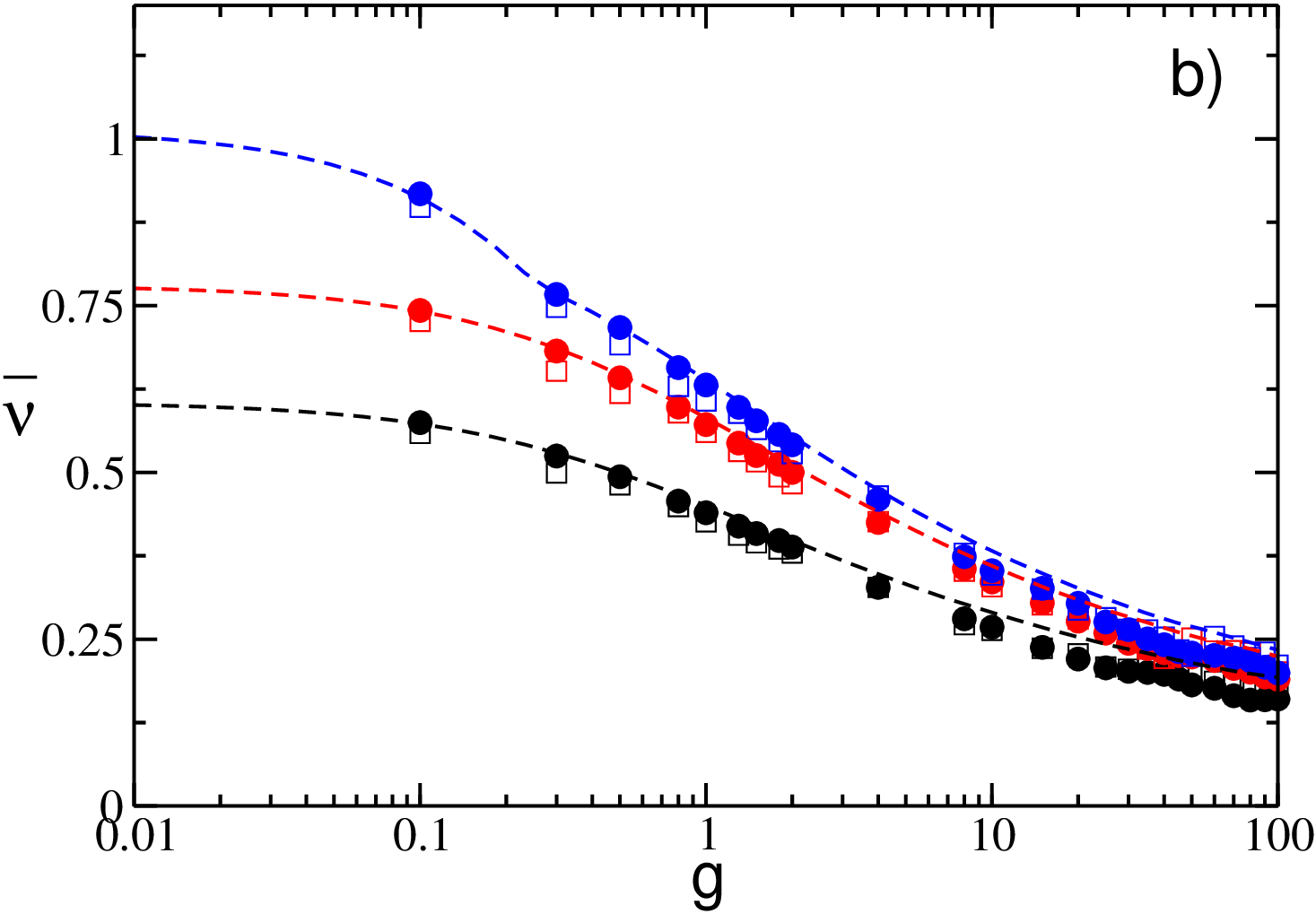}
\caption{Fraction of active neurons $n_A$ (a) and average network's frequency $\bar \nu$ (b)
as a function of the synaptic strength $g$ for uniform distributions $P(I)$ with different supports. Inset: average neuronal excitability of the active neurons $\bar I_A$ versus $g$.
Empty (filled) symbols refer to numerical simulation with $N = 400$ ($N = 1400$) and dashed lines 
to the corresponding analytic solution. Symbols and lines correspond from bottom to top 
to $[l_1, l_2] = [1.0 , 1.5]$ (black); $[l_1,l_2] = [1.0, 1.8]$  (red) and $[l_1, l_2 ] = [1.2 , 2.0]$ (blue). The data have been averaged over a time interval $t_S=1\times 10^6$
after discarding a transient of $10^6$ spikes.} 
\label{fig:nStar_Freq_FC}
\end{figure}

The analytic results for these quantities compare quite well with the 
numerical findings estimated for different distribution intervals 
$[l_1,l_2]$, different coupling strengths and system sizes, as shown in 
Fig.~\ref{fig:nStar_Freq_FC}.  Apart for definitely large coupling $g > 10$
where some discrepancies among the mean field estimations and 
the simulation results for $\bar \nu$ are observable (see Fig.~\ref{fig:nStar_Freq_FC} (b)).
These differences are probably due to the discreteness of the pulses,
which cannot be neglected for very large synaptic strengths.

As a general feature we observe that $n_A$ is steadily decreasing with $g$,
thus indicating that a group of neurons with higher effective inputs ({\it  winners}) silence the 
other neurons ({\it losers}) and that the number of
{\it winners} eventually vanishes for sufficiently large coupling in the limit of large system sizes.
Furthermore, the average excitability of the active neurons (the {\it winners}) $\bar I_A$ 
increases with $g$, as shown in the inset of Fig.~\ref{fig:nStar_Freq_FC} (a), 
thus revealing that only the neurons with higher excitabilities  survive to the
silencing action exerted by the other neurons. At the same time, as an effect of the 
growing inhibition the average firing rate of the {\it winners} dramatically slows down. 
Therefore despite the increase of $\bar I_A$ the average effective input $\bar \mu$ indeed decreases
for increasing inhibition.  
This represents a clear example of the winner takes all (WTA) mechanism
obtained via (lateral) inhibition, which has been shown to have biological
relevance for neural systems~\cite{yuille1989,ermentrout1992,fukai1997,plenz2003}.

\begin{figure}
\centering
\includegraphics[width=0.95\linewidth]{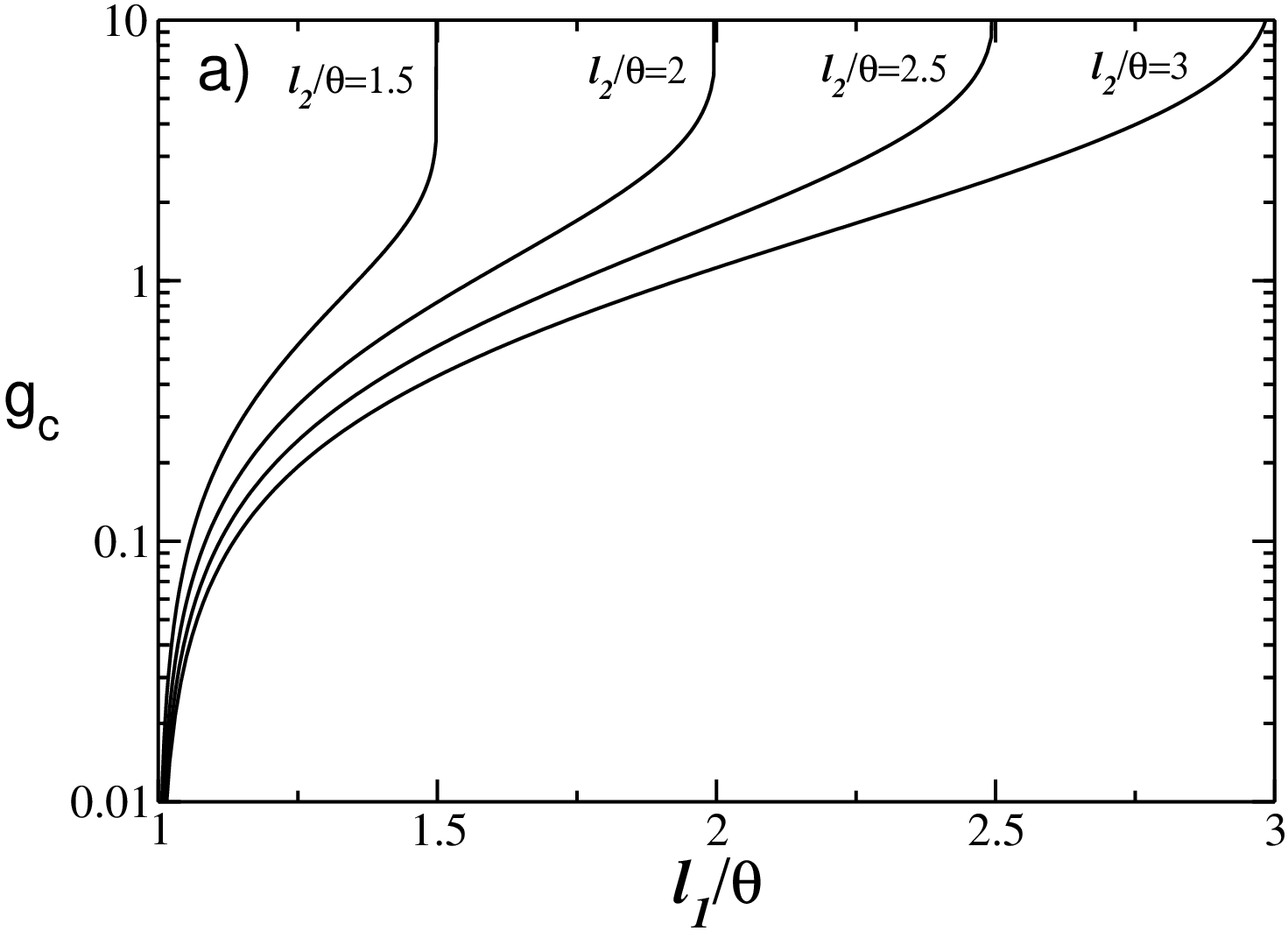}
\includegraphics[width=0.95\linewidth]{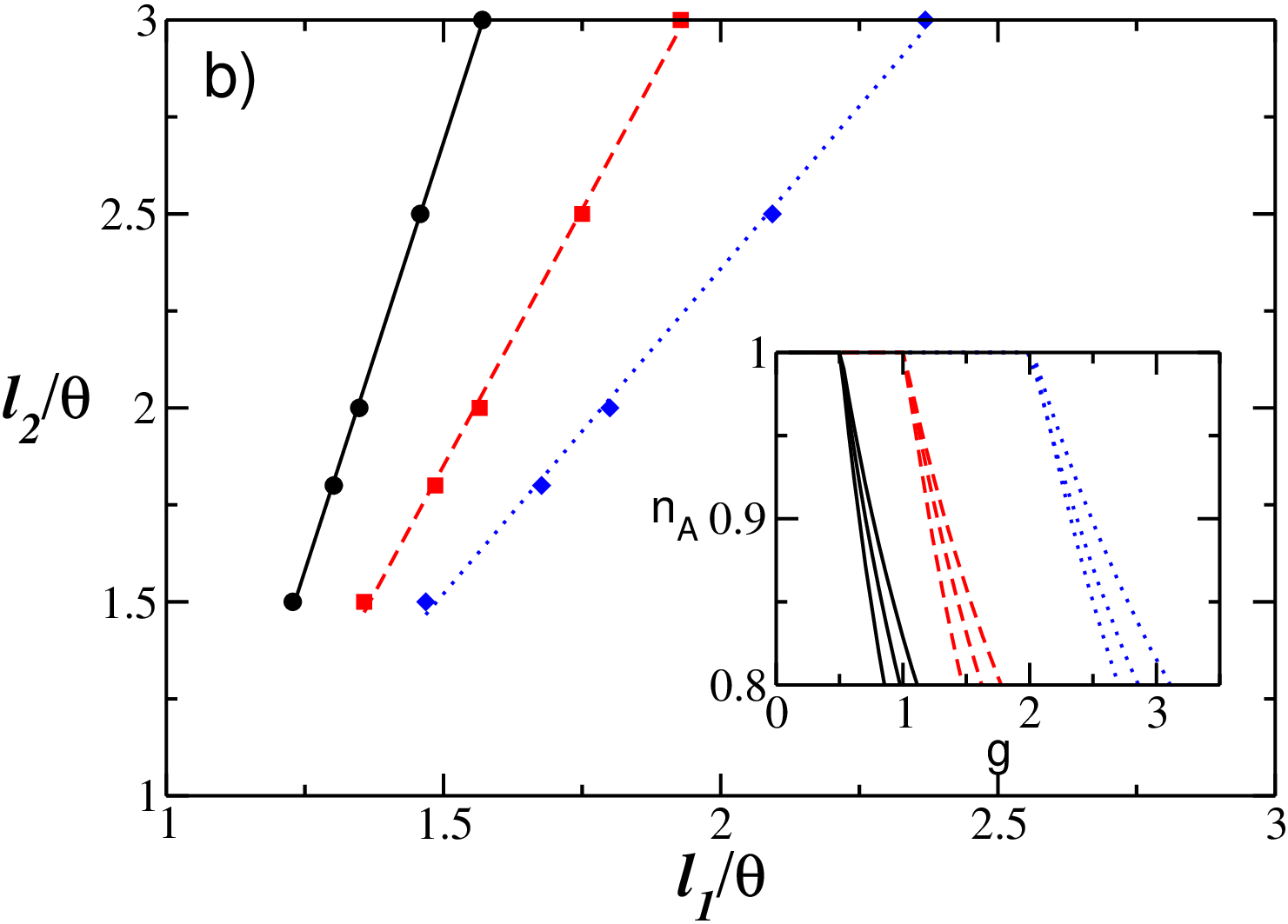}
\caption{a) Critical value $g_c$ as a function of the lower value of
the excitability $l_1$  for several choices of the upper limit
$l_2$. b) Isolines corresponding to constant values of $g_c$
in the $(l_1,l_2)$-plane: namely, $g_c = 0.5$ (black solid line), 
$g_c = 1.0$ (red dashed line), $g_c = 2.0$ (blue dotted line).
Inset: Dependence of $n_A$ on $g$ for three couples of values $(l_1,l_2)$
chosen along each of the isolines reported in the main figure.
} 
\label{fig:gCritic_FC}
\end{figure}

It is important to understand which is the minimal coupling value $g_c$ for which
the firing neurons start to die. In order to estimate $g_c$
it is sufficient to set $n_A=1$ in Eqs. \eqref{eq:unif_nu} and \eqref{eq:nStar_FC}.
In particular, one gets
\begin{equation}
g_c = (l_1-\theta)/\bar \nu \quad ,
\label{gc}
\end{equation}
thus for $l_1=\theta$ even an infinitesimally small coupling is in principle 
sufficient to silence some neurons.  
Furthermore, from Fig.~\ref{fig:gCritic_FC} (a) it is evident that
whenever the excitabilities become homogeneous, i.e. for $l_1 \to l_2$, the critical
synaptic coupling $g_c$ diverges towards infinity. 
Thus heterogeneity in the excitability distribution is a necessary condition in order 
to observe a gradual neurons' death, as shown in Fig.~\ref{fig:nStar_Freq_FC} (a).

This is in agreement with the results reported in ~\cite{bressloff2000},
where homogeneous fully coupled networks of inhibitory LIF neurons have
been examined. In particular, for finite systems and slow synapses 
the authors in~\cite{bressloff2000} 
reveal the existence of a sub-critical Hopf bifurcation from a fully 
synchronized state to a regime characterized by oscillator death 
occurring at  some critical  $g_c$. However, in the thermodynamic limit 
$g_c \to \infty$ for fast as well as slow synapses, in agreement
with our mean field result for instantaneous synapses.

We also proceed to investigate the isolines corresponding to the same critical $g_c$ 
in the $(l_1,l_2)$-plane, 
the results are reported in Fig.~\ref{fig:gCritic_FC} (b) for three selected values of $g_c$. 
It is evident that the $l_1$ and $l_2$-values associated to the isolines
display a direct proportionality among them. 
However, despite lying on the same $g_c$-isoline, different 
parameter values induce a completely different 
behaviour of $n_A$ as a function of the synaptic strength, as shown in the inset
of Fig.~\ref{fig:gCritic_FC} (b).

Direct simulations of the network at finite sizes, namely for $N =400$ and $N=1400$,
show that for sufficiently large coupling neurons with similar
excitabilities tend to form clusters, similarly to what reported 
in~\cite{Luccioli2010Irregular}, for the same model here studied,
but with a delayed pulse transmission.  However, at variance with
\cite{Luccioli2010Irregular}, the overall macroscopic dynamics is asynchronous 
and no collective oscillations can be
detected for the whole range of considered synaptic strengths.

\section{Sparse Networks : \\ Neurons' rebirth}
\label{sec:sparse}

In this Section we will consider a network with sparse connectivity, namely each neuron 
is supra-threshold and it receives instantaneous IPSPs from 
$K << N$ randomly chosen neurons in the network. 
Due to the sparseness, the input spike trains can be considered as uncorrelated and 
at a first approximation it can be assumed that each spike train is Poissonian 
with a frequency $\bar \nu$ correspondent to the average firing rate of the neurons
in the network~\cite{BrunelHakim1999,Brunel2000Sparse}. Usually, the mean 
activity of a LIF neural network has been estimated in the context of 
the diffusion approximation~\cite{ricciardi2013diffusion,tuckwell2005}.
This approximation is valid whenever the arrival frequency of the IPSPs
is high with respect to the firing emission, while 
the amplitude of each IPSPs (namely, $G =g/K$) is small 
with respect to the firing threshold $\theta$. This latter hypothesis 
in our case is not valid for sufficiently large (small) synaptic strength $g$
(in-degree $K$), as it can be appreciated by the 
comparison shown in Fig.~\ref{fig:shot_diffusion} in Appendix B.
Therefore the synaptic inputs should be treated as shot noise.
In particular, here we apply an extended version of the analytic approach derived by Richardson and Swabrick 
in~\cite{richardson2010firing} to estimate the average firing rate and the 
average coefficient of variation for LIF neurons with instantaneous
synapses subject to inhibitory shot noise 
of constant amplitude (for more details see Appendices B and C).
 
At variance with the fully coupled case, the fraction of active neurons
$n_A$ does not saturate to a constant value for sufficiently short times.
Instead, $n_A$ increases in time, due to the rebirth of losers 
which have been previously silenced by the firing activity 
of the winners, as shown in  in Fig.~\ref{fig:SilecingNeuron_FC_Sparse}(b). 
This effect is clearly illustrated by considering the probability distributions 
$P(\nu)$
of the firing rates of the neurons at successive integration times $t_S$. These
are reported in the insets of Fig~\ref{fig:SilecingNeuron_FC_Sparse}(b)
for two coupling strengths and two times: namely, $t_S = 985$ (red lines)
and $t_S = 1 \times 10^6$ (green lines).
From these data is evident that the fraction of neurons with low
firing rate (the losers) increases in time, while the fraction of high
firing neurons remains almost unchanged. Moreover,
the variation of $n_A$ slows down for increasing  $t_S$
and $n_A$  approaches some apparently asymptotic profile
for sufficiently long integration times. Furthermore, $n_A$ has a non 
monotonic behaviour with $g$, as opposite to the fully coupled case. 
In particular, $n_A$ reveals a minimum $n_{A_m}$ at some intermediate
synaptic strength $g_m$ followed by an increase towards $n_A = 1$
at large $g$. As we have verified, as far as $ 1 < K << N$ finite size
effects are negligible and the actual value of $n_A$ depends only on 
the in-degree $K$ and the considered simulation time $t_S$.
In the following we will try to explain the origin of such a behaviour.

\begin{figure}
\centering
\includegraphics[width=0.95\linewidth]{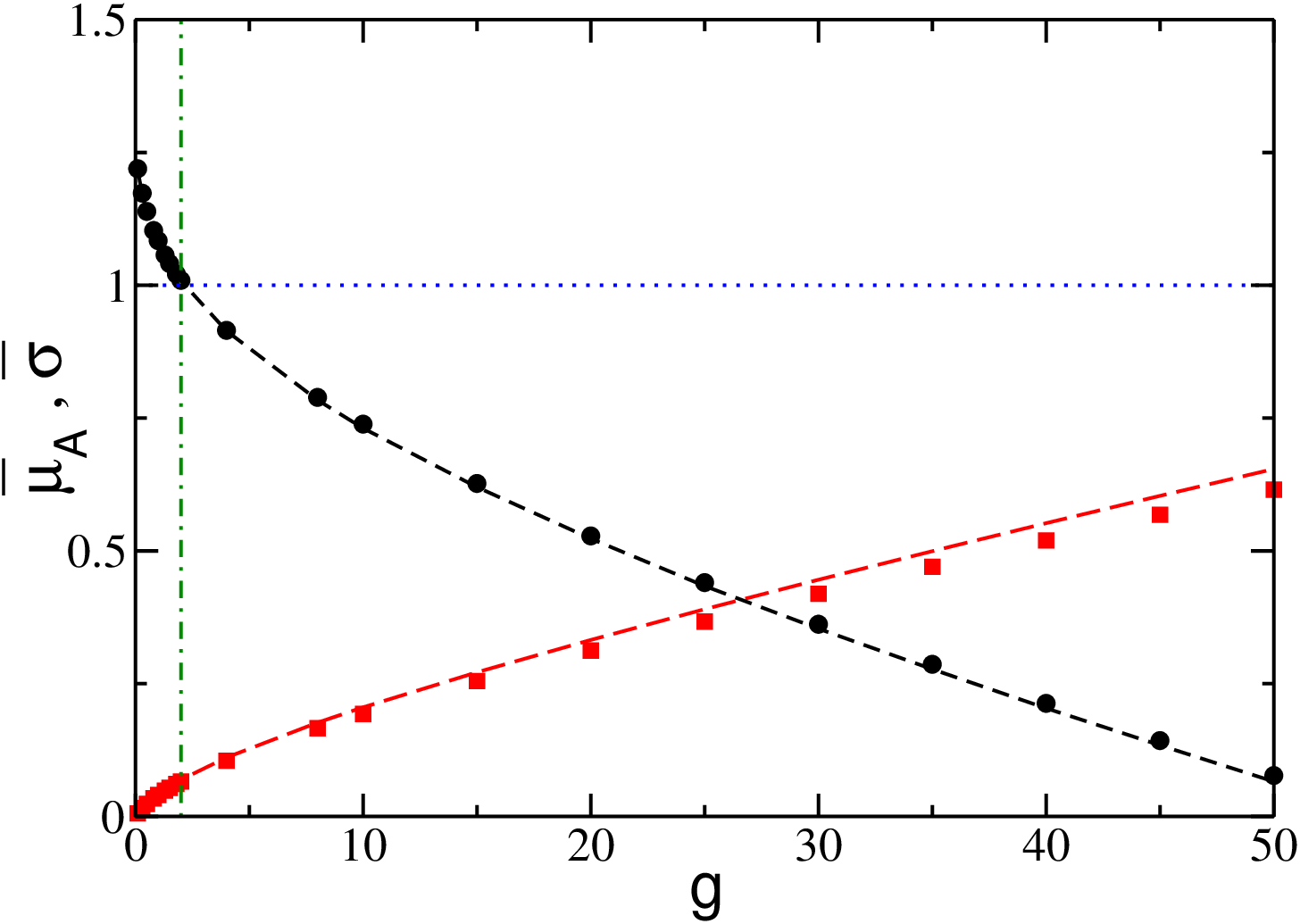}
\caption{Effective average input of the active neurons ${\bar \mu_A }$ (black circles) 
and average fluctuations of the synaptic currents ${\bar \sigma}$ (red squares) as a function
of the inhibitory coupling $g$. The threshold potential $\theta = 1$ 
is marked by the (blue) horizontal dotted line and $g_{m}$ by the (green) vertical dash-dotted line. 
The dashed black (red) line refer to the theoretical estimation for $\mu_A$ ($\sigma$)
reported in  Eq.~\eqref{eq:eff_input_GC}
 (Eq.~\eqref{eq:sigma}) and averaged only over the active neurons.
The data refer to $N = 1400$, $K = 140$, $[l_1,l_2]=[1.0:1.5]$ and to a simulation
time $t_S = 1 \times 10^6$.} 
\label{fig:neuron_react}
\end{figure}

\begin{figure}
\centering
\includegraphics[width=0.48\linewidth]{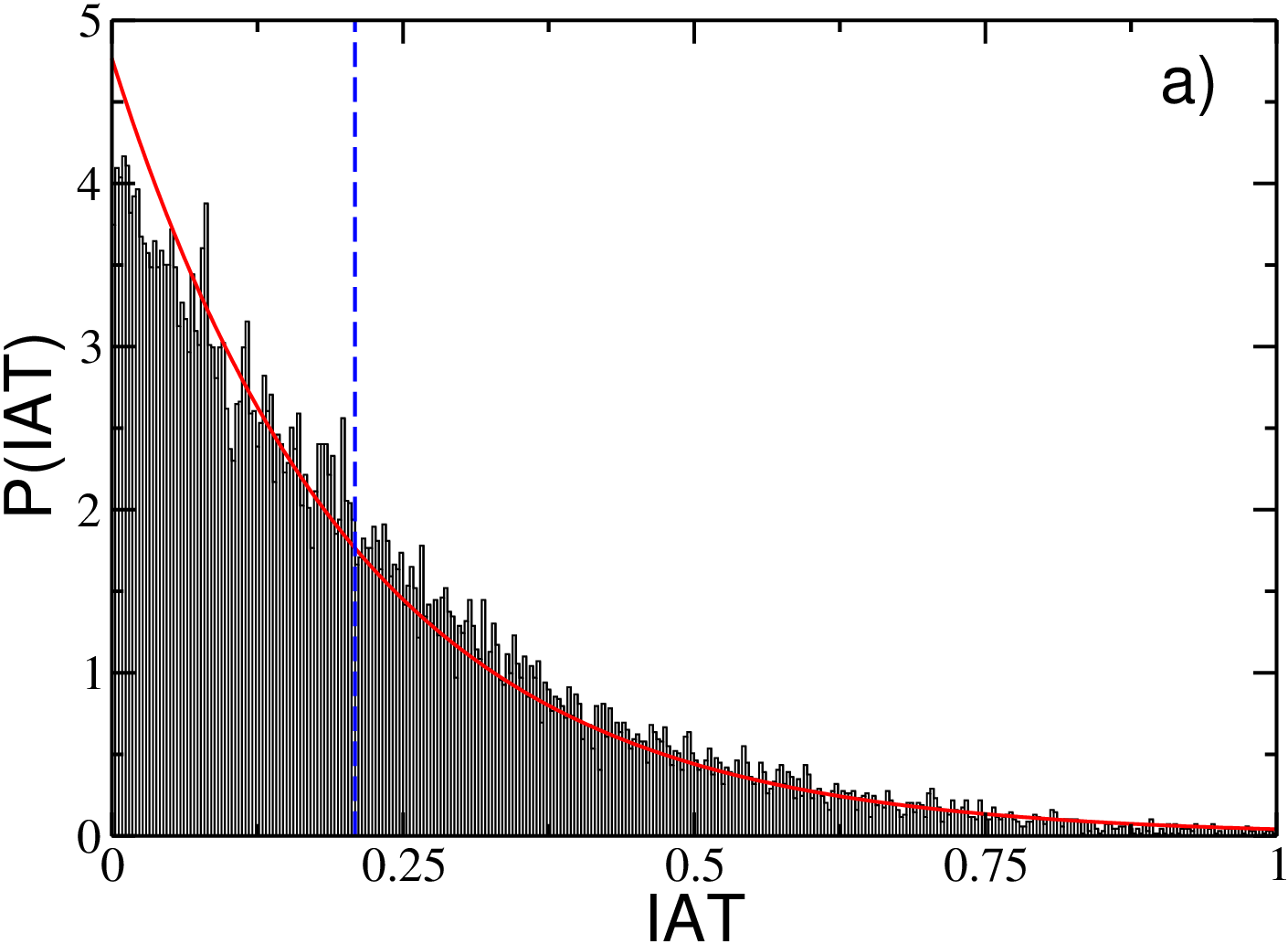}
\includegraphics[width=0.48\linewidth]{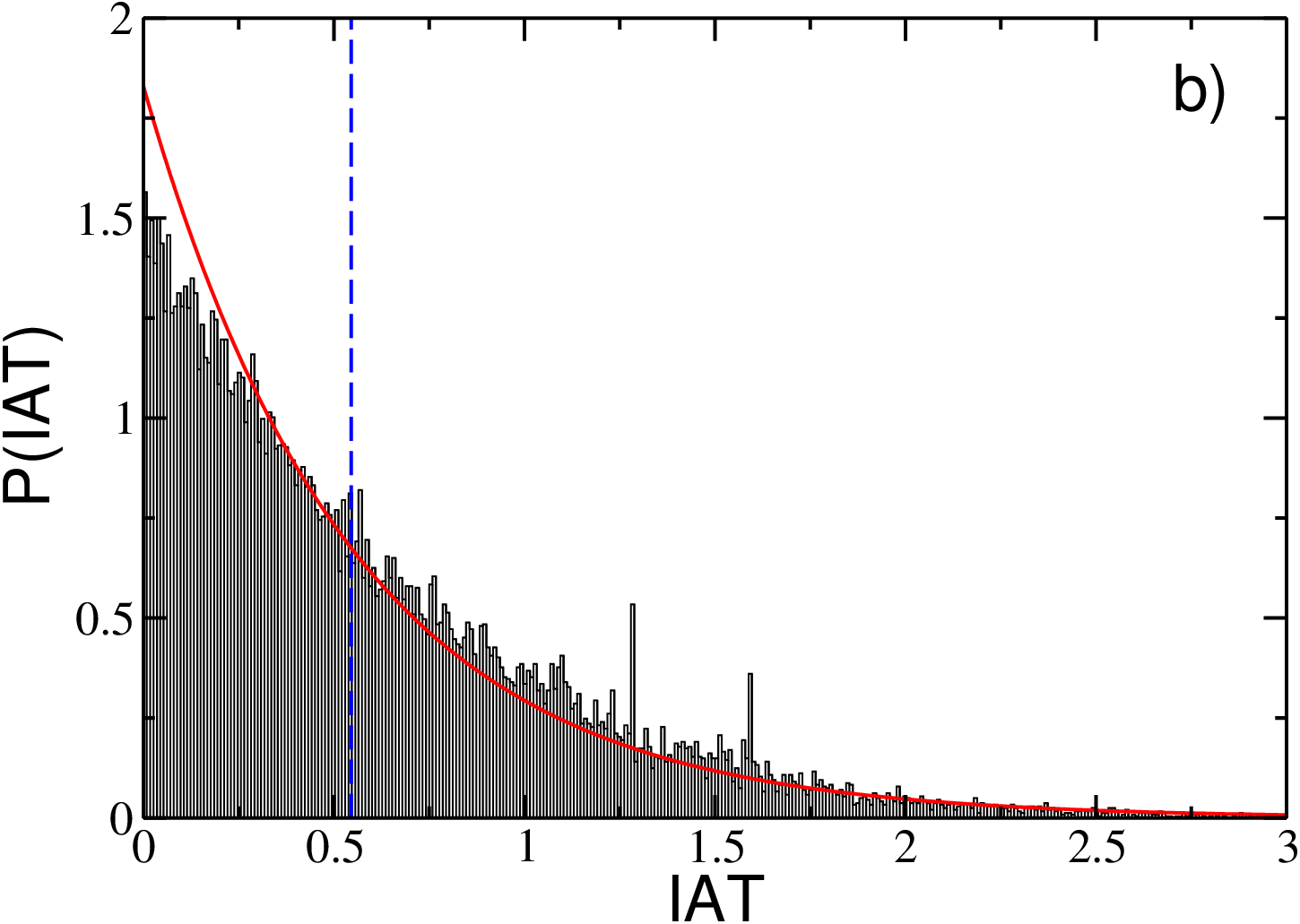}
\caption{Probability distributions of the inter-arrival times (IATs) for a 
generic neuron in 
the network for a) $g=1.3$ and b) $g = 10$. In both panels, the red continuous 
line indicate the exponential distribution corresponding to a purely Poissonian process 
with arrival rate given by $R = n_A K {\bar \nu}$ and the dashed blue vertical lines refer to 
the average IAT for the Poissonian distribution, namely $1/R$.
The distributions have been evaluated for the arrival of $5\times 10^5$ IPSPs. 
Other parameters used for the simulation as in 
Fig.~\ref{fig:SilecingNeuron_FC_Sparse} (b).}
\label{fig:IAI_distribution}
\end{figure}

Despite the model is fully deterministic, due to the random connectivity 
the rebirth of silent neurons can be interpreted in the framework of activation
processes induced by random fluctuations. In particular, we can assume that each neuron in the network will receive 
$n_A K$ independent Poissonian trains of inhibitory kicks of constant amplitude $G$ 
characterized by an average frequency $\bar \nu$ , thus each synaptic input
can be regarded as a single Poissonian train with total frequency $R=n_A K \bar \nu$.
Therefore,  each neuron, characterized by its own excitability $I$,
will be subject to an average effective input $\mu(I)$
(as reported in Eq.~ \eqref{eq:eff_input_GC}) plus fluctuations in the synaptic current 
of intensity
\begin{equation}
\label{eq:sigma}
\sigma = g \sqrt{\frac{n_A \bar{\nu}}{K}} \qquad .
\end{equation}
Indeed, we have verified that \eqref{eq:sigma} gives a quantitatively
correct estimation of the synaptic current fluctuations
over the whole range of synaptic coupling here considered
(as shown in Fig.~\ref{fig:neuron_react}).
A closer analysis of the probability distributions $P(IAT)$ of the inter-arrival times (IATs)
indicates that these are essentially exponentially distributed, as expected for Poissonian processes,
with a decay rate given by $R$, as evident from Fig.~\ref{fig:IAI_distribution} for two
different synaptic strenghts.

For instantaneous IPSP, the current fluctuations are due to stable chaos~\cite{politi2010stable},
since the maximal Lyapunov exponent is negative for the whole range of coupling, as we have verified. 
Therefore, as reported by many authors, erratic fluctuations in  inhibitory neural networks with instantaneous synapses
are due to finite amplitude instabilities, while at the infinitesimal level the system
is stable~\cite{Zillmer2006,JhankeTimme2008PRL,Timme2009ChaosBalance,Luccioli2010Irregular,
MonteforteBalanced2012,angulo2014,ullner2016self}.

In this picture, the silent neurons stay in a quiescent state corresponding 
to the minimum of the effective potential ${\cal U}(v) = v^2/2- \mu v$
and in order to fire they should overcome a barrier $\Delta {\cal U} = (\theta - \mu)^2/2$.
The average time $t_A$ required to overcome such barrier can
be estimated accordingly to the Kramers'  theory for 
activation processes~\cite{hanggi1990reaction, tuckwell2005}, namely
\begin{equation}
\label{eq:kramer_eq1}
t_A \simeq  \tau_0 \exp\left({\frac{\left(\theta - \mu({I})\right)^2}{\sigma^2}}\right) \; ;
\end{equation}
where $\tau_0$ is an effective time scale taking in account the intrinsic non stationarity of the process, i.e. 
the fact that the number of active neurons increases during the time evolution.

It is important to stress that the expression \eqref{eq:kramer_eq1} will remain valid also in the limit of large synaptic couplings,
because not only $\sigma^2$, but also the barrier height will increase with $g$.
Furthermore, both these quantities grow quadratically with $g$ at sufficiently large
synaptic strength, as it can be inferred from Eqs.~\eqref{eq:eff_input_GC} and
\eqref{eq:sigma}.
 
It is reasonable to assume that at a given time $t_S$ all neurons with $t_A < t_S$ 
will have fired at least once and that the more excitable will fire first.
Therefore by assuming that the fraction
of active neurons at time $t_S$ is $n_A(t_S)$, the last neuron which has fired
should be characterized  by the following excitability 
\begin{equation}
\label{eq:nStar_Ihat}
\hat{I} = l_2 -n_A(t_S) (l_2 - l_1)\;  ;
\end{equation}
for excitabilities $I$ uniformly distributed in the interval $[l_1,l_2]$.
In order to obtain an explicit expression for the fraction of active neurons at time $t_S$,
one should solve the equation \eqref{eq:kramer_eq1}  for the neuron
with excitability $\hat I$ by setting $t_S = t_A$, thus obtaining
the following solution

\begin{equation}
\label{eq:kramer_eqNstar}
n_A(t_S) = \left\{\begin{array}{lr}
\dfrac{\phi - 2\beta\gamma + \sqrt{\phi^2 - 4 \phi \beta \gamma}}{2 \gamma^2} & \mbox{if } n_A<1 \\
1 &	\text{otherwise}
\end{array}
\right.
\end{equation}
where
$$
\gamma=(l_2-l_1)+g \bar{\nu} \quad \phi = \frac{g^2}{K} \bar{\nu}\ln(t_S/\tau_0) \quad \beta = \theta - l_2 \enskip .
$$ 

\begin{figure}
\centering
\includegraphics[width=0.95\linewidth]{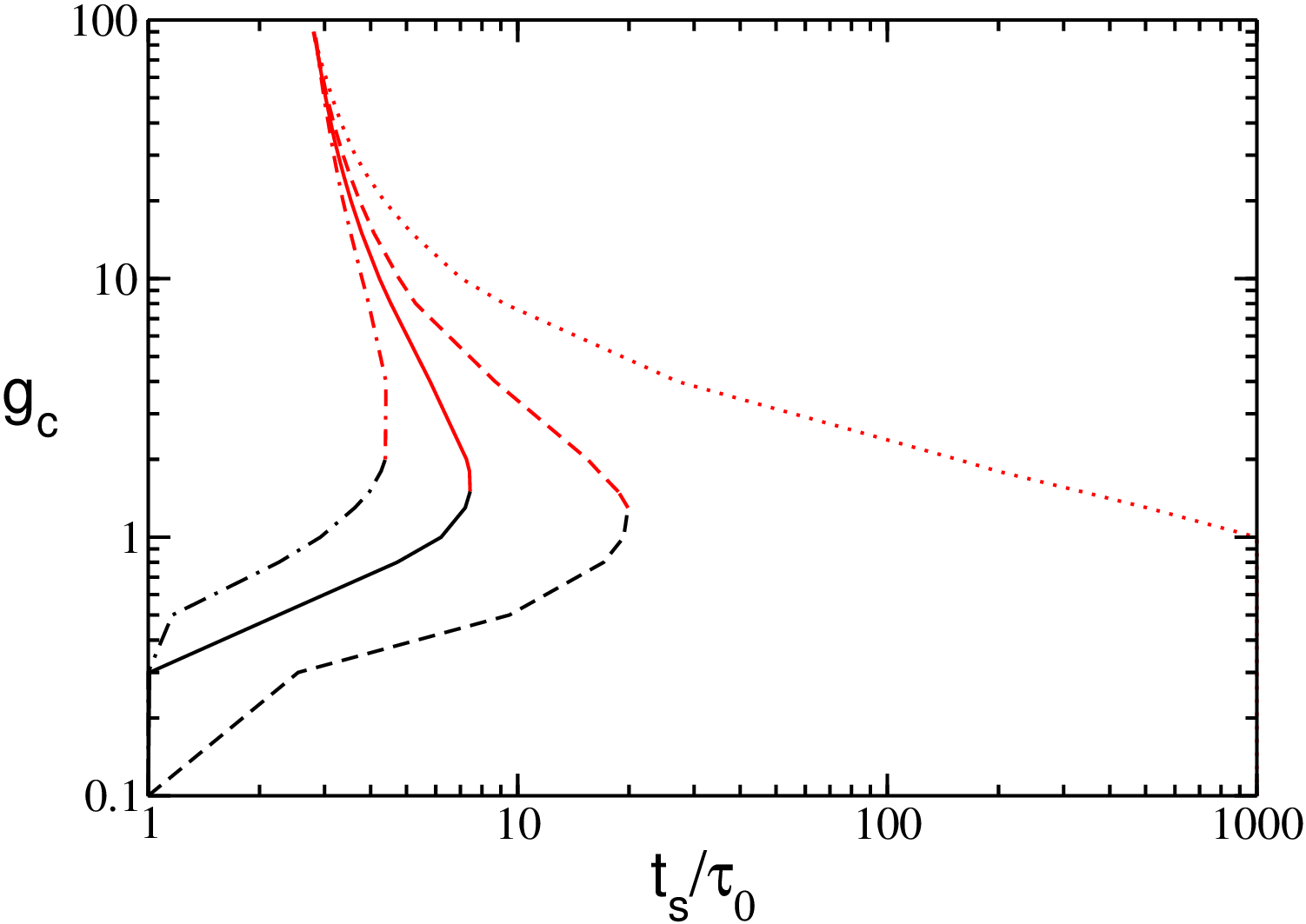}
\caption{Critical values $g_{c1}$ (black) and $g_{c2}$ (red) as calculated from Eq. 
\eqref{eq:kramer_eqNstar} for $l_1 = 1.2$ (dash-dotted), $l_1 = 1.15$ (continuous), $l_1 = 1.1$ 
(dashed) and $l_1 = 1.0$ (dotted line). All the values entering in Eq. \eqref{eq:kramer_eqNstar} are 
taken from simulation. 
All other parameters used for the simulation as in Fig.~\ref{fig:SilecingNeuron_FC_Sparse} (b).}
\label{fig:gc_sparse}
\end{figure}

Equation \eqref{eq:kramer_eqNstar} gives the dependence of $n_A$
on the coupling strength $g$ for a fixed integration time $t_S$ and time scale $\tau_0$, 
whenever we can provide the value of the average frequency $\bar \nu$.
A quick inspection to Eqs. \eqref{eq:kramer_eq1} and \eqref{eq:kramer_eqNstar} 
shows that, setting $n_A = 1$, we obtain two solutions for the critical couplings  $g_{c1}$ ($g_{c2}$) 
below (above) which all neurons will fire at least once in the cosidered time interval.
The solutions are reported in Fig~\ref{fig:gc_sparse}, in particular
we observe that whenever $l_1 \to v_{th}$ the critical coupling $g_{c1}$ will vanish,
analogously to the fully coupled situation.
These results clearly indicate that $n_A$ should display a minimum for some finite coupling
strenght $g_m \in (g_{c1},g_{c2})$. Furthermore,
as shown in Fig~\ref{fig:gc_sparse} the two critical couplings
approach one another for increasing $t_S$ and finally merge, indicating
that at sufficiently long times all the neurons will be active at any synaptic coupling strength $g$.

The average frequency $\bar \nu$ can be obtained analytically by following the 
approach described in Appendix B for LIF neurons with instantaneous
synapses subject to inhibitory Poissonian spike trains.
In particular, the self-consistent expression for the average frequency reads as
\begin{equation}
\label{eq:nu_sparse}
\bar{\nu} =  \int_{\{I_A\}} dI P(I) \nu_0(I,G,n_A,\bar \nu) \quad ;
\end{equation}
where the explicit expression of $\nu_0$ is given by Eq.~\eqref{eq:nu0} in Appendix B.  

\begin{figure*}
\centering
\includegraphics[width=0.45\linewidth]{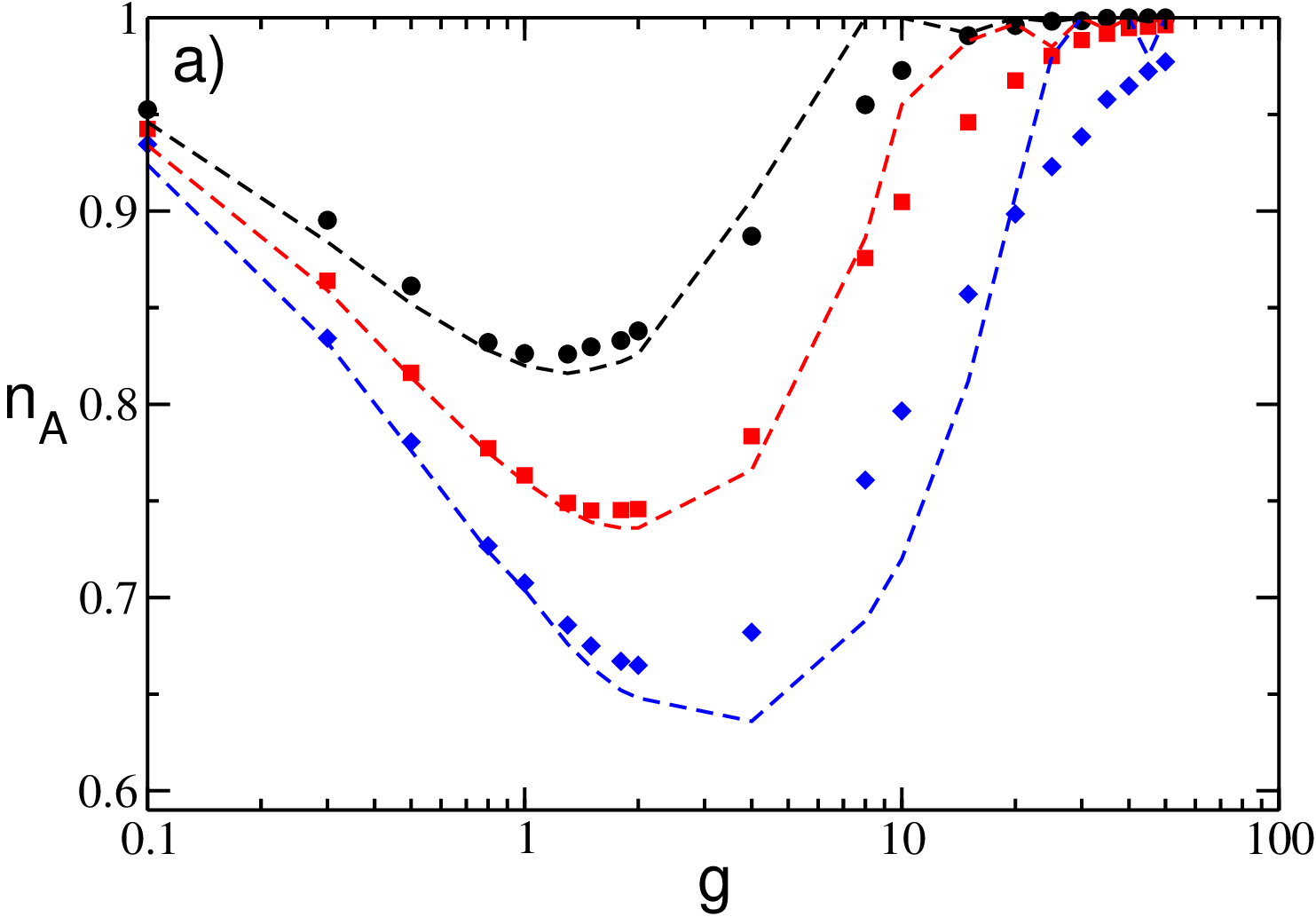}
\includegraphics[width=0.45\linewidth]{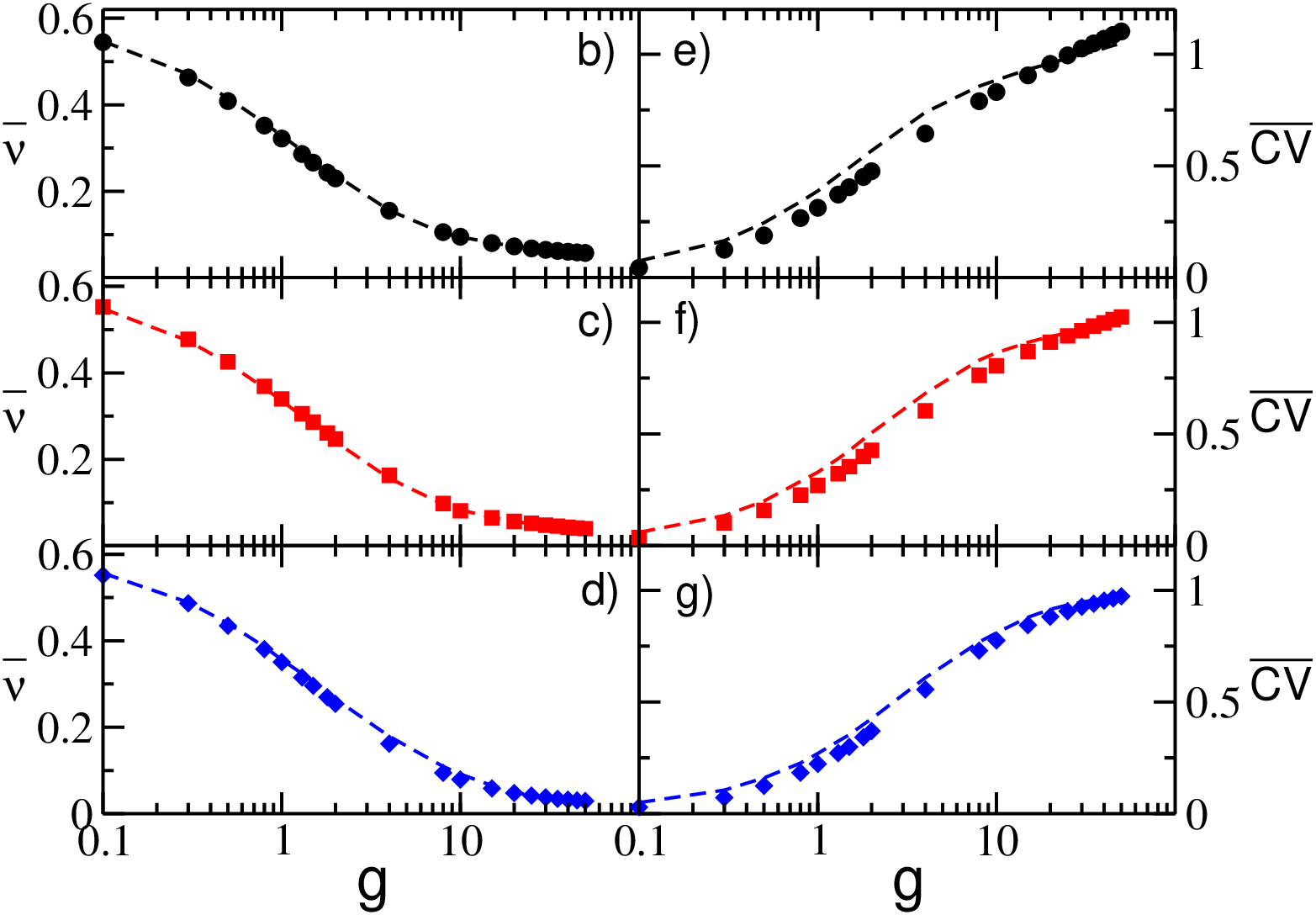}
\includegraphics[width=0.3\linewidth]{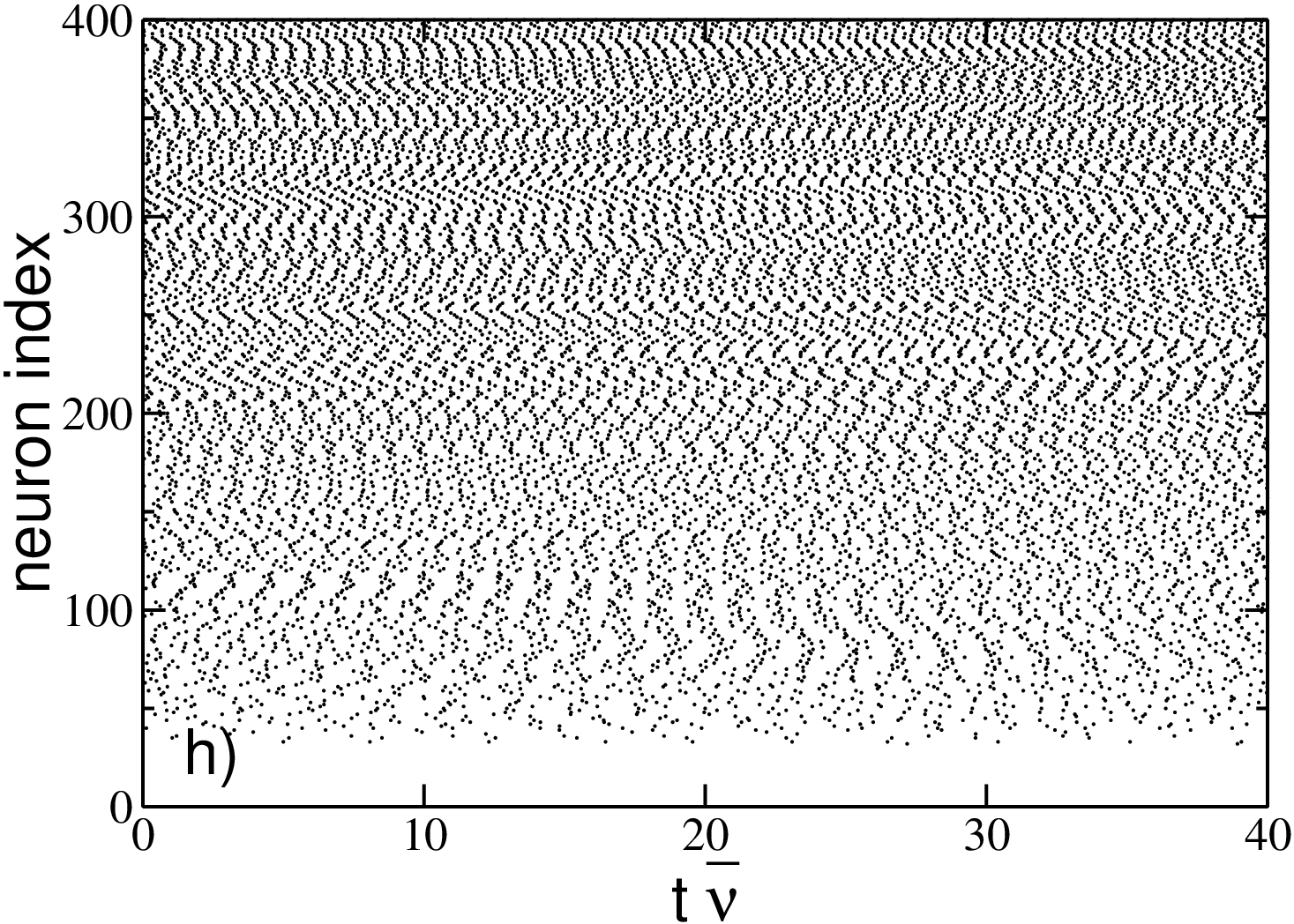}
\includegraphics[width=0.3\linewidth]{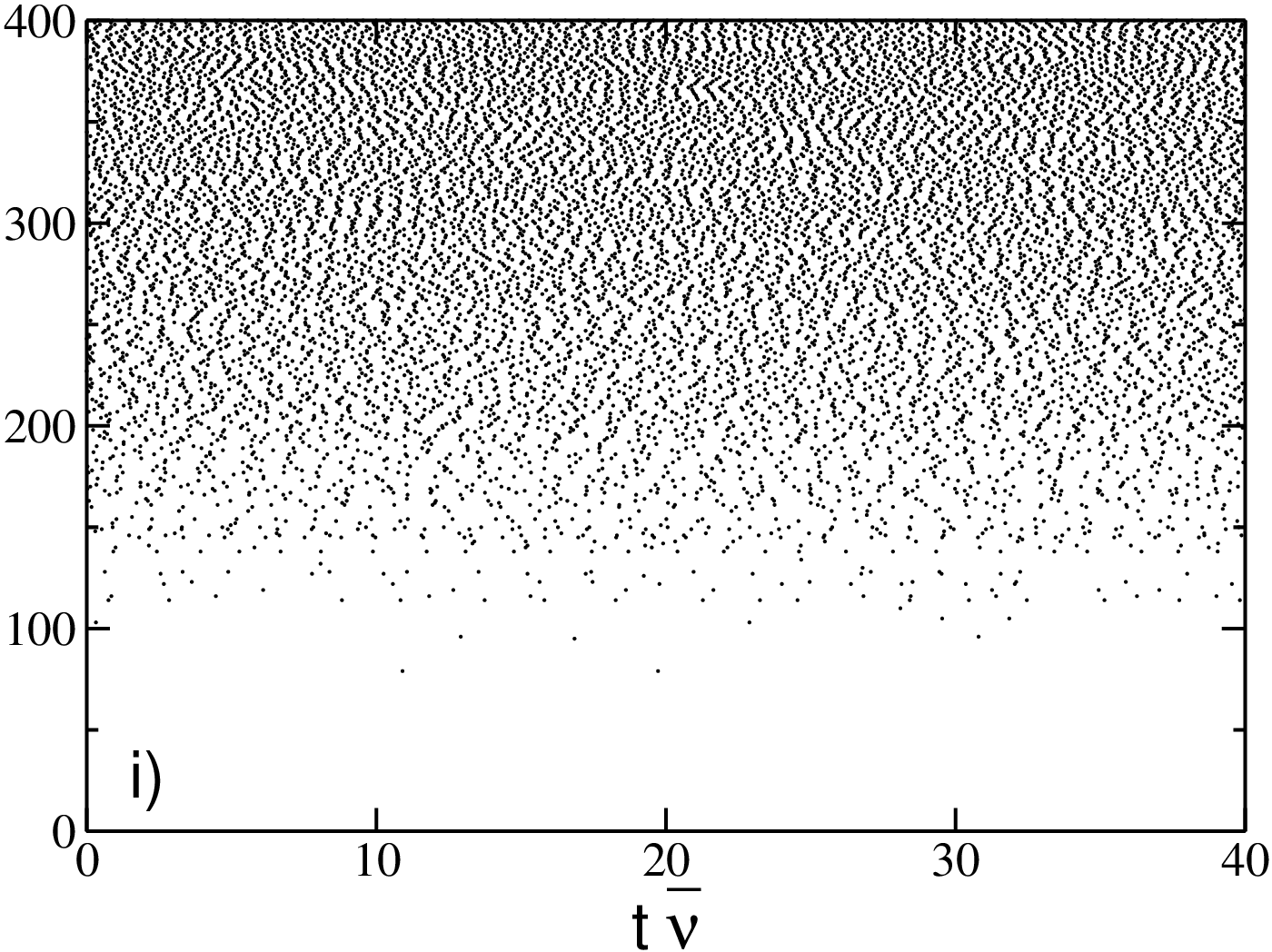}
\includegraphics[width=0.3\linewidth]{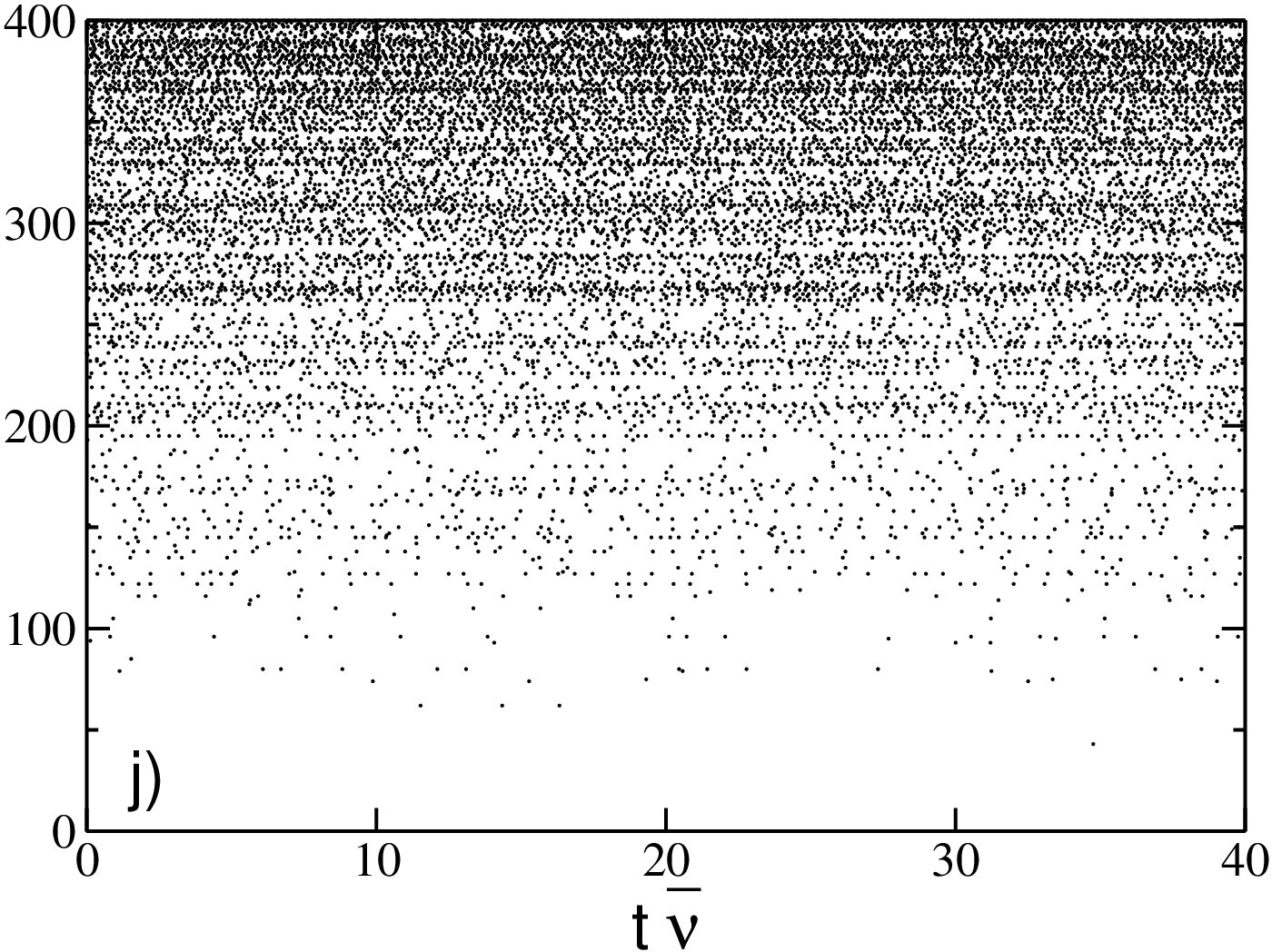}
\includegraphics[width=0.3\linewidth]{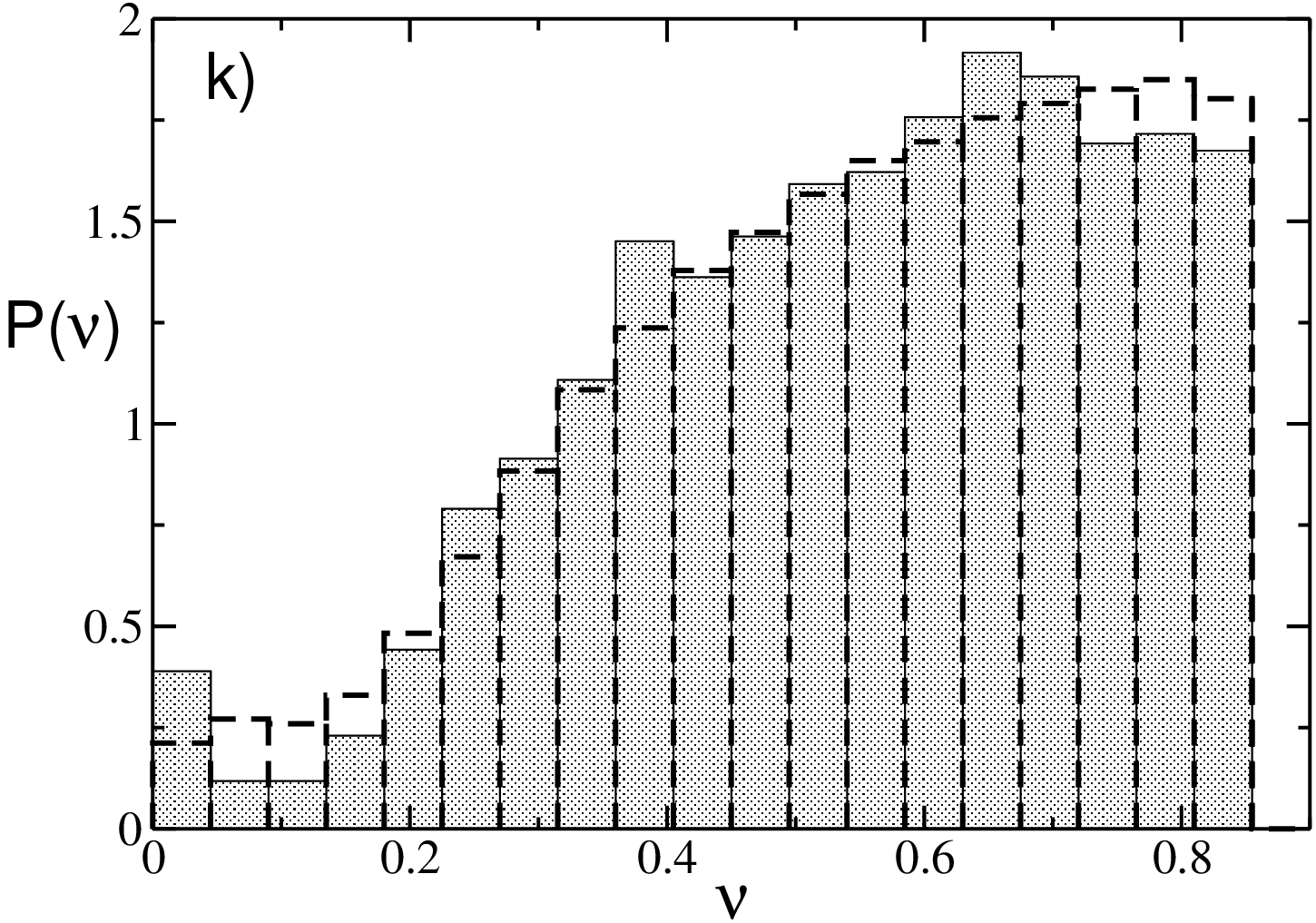}
\includegraphics[width=0.3\linewidth]{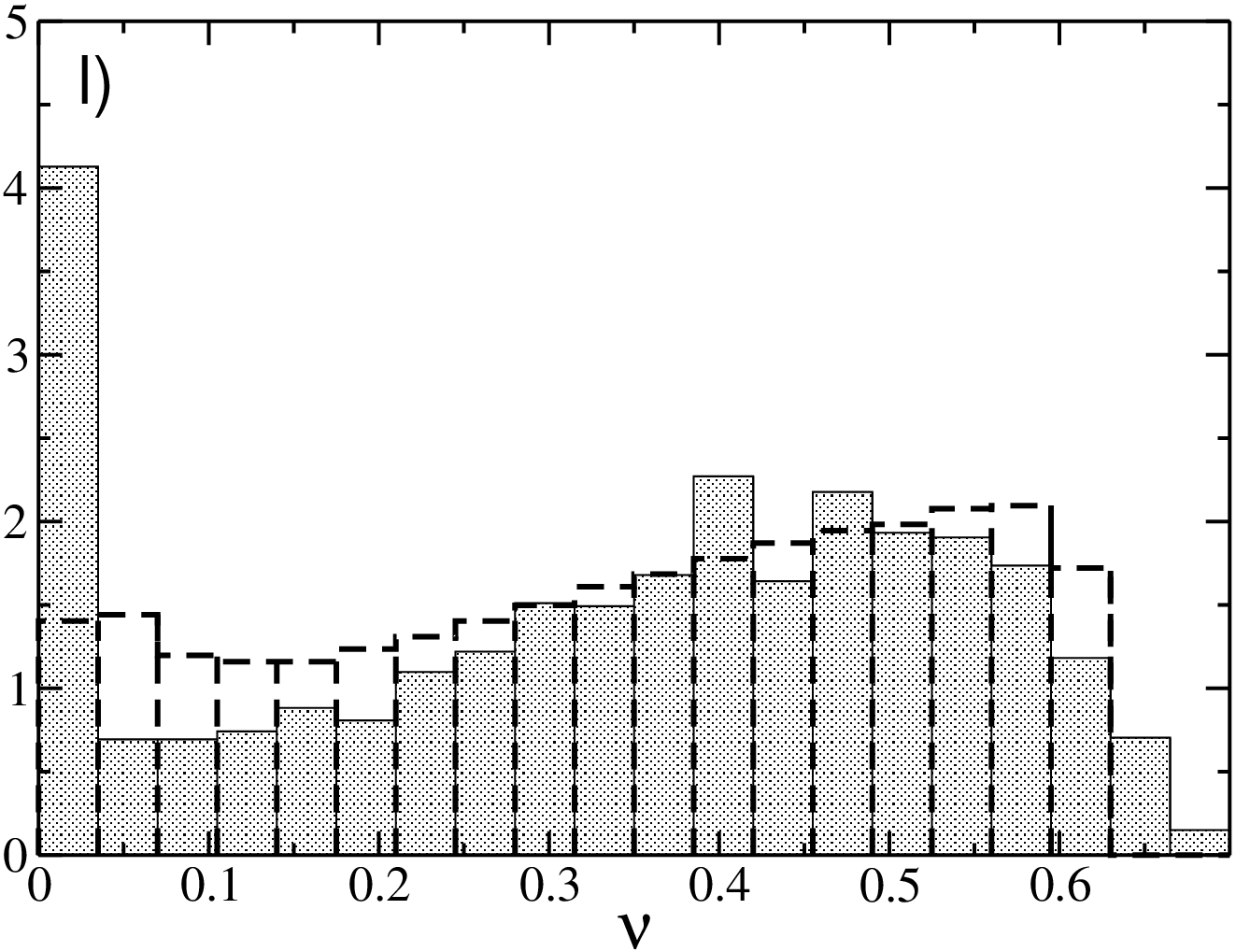}
\includegraphics[width=0.3\linewidth]{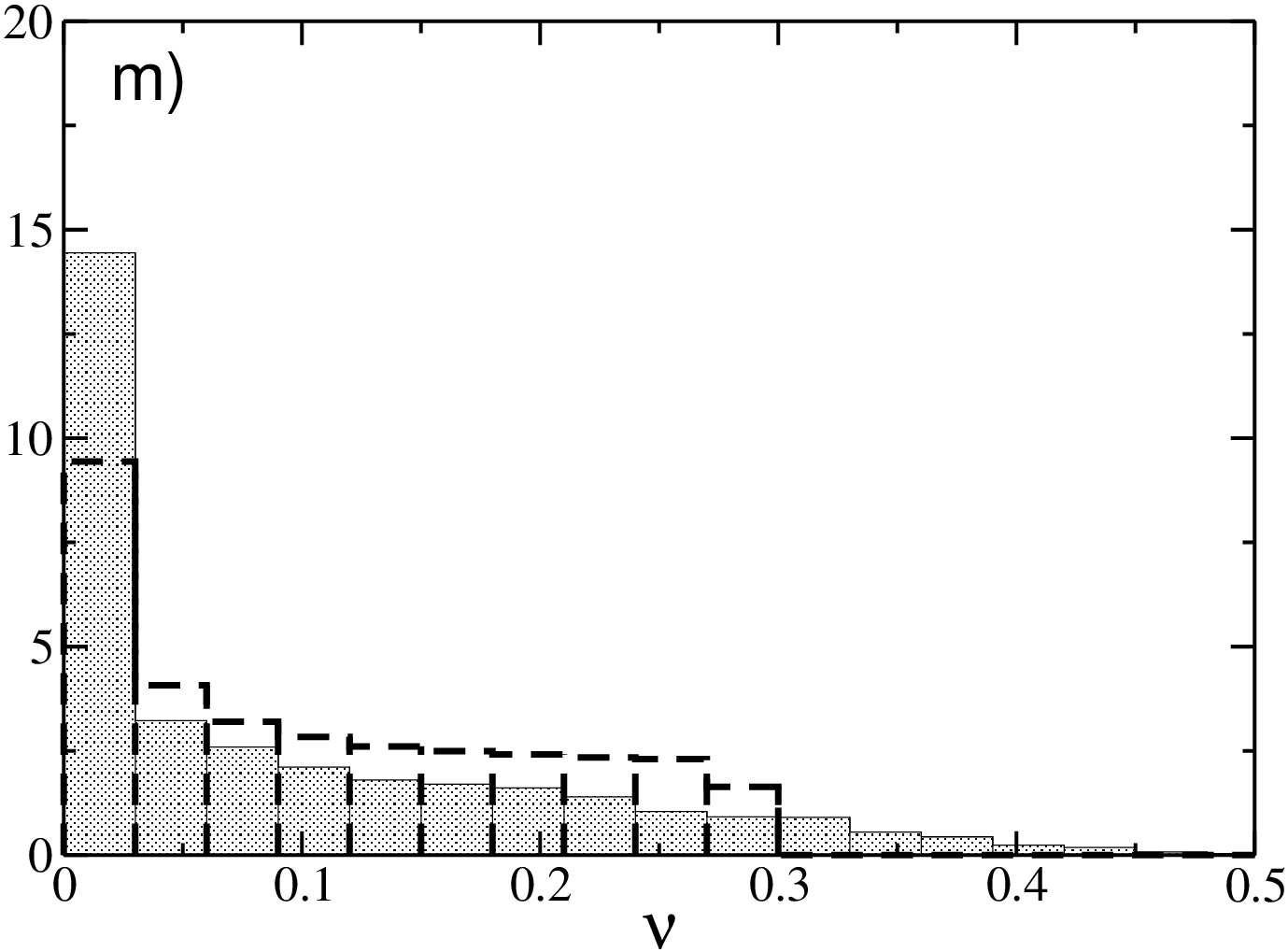}
\caption{a) Fraction of active neurons $n_A$ as a function of inhibition for several values of $K$. 
b-d) Average network's firing rate $\bar \nu$ for the same cases depicted in a),  
and the corresponding $\overline{CV}$ (e-f). In all panels, filled symbols correspond to numerical data and dashed lines to 
semi-analytic values: black circles correspond to $K=20$ ($t_s/\tau_0 = 11$), red squares to $K=40$ ($t_s/\tau_0 = 19$) 
and blue diamonds to $K=80$ ($t_s/\tau_0 = 26.6$). 
The data are averaged over a time interval $t_S=1 \times 10^6$ and 10 different realizations of the random network. h-j) 
Raster plots for three different synaptic strengths for $N=400$ and $K=40$: namely, h) $g=0.1$; i) $g=1$ and j) $g=8$. 
The corresponding value for the fraction of active neurons, average frequency and average coefficient of variation are
$n_A =  (0.94,0.76,0.88)$, $\bar \nu = (0.55,0.34,0.10)$ and $\overline{CV} = (0.04,0.27,0.76)$,
respectively. The neurons are ordered in terms of their intrinsic excitability and the time is rescaled by the average 
frequency $\bar \nu$. k-l) Probability distributions $P(\nu)$ of the the single neuron firing rate $\nu$, for the same 
values of $g$ in the panels above. 
Empty-discontinuous bars correspond to the theoretical prediction while the filled bars indicate the histogram calculated with
the simulation. The remaining parameters as in Fig. \ref{fig:SilecingNeuron_FC_Sparse} (b).
} 
\label{fig:allSparse_Several_K}
\end{figure*}

The simultaneous solution of Eqs.~\eqref{eq:kramer_eqNstar} and
\eqref{eq:nu_sparse} provides a theoretical estimation of $n_A$ and
$\bar \nu$ for the whole considered range of synaptic strength,
once the unknown time scale $\tau_0$ is fixed. 
This time scale has been determined via an optimal fitting procedure for sparse networks
with $N=400$ and $K=20$, 40 and 80 at a fixed integration time $t_S=1 \times 10^6$.
The results for $n_A$ are reported in Fig.~\ref{fig:allSparse_Several_K} (a), 
the estimated curves reproduce reasonably well the numerical data for $K=20$
and 40, while for $K=80$ the agreement worsens at large coupling strengths.
This should be due to the fact that by increasing $g$ and $K$ the spike trains
stimulating each neuron cannot be assumed to be completely independent,
as done for the derivation of Eqs.~\eqref{eq:kramer_eqNstar} and
\eqref{eq:nu_sparse}. Nevertheless,  the average frequency $\bar \nu$ is quantitatively well reproduced for the considered $K$ values over the entire range of the synaptic strengths, as it is evident from Figs.~\ref{fig:allSparse_Several_K} (b-d). A more detailed comparison 
between the theoretical estimations and the numerical data 
can be obtained by considering the distributions $P(\nu)$ of
the single neuron firing rate for different coupling strengths reported
in Figs.~\ref{fig:allSparse_Several_K} (k-m) for $K=40$. The overall
agreement can be considered as more than satisfactory, the observable 
discrepancies are probably due to the fact that our approach neglect 
a further source of disorder present in the system and 
related to the heterogeneity in the number of active
pre-synaptic neurons~\cite{BrunelHakim1999}.

We have also estimated analytically the average coefficient of variation of the
firing neurons $\overline{CV}$ by extending the method derived in~\cite{richardson2010firing}
to obtain the response of a neuron receiving synaptic shot noise inputs.
The analytic expressions  of the coefficient of variation for LIF neurons
subject to inhibitory shot noise with fixed post-synaptic amplitude are obtained 
by estimating the second moment of the associated first-passage-time distribution,
the details are reported in Appendix C. The coefficient of variation can be estimated,
once  the self consistent values for $n_A$ and $\bar{\nu}$ have been obtained 
via Eqs.~\eqref{eq:kramer_eqNstar} and \eqref{eq:nu_sparse}.
The comparison with the numerical data, reported
in Figs~\ref{fig:allSparse_Several_K} (e-g), reveals a good agreement over the whole 
range of  synaptic strengths for all the considered in-degrees.

At sufficiently small synaptic coupling the neurons fire tonically and almost independently, as it emerges clearly from the raster plot in Fig~\ref{fig:allSparse_Several_K} (h) and by the 
fact that $\bar \nu$ approaches the average value for the uncoupled 
system (namely, $0.605$) and $\overline{CV} \to 0$. 
Furthermore, the neuronal firing rates are distributed towards finite
values indicating that the inhibition as a minor role in this case, as
shown in Fig~\ref{fig:allSparse_Several_K} (k).
By increasing the coupling, $n_A$ decreases, as an effect of the inhibition more and more neurons are silenced (as evident from Fig. ~\ref{fig:allSparse_Several_K} (l)) and the average firing rate decrease,
at the same time the dynamics becomes slightly more irregular
as shown in Fig~\ref{fig:allSparse_Several_K} (i).
At large coupling $g > g_m$, a new regime appears, where almost all neurons become
active but with an extremely slow dynamics which is essentially stochastic with $\overline{CV} \simeq 1$, as testified also by the raster plot reported in Fig~\ref{fig:allSparse_Several_K} (j) and by by the firing rate distribution shown in  Fig~\ref{fig:allSparse_Several_K} (m).

\begin{figure}
\centering
\includegraphics[width=0.9\linewidth]{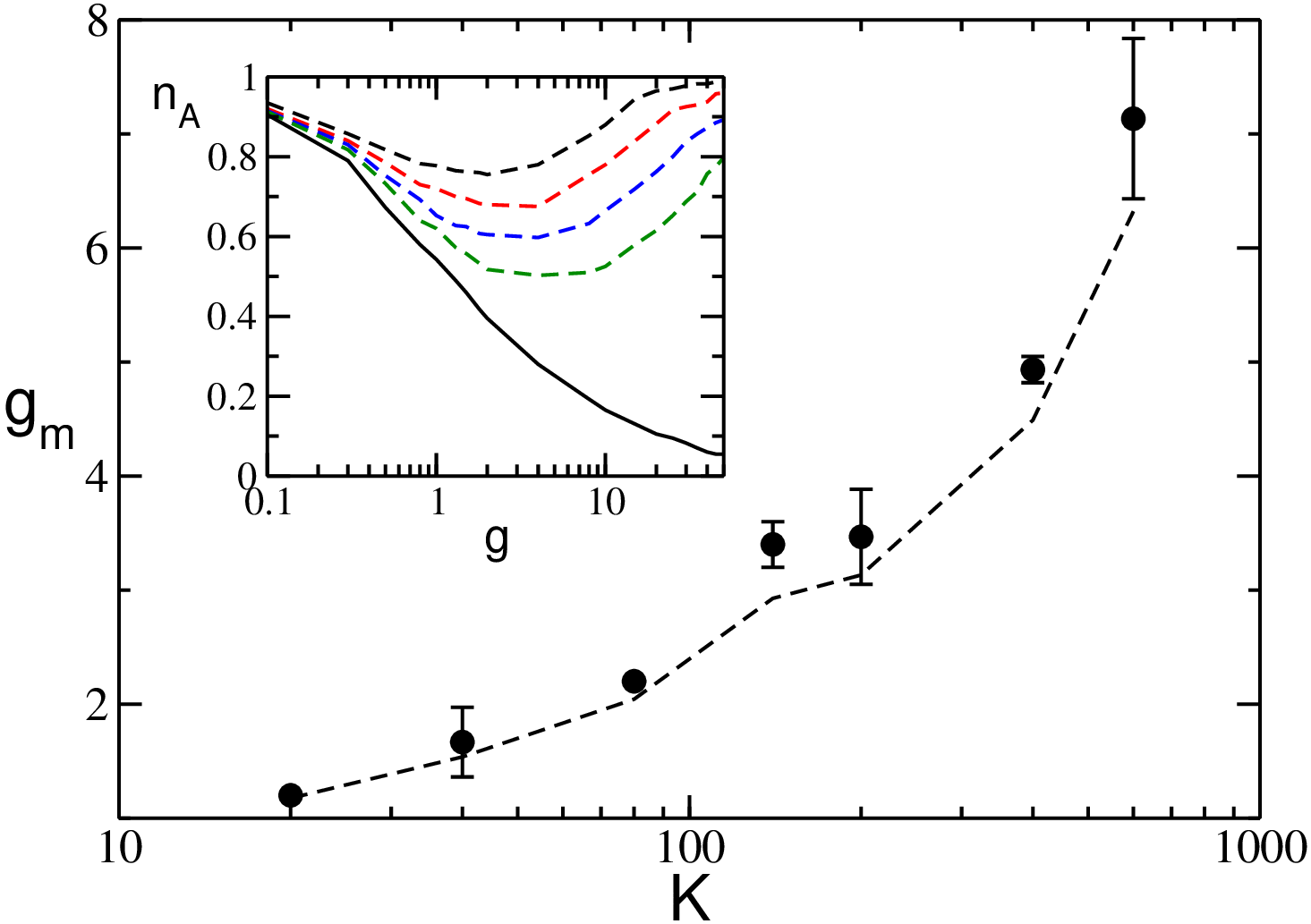}
\caption{$g_m$ as a function of the in-degree $K$.
The symbols refer to numerical data, while the dashed line to the 
expression \eqref{gmin}. Inset: $n_A$ versus $g$ for the fully coupled
case (solid black line) and for diluted networks (dashed lines), 
from top to bottom $K=20$, 40, 80, 140.
A network of size $N=1400$ is evolved during a period $t_S = 1 \times 10^5$ after discarding a transient of $10^6$ spikes, the data are averaged over 5 different random realizations of the network. 
Other parameters as in Fig.~\ref{fig:SilecingNeuron_FC_Sparse}.
} 
\label{fig:Gmin_vs_K}
\end{figure}

Furthermore, from Fig. \ref{fig:allSparse_Several_K}(a) it is clear that the value of 
the minimum of the fraction of active neurons ${n_A}_m$ decreases by increasing
the network in-degree $K$, while $g_{m}$ increases with $K$. 
This behaviour is further investigated in a larger network, namely $N=1400$,
and reported in the inset of Fig.~\ref{fig:Gmin_vs_K}. It is evident that
$n_A$ stays close to the globally coupled solutions over larger and
larger intervals for increasing $K$. 
This can be qualitatively understood by the fact that the current
fluctuations  Eq.~\eqref{eq:sigma}, responsible for the rebirth of silent neurons,
are proportional to $g$ and scales as $1/\sqrt{K}$, therefore at larger in-degree 
the fluctuations have similar intensities only for larger synaptic coupling. 

The general mechanism behind neurons' rebirth can be understood 
by considering the value of the effective neuronal input 
and of the current fluctuations as a function of $g$. 
As shown in Fig.~\ref{fig:neuron_react}, the effective
input current $\bar \mu_A$, averaged over the
active neurons, essentially coincide with the average of the 
excitability ${\bar I}_A$ for $g \to 0$, where the neurons can
be considered as independent one from the others. The inhibition leads
to a decrease of ${\bar \mu}_A$, and to a crossing of the threshold
$\theta$ exactly at $g=g_m$. This indicates that at $g < g_m$ the active neurons, being
on average supra-threshold, fire almost tonically inhibiting the losers via a WTA mechanism.
In this case the firing neurons are essentially mean-driven and the current fluctuations 
play a role on the rebirth of silent neurons only on extremely long time scales, 
this is confirmed by the low values of $\bar \sigma$ in such a range, as evident from
Fig.~\ref{fig:neuron_react}. On the other hand, 
for $g > g_m$, the active neurons are now on average below threshold
while the fluctuations dominate the dynamics. In particular, the firing
is now extremely irregular being due mainly to reactivation processes.
Therefore the origin of the minimum
in $n_A$ can be understood as a transition from a mean-driven to a fluctuation-driven
regime~\cite{renart2007}. 

A quantitative definition of $g_m$ can be given by requiring that the average  
input current of the active neurons $\bar \mu_A$ crosses the threshold $\theta$ at $g=g_m$, namely
\begin{equation*}
\bar{\mu}_A(g_m) = \bar{I}_A - g_m \bar{\nu}_m {n_A}_m = \theta \;;
\end{equation*}
where $\bar{I}_A$ is the average excitability of the firing neurons,
while ${n_A}_m$ and $\bar{\nu}_m$ are the number of active neurons and the average frequency at the minimum.

For an uniform distribution $P(I)$, a simpler expression for $g_m$ can be derived,
namely
\begin{equation}
g_{m} = \bar{\nu}^{-1}_{m} \left[ \frac{l_2 - \theta}{n_{A_{m}}} + \frac{1}{2}( l_1 - l_2  ) \right]\;.
\label{gmin}
\end{equation}
We have compared the numerical measurements of
$g_m$ with the estimations obtained by employing Eq.~\eqref{gmin},
where  $n_{A_{m}}$ and $\bar{\nu}$ are obtained from the simulations.
As shown in Fig.~\ref{fig:Gmin_vs_K} for a network of size $N=1,400$, the overall 
agreement is more than satisfactory for in-degrees ranging
over almost two decades (namely, for $20 \le K \le 600$).
This confirms that our guess (that the minimum $n_{A_{m}}$ occurs exactly at the transition from mean-driven 
to fluctuation-driven dynamics) is consistent with the numerical data for a wide range of in-degrees.

It should be stressed that, as we have verified for various
system sizes (namely, $N=700$,1400 and 2800) and for a constant
average in-degree $K=140$, for instantaneous synapses the network is 
in an heterogeneous asynchronous state for all the considered values of the synaptic coupling.
This is demonstrated  by the fact that the intensity of the fluctuations of the average firing activity, 
measured by considering the low-pass filtered linear super-position of all the spikes emitted in 
the network, vanishes as $1/\sqrt{N}$~\cite{wang1996gamma}.
Therefore, the observed transition at $g=g_m$ is not associated to the emergence of
irregular collective behaviours as reported for globally coupled heterogeneous inhibitory networks of 
LIF neurons with delay~\cite{Luccioli2010Irregular} and of pulse-coupled phase oscillators~\cite{ullner2016self}.

%\textbf{Equations of the fitting:}
%\begin{eqnarray}
%g(1400)_{min} = 0.28394 K^{0.4902} \\
%g(400)_{min} = 0.37773 K^{0.5045} \\
%n(1400)_{min} = 1.1811 - 0.1203 \log (K) \\
%n(400)_{min} = 1.1458 - 0.1258 \log (K) \\
%\nu (1400)_{min} = 0.52042 K^{-0.181} \\
%\nu (400)_{min} = 0.46848 K^{-0.188} \\
%\end{eqnarray}

\section{Effect of synaptic filtering}
\label{sec:alpha}

In this Section we will investigate how synaptic filtering 
can influence the previously reported results. In particular, we will consider non instantaneous IPSP
with $\alpha$-function profile~\eqref{eq:alpha}, whose evolution is ruled by a single time scale $\tau_\alpha$.

\subsection{Fully Coupled Networks}
\label{sec:FC_Alpha}

\begin{figure*}
\centering
\includegraphics[width=0.45\linewidth]{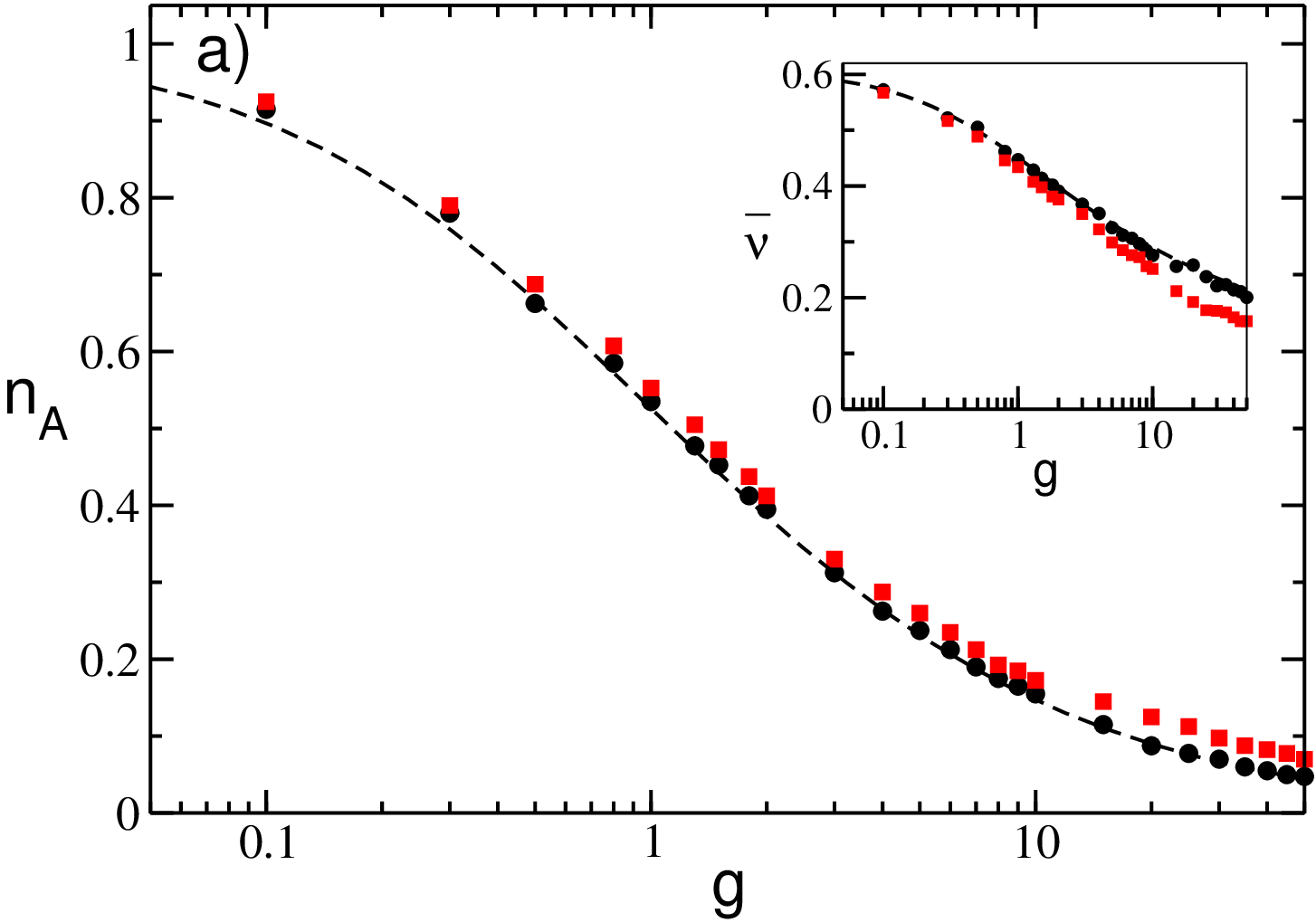}
\includegraphics[width=0.45\linewidth]{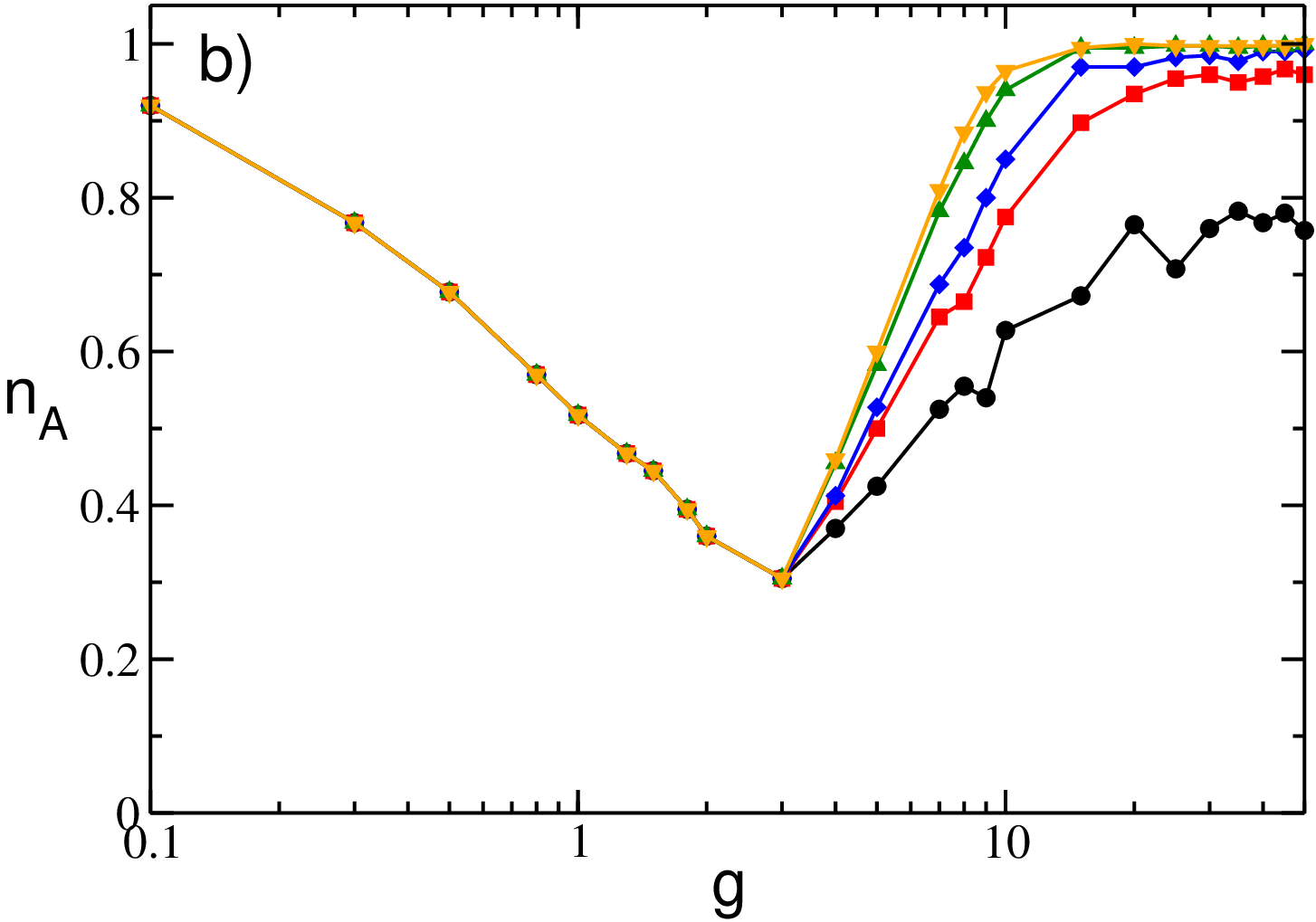}
\caption{Fraction of active neurons $n_A$ as a function of the inhibition 
with IPSPs with $\alpha$-profile for a fully coupled topology (a) and
a sparse network (b) with $K=20$.
a) Black (red)  symbols correspond to $\tau_{\alpha} = 10$ ($\tau_{\alpha} = 0.125$), 
while the dashed lines are the theoretical predictions \eqref{eq:unif_nu} and \eqref{eq:nStar_FC}
previously reported for instantaneous synapses. The data are averaged over
a time window $t_S = 1 \times 10^5$. Inset: average frequency $\bar \nu$ as a function of $g$.
b) $n_A$ is measured at successive times : from lower to upper curve the considered times are 
$t_S = \{1000, \, 5000, \,10000, \, 50000, \,100000\}$, while
$\tau_{\alpha} = 10$. The system size is $N = 400$ in both cases,
the distribution of excitabilities is uniform with  $[l_1,l_2] = [1.0,1.5]$ and $\theta=1$.} 
\label{fig:alfa_Nstar_vs_Time}
\end{figure*}

Let us first examine the fully coupled topology, in this case
we observe analogously to the $\delta$-pulse coupling that
by increasing the inhibition, the number of active neurons steadily
decreases towards a limit where only few neurons (or eventually only one) will survive.
At the same time the average frequency also decreases monotonically,
as shown in Fig.~\ref{fig:alfa_Nstar_vs_Time} for two different $\tau_\alpha$
differing by almost two orders of magnitude. Furthermore, the mean field
estimations \eqref{eq:unif_nu} and \eqref{eq:nStar_FC} obtained 
for $n_A$ and $\bar {\nu}$ represent a very good
approximation also for $\alpha$-pulses (as shown in Fig.~\ref{fig:alfa_Nstar_vs_Time}).
In particular, the mean field estimation essentially coincides with the numerical values for
slow synapses, as evident from the data 
reported in Fig.~\ref{fig:alfa_Nstar_vs_Time} for $\tau_{\alpha} = 10$ (black filled circles).
This can be explained by the fact that for sufficiently slow synapses,
with $\tau_P > {\bar T}_{ISI}$, the neurons feel the synaptic input current as continuous, because each input 
pulse has essentially no time to decay between two firing events. Therefore the mean field approximation for the input current~\eqref{eq:eff_input_GC} works definitely well in this case. This is
particularly true for $\tau_{\alpha} = 10$, where $\tau_P =20$ and ${\bar T}_{ISI} \simeq 2 - 6$
in the range of the considered coupling. 
While for $\tau_{\alpha} = 0.125$, we observe some deviation from the mean field results (red squares in  Fig.~\ref{fig:alfa_Nstar_vs_Time}) 
and the reason for these discrepancies reside in the fact that $\tau_P < {\bar T}_{ISI}$ for any coupling strength, therefore the discreteness of the pulses cannot be completely neglected
in particular for large amplitudes (large synaptic couplings) analogously to what observed for
instantaneous synapses.

\subsection{Sparse Networks}
\label{sec:sparse_alfa}

\begin{figure*}
\centering
\includegraphics[width=0.45\linewidth]{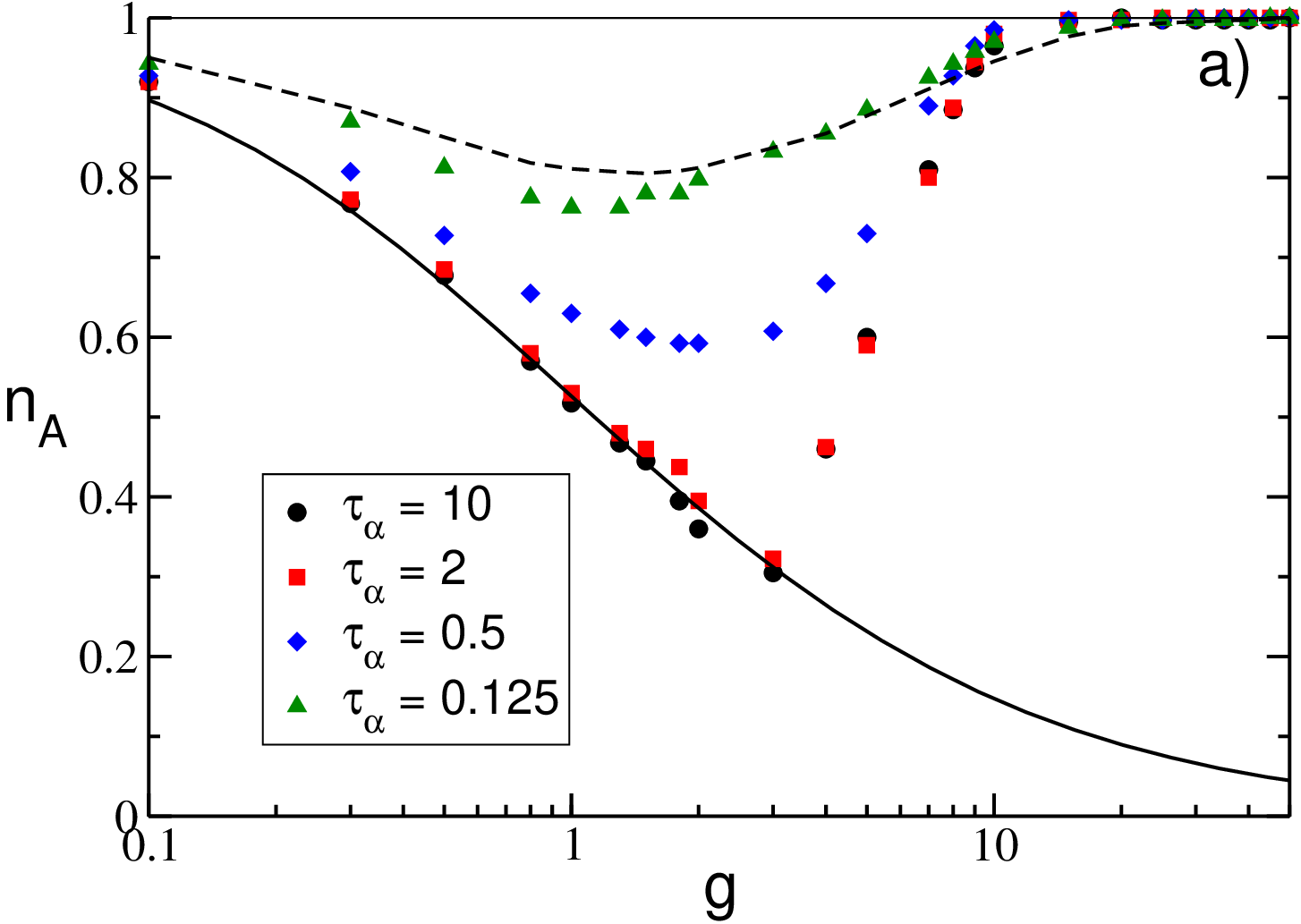}
\includegraphics[width=0.45\linewidth]{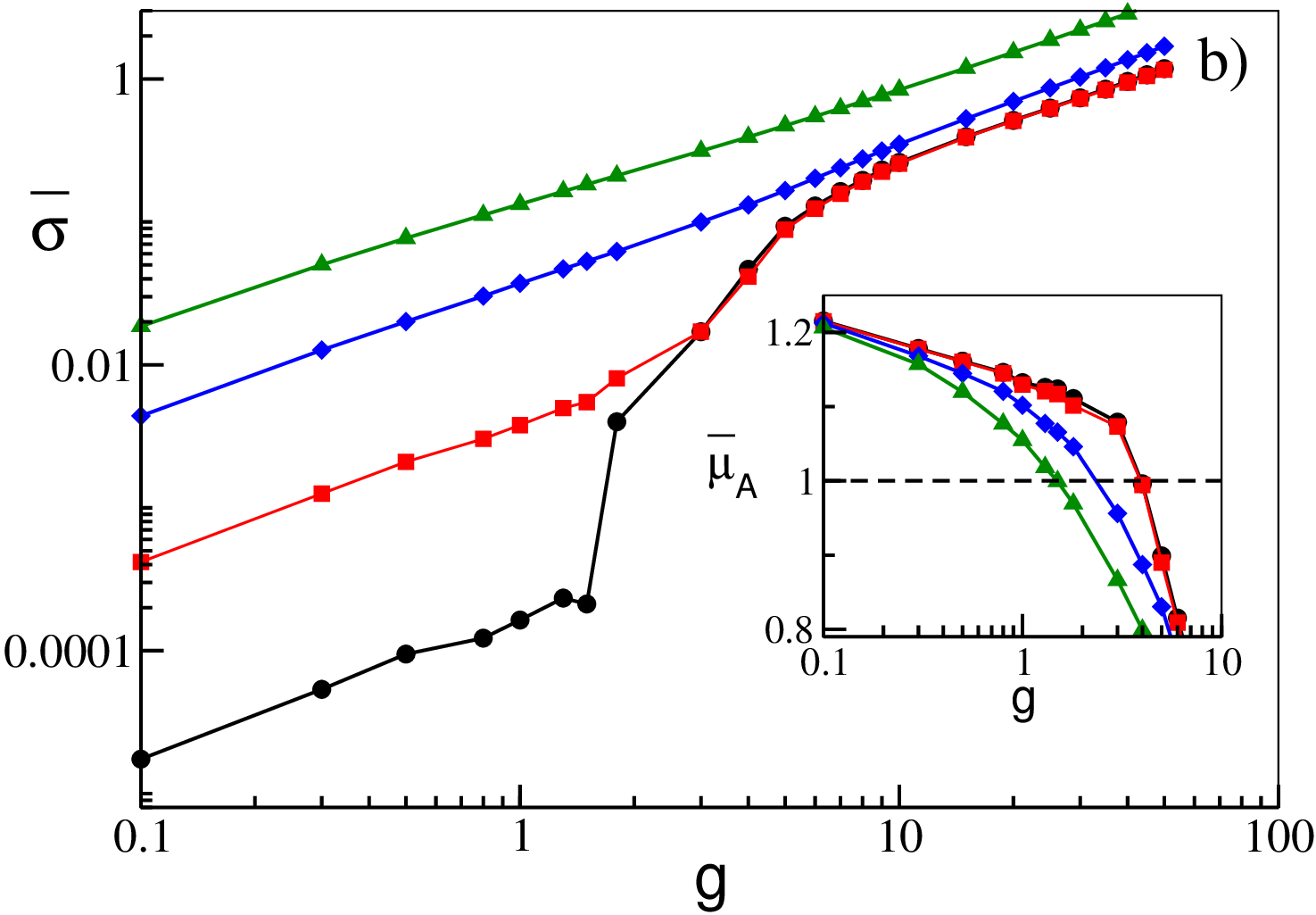}
\includegraphics[width=0.45\linewidth]{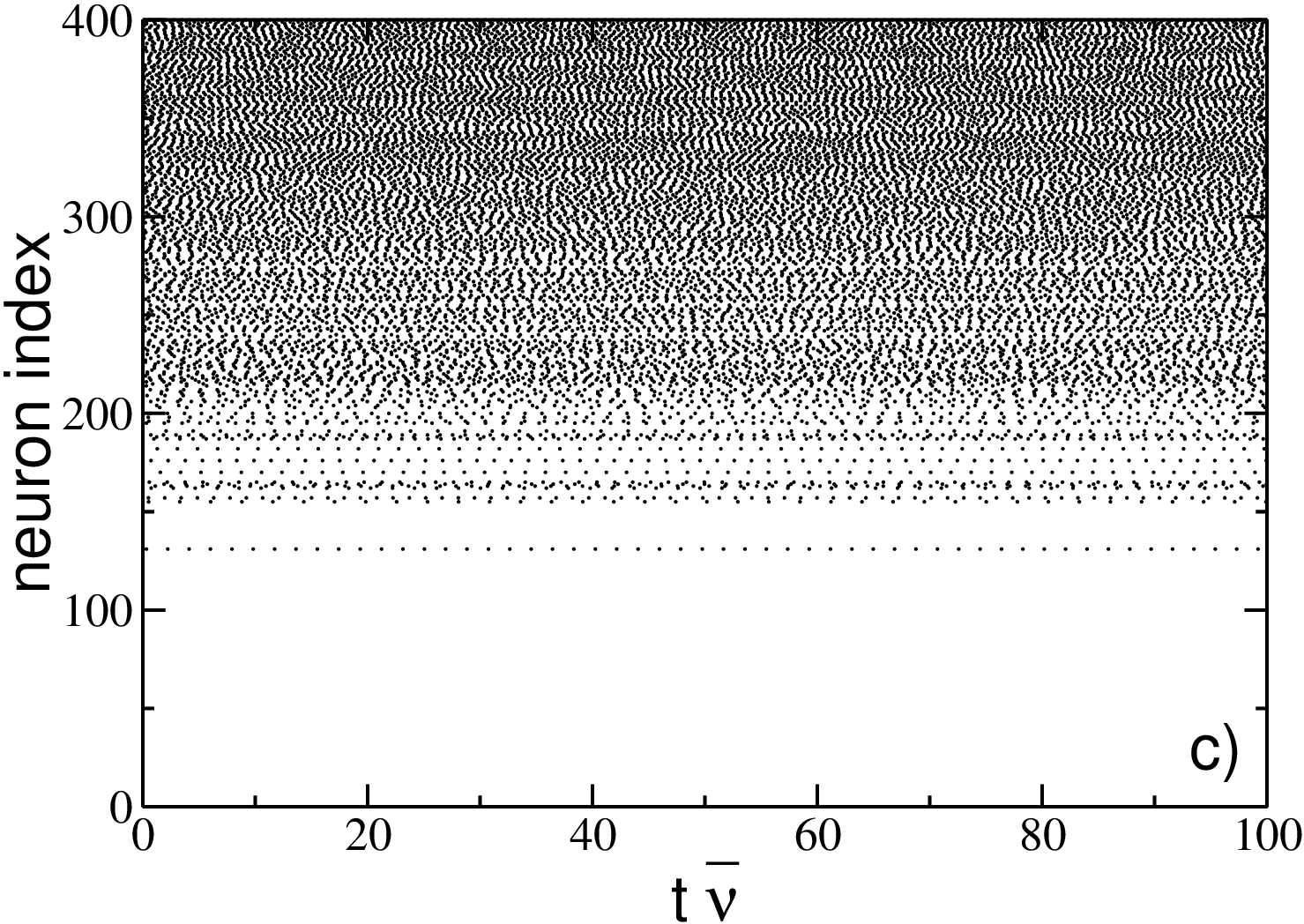}
\includegraphics[width=0.45\linewidth]{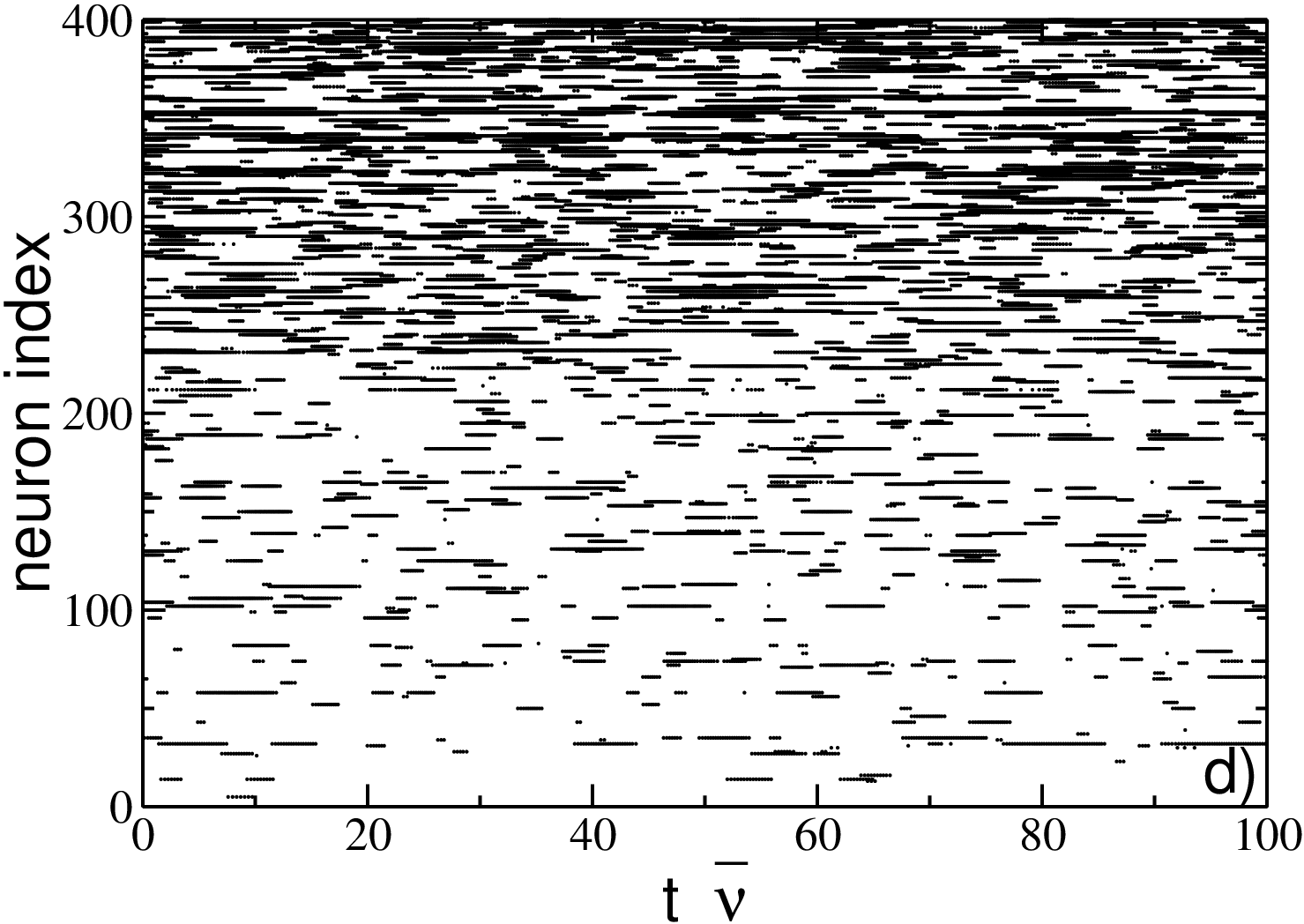}
\caption{a) Fraction of active neurons for a network of $\alpha$-pulse coupled neurons as  a function
of $g$ for various $\tau_{\alpha}$: namely,  $\tau_{\alpha} = 10$ (black circles),  $\tau_{\alpha} = 2$ (red squares),
$\tau_{\alpha} = 0.5$ (blue diamonds) and  $\tau_{\alpha} = 0.125$ (green triangles).
For instantaneous synapses, the fully coupled analytic solution is reported (solid line), 
as well as the measured $n_A$ for the sparse network with same level of dilution and estimated
over the same time interval (dashed line). 
b) Average fluctuations of the synaptic current  ${\bar \sigma}$ versus $g$ for ISPS with $\alpha$-profile  the symbols refer to the same $\tau_\alpha$ as in panel (a). Inset: Average input current ${\bar \mu}_A$ of the active
neurons vs $g$, the dashed line is the threshold value $\theta=1$.
The simulation time has been fixed to $t_S = 1 \times 10^5$.
c-d) Raster plots for two different synaptic strengths for $\tau_\alpha =10$: namely, c) $g=1$ corresponds to $n_A \simeq  0.52$, $\bar \nu \simeq 0.45$ and $\overline{CV} \simeq 3 \times 10^{-4}$; while i) $g=10$  to $n_A \simeq 0.99 $, 
$\bar \nu \simeq 0.06$ and $\overline{CV} \simeq 4.1$. The neurons are ordered according to their intrinsic excitability and the time is rescaled by the average frequency $\bar \nu$.
The data have been obtained for a system size $N = 400$ and $K = 20$, other parameters as
in Fig.~\ref{fig:alfa_Nstar_vs_Time}.}
\label{fig:alfa_fluctuations}
\end{figure*}

\begin{figure*}
\centering
\includegraphics[width=0.32\linewidth]{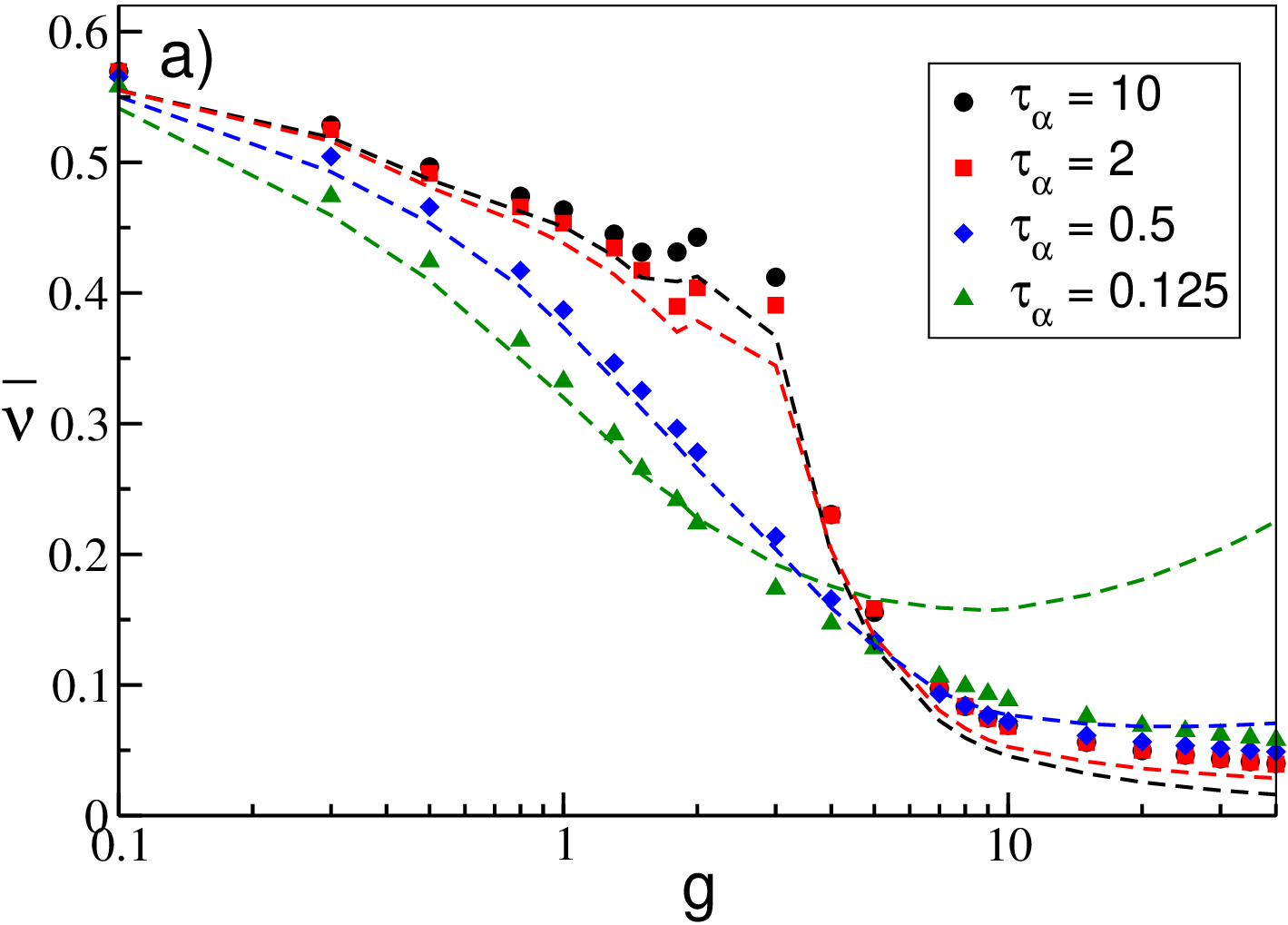}
\includegraphics[width=0.32\linewidth]{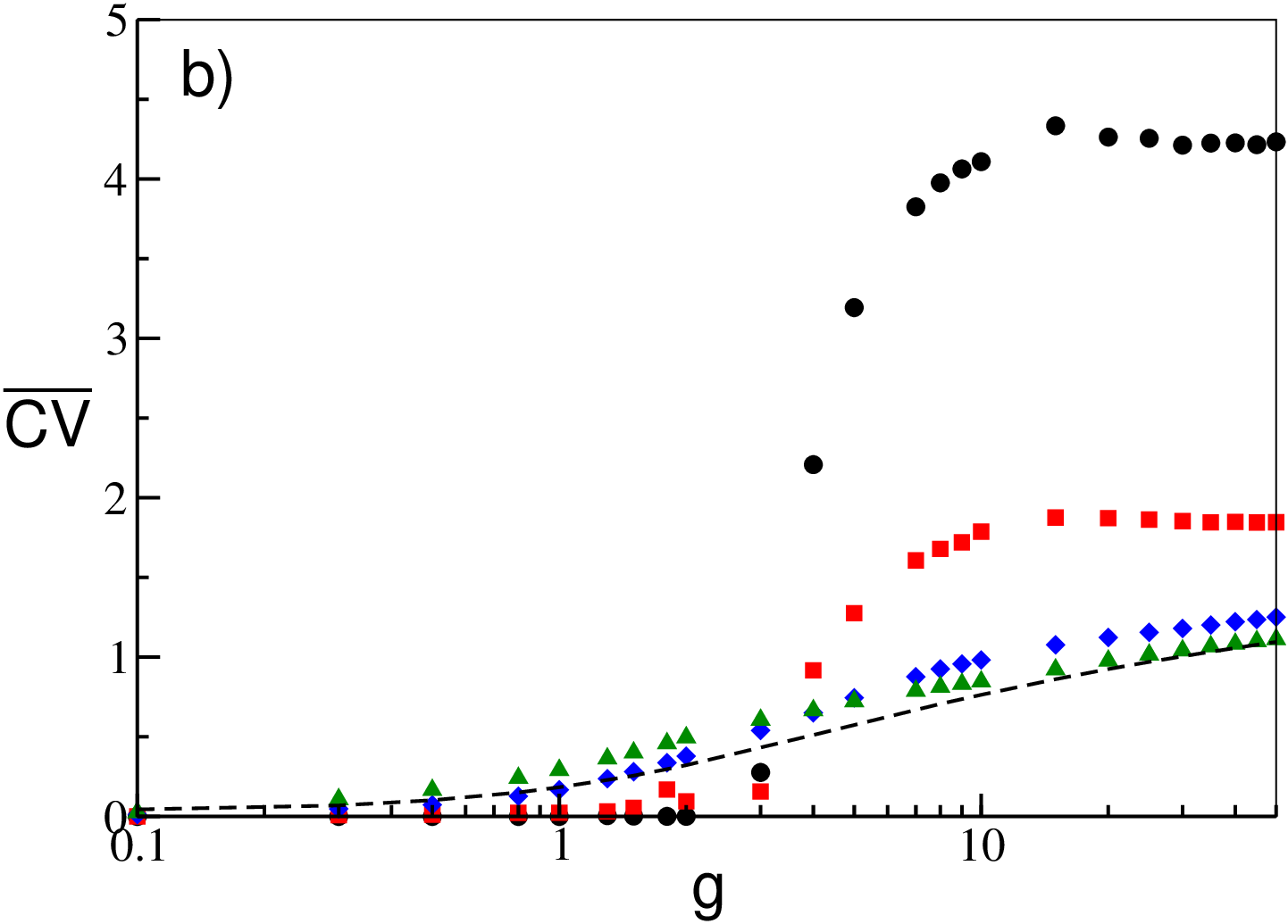}
\includegraphics[width=0.32\linewidth]{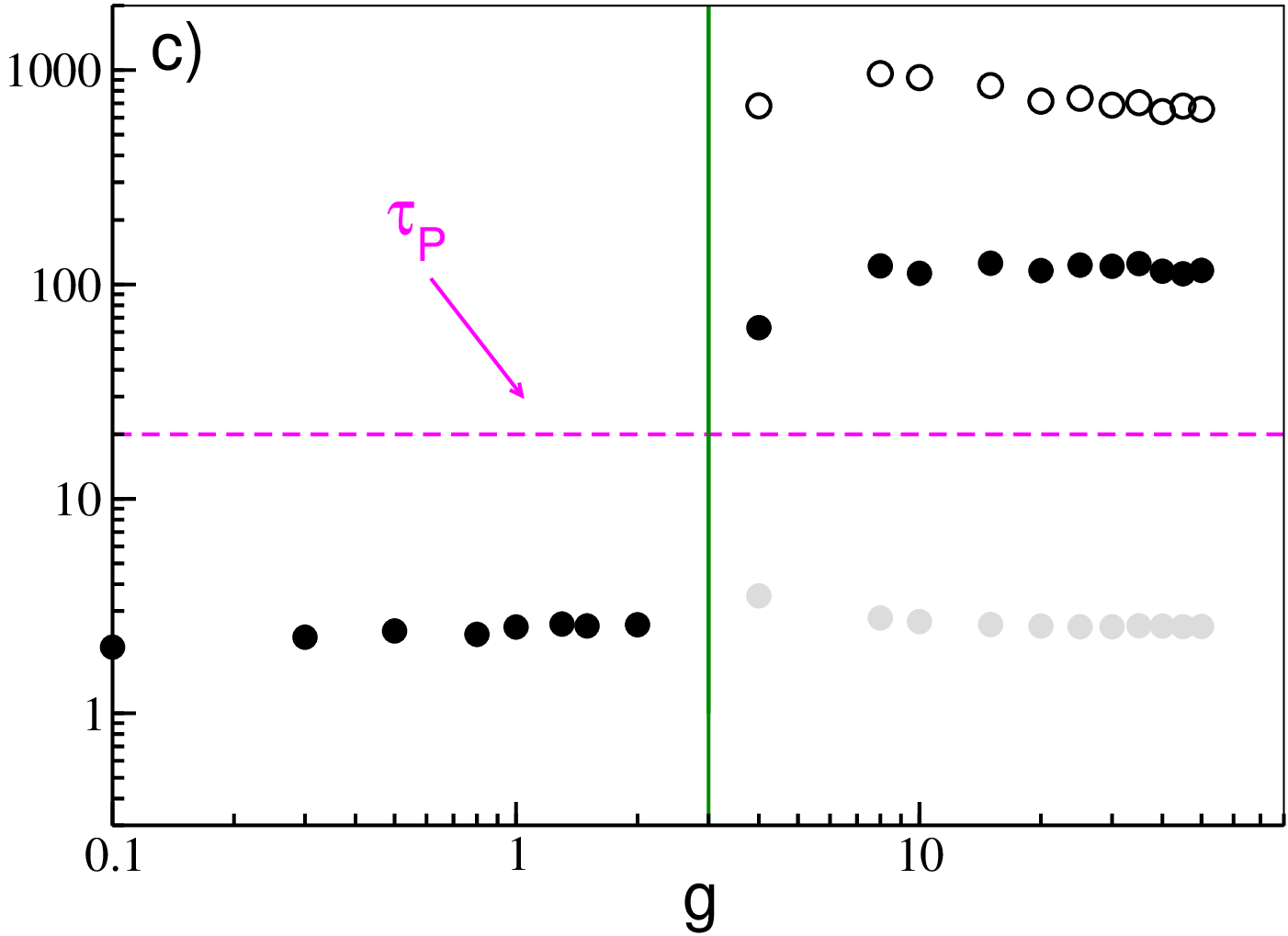}
\caption{a) Average firing rate $\bar \nu$  vs $g$ for a network of $\alpha$-pulse 
coupled neurons, for four values of $\tau_{\alpha}$. Theoretical estimations for $\bar \nu$ calculated with the adiabatic 
approach  \eqref{eq:nu_adiabatic_appendix_Self} are reported as dashed lines of colors
corresponding to the relative symbols.
b) Average coefficient of variation $\overline{CV}$  for four values of $\tau_{\alpha}$
as a function of the inhibition. The dashed line refers to the values obtained for instantaneous 
synapses and a sparse network with the same value of dilution. 
c): Average inter-spike interval ${\bar T}_{ISI}$ (filled black circles)
as a function of $g$ for  $\tau_\alpha=10$. For $g > g_m$ the
average inter-burst interval (empty circles) and the average ISI measured within bursts (gray circles) are also shown,
together with the position of $g_m$ (green veritical line).  
The symbols and colors denote the same $\tau_{\alpha}$ values as in Fig.~\ref{fig:alfa_fluctuations}.
All the reported data  were calculated for
a system size $N = 400$ and $K = 20$ and for a fixed simulation time of $t_S = 1 \times 10^5$.} 
\label{fig:alfa_sparse_nu_nStar}
\end{figure*}

  For the sparse networks $n_A$ has the same qualitative behaviour as a function of the synaptic inhibition observed for instantaneous IPSPs, as shown in Fig.~\ref{fig:alfa_Nstar_vs_Time} (b) and
Fig.~\ref{fig:alfa_fluctuations} (a).  The value of $n_A$ decreases with $g$ and reaches 
a minimal value at $g_m$, afterwards it increases towards $n_A=1$ at larger coupling.
The origin of the minimum of $n_A$ as a function of $g$ is the same as
for instantaneous synapses,  for $g < g_m$ the active neurons are subject 
on average to a supra-threshold effective input ${\bar \mu}_A$,
while at larger coupling  ${\bar \mu}_A  < \theta$, as shown in the inset of 
Fig.~\ref{fig:alfa_fluctuations} (b).  This is true for any value of $\tau_\alpha$,
however, this transition from mean- to fluctuation-driven becomes dramatic
for slow synapses . As evidenced from the data for the average output firing rate $\bar \nu$ and 
the average coefficient of variation $\overline{CV}$, these quantities have almost discontinuous jumps 
at $g=g_m$, as shown in Fig.~\ref{fig:alfa_sparse_nu_nStar} .

Therefore, let us first concentrate on slow synapses with $\tau_\alpha$
larger than the membrane time constant, which is one for adimensional units.
For $g < g_m$
the fraction of active neurons is frozen in time, at least 
on the considered time scales, 
as revealed by the data in Fig.~\ref{fig:alfa_Nstar_vs_Time} (b). Furthermore, for $g < g_m$, 
the mean field approximation obtained for the fully coupled case 
works almost perfectly both for $n_A$ and ${\bar \nu}$,
as reported in  Fig.~\ref{fig:alfa_fluctuations} (a).  
The frozen phase is characterized by extremely small values of the current fluctuations ${\bar \sigma}$ 
(as shown Fig.~\ref{fig:alfa_fluctuations} (b)) and a quite high firing rate ${\bar \nu} \simeq 0.4 - 0.5$  with an associated average coefficient of variation $\overline{CV}$ almost zero (see  
black circles and red squares in Fig.~\ref{fig:alfa_sparse_nu_nStar}). 
Instead, for $g > g_m$ the number of active 
neurons increases in time similarly to what observed for the instantaneous synapses,
while the average frequency becomes extremely small $\bar \nu \simeq 0.04 - 0.09$
and the value of the coefficient of variation becomes definitely larger than one.

These effects can be explained by the fact that,  below $g_m$ the active
neurons (the winners) are subject to an effective input ${\bar \mu}_A  > \theta$ that
induces a quite regular firing, as testified by  the raster plot displayed in Fig.~\ref{fig:alfa_fluctuations} (c).
The supra-threshold  activity of the winners  joined together with the filtering action of the synapses guarantee that on average each neuron in the network receive an  almost continuous current, with 
small fluctuations in time. These results explain why the mean field approximation still works
in the frozen phase, where the fluctuations in the synaptic currents are essentially
negligible and unable to induce any neuron's rebirth, at least on realistic time scales.
In this regime the only mechanism in action is the WTA, fluctuations begin
to have a role for slow synapses only for $g > g_m$. Indeed, as shown in  Fig.~\ref{fig:alfa_fluctuations} (b),
the synaptic fluctuations $\bar \sigma$  for $\tau_\alpha = 10$ (black circles) are almost negligible 
for $g < g_m$ and show an enormous increase at $g=g_m$ of almost
two orders of magnitude. Similarly at $\tau_\alpha = 2$ (red square) a noticeable increase 
of  ${\bar \sigma}$ is observable at the transition.
 
In order to better understand the abrupt changes in $\bar \nu$ and in $\overline{CV}$ observable for
slow synapses at $g=g_m$, let us consider  the case $\tau_\alpha = 10$.
As shown in  Fig.~\ref{fig:alfa_sparse_nu_nStar} (c), 
$\tau_P > {\bar T}_{ISI} \simeq  2- 3$  for $g < g_m$, 
therefore for these couplings the IPSPs have no time to decay between a firing emission and the next one, thus the 
synaptic fluctuations $\bar \sigma$ are definitely small in this case, as already shown. At $g_m$ an abrupt jump is observable to large values  ${\bar T}_{ISI}  >  \tau_P$, this is
due to the fact that now the neurons display bursting activities, as 
evident from the raster plot shown in Fig.~\ref{fig:alfa_fluctuations} (d).
The bursting is due to the fact that, for $g > g_m$, the active neurons are subject
to an effective input which is on average sub-threshold, 
therefore the neurons preferentially tend to be silent. However, due to current 
fluctuations a neuron can pass the threshold and the silent periods can be interrupted
by  bursting phases where the neuron fires almost regularly. 
As a matter of fact, the silent (inter-burst) periods are very long $\simeq 700-900$,  
if compared to the duration of the bursting periods, namely $\simeq 25-50$,
as shown in Fig.~\ref{fig:alfa_sparse_nu_nStar} (c).
This explains the abrupt decrease of the average firing rate reported in Fig.~\ref{fig:alfa_sparse_nu_nStar} (a).
Furthermore, the inter-burst periods are exponentially distributed with an associated coefficient of variation $\simeq 0.8 - 1.0$, which clearly indicates the stochastic nature of the  switching from the silent to the bursting phase. The firing periods within the bursting phase are instead
quite regular, with an associated coefficient of variation
$\simeq 0.2$, and with a duration similar to  ${\bar T}_{ISI}$  measured in the frozen phase
(shaded gray circles in Fig.~\ref{fig:alfa_sparse_nu_nStar} (c)).
Therefore, above $g_m$ the distribution of the ISI exhibits a long exponential
tail associated to the bursting activity and this explains the very large
values of the measured coefficient of variation.
By increasing the coupling the fluctuations in the input current becomes larger thus
the fraction of neurons that fires at least once within a certain time interval increases.
At the same time, $\bar \nu$,  the average inter-burst periods and the firing periods 
within the bursting phase remain almost constant at  $g >10$, 
as shown in Fig.~\ref{fig:alfa_sparse_nu_nStar} (a). This indicates that
the decrease of ${\bar \mu}_A$ and the increase of $\bar \sigma$ due to
the increased inhibitory coupling essentially compensate each other in this
range. Indeed, we have verified that for $\tau_\alpha=10$ and $\tau_\alpha = 2$
${\bar \mu}_A$ ($\bar \sigma$) decreases (increases) linearly with $g$ with
similar slopes, namely ${\bar \mu}_A \simeq 0.88  - 0.029 g$ while
$\bar \sigma \simeq  0.05 + 0.023 g$.

For faster synapses, the frozen phase is no more present. 
Furthermore, due to rebirths induced by current flutuations, $n_A$ is always
larger than the fully coupled mean field result  \eqref{eq:nStar_FC}, even at $g < g_m$.
It is interesting to notice that by decreasing $\tau_\alpha$, we are now approaching the
instantaneous limit, as indicated by the results reported for $n_A$ 
in Fig.~\ref{fig:alfa_fluctuations} (a) and $\overline{CV}$ 
in Fig.~\ref{fig:alfa_sparse_nu_nStar} (b). 
In particular, for 
$\tau_\alpha = 0.125$ (green triangles) the data 
almost collapse on the corresponding values measured for 
instantaneous synapses in a sparse networks with the same
characteristics and over a similar time interval (dashed line).  Furthermore, for fast
synapses with $\tau_\alpha < 1$ the bursting activity is no more present as it can be 
appreciated by the fact that at most $\overline{CV}$ approaches one in the very large coupling limit.

%\begin{figure}
%\centering
%\includegraphics[width=0.9\linewidth]{Figures/nStar_first_order.eps}
%\caption{Fraction of active neurons measured at successive simulation times $t_s$ as a 
%function of the inhibition. 
%From lower to upper curve the considered times are 
%$t_s = \{1000, \, 5000, \,10000, \, 50000, \,100000\}$. 
%%Inset, $n_A(t_s)$ as a function of the simulation time for 
%%three selected values of the coupling: namely $g = 0.1$ (black solid curve);
%%$g = 3$ (red dashed line) and $g=10$ (blue dotted line).
%Parameters of the simulation $\tau_{\alpha} = 10$, $N = 400$, $K = 20$.} 
%\label{fig:alfa_fully}
%\end{figure}
%

For sufficiently slow synapses, the average firing rate $\bar \nu$
can be estimated analytically by applying the so-called adiabatic approach
developed by Moreno-Bote and Parga in~\cite{moreno2004,moreno2010response}.
This method applies to LIF neurons with a synaptic time scale definitely longer 
than the membrane time constant.
In these conditions, the output firing rate can be reproduced by assuming that
the neuron is subject to an input current with time correlated fluctuations,
which can be represented as colored noise 
with a correlation time given by the pulse duration  $\tau_P = 2 \tau_\alpha$ (for more details see Appendix D).
In this case we are unable to develop a self-consistent approach to obtain at
the same time $n_A$ and the average frequency. However,  once $n_A$ is provided 
by simulations the analytic estimated  \eqref{eq:nu_adiabatic_appendix_Self} obtained with the adiabatic approach
gives very good agreement with the numerical data for sufficiently slow synapses,
namely for $\tau_P \ge 1$, as shown in Fig.~\ref{fig:alfa_sparse_nu_nStar}(a)
for $\tau_\alpha =10$, 2 and 0.5. The theoretical expression  \eqref{eq:nu_adiabatic_appendix_Self} is even able 
to reproduce the jump in the average frequencies observable at $g_m$ and therefore to 
capture the bursting phenomenon. By considering  $\tau_P < 1$, as expected,
the theoretical expression fails  to reproduce the numerical data in particular at large coupling
(see the dashed green line in Fig.~\ref{fig:alfa_sparse_nu_nStar}(a) corresponding to $\tau_\alpha = 0.125$). 

By following the arguments reported in~\cite{moreno2004},
the bursting phenomenon observed for $\tau_\alpha > 1$ and $g > g_m$ 
can be interpreted at a mean field level as the response of a sub-threshold
LIF neuron subject to colored noise with correlation $\tau_P$.
In this case, the neuron is definitely sub-threshold, but in presence of a large fluctuation
it can be lead to fire and due to the finite correlation time
it can remain supra-threshold regularly firing for a period $\simeq \tau_P$. 
The validity of this interpretation is confirmed by the fact that 
the measured average bursting periods are of the order of the correlation
time $\tau_P = 2 \tau_\alpha$,
namely, $\simeq 27-50$ ($\simeq 7-14$) for $\tau_\alpha =10$ 
($\tau_\alpha =2$).

\begin{figure*}
\centering
\includegraphics[width=0.32\linewidth]{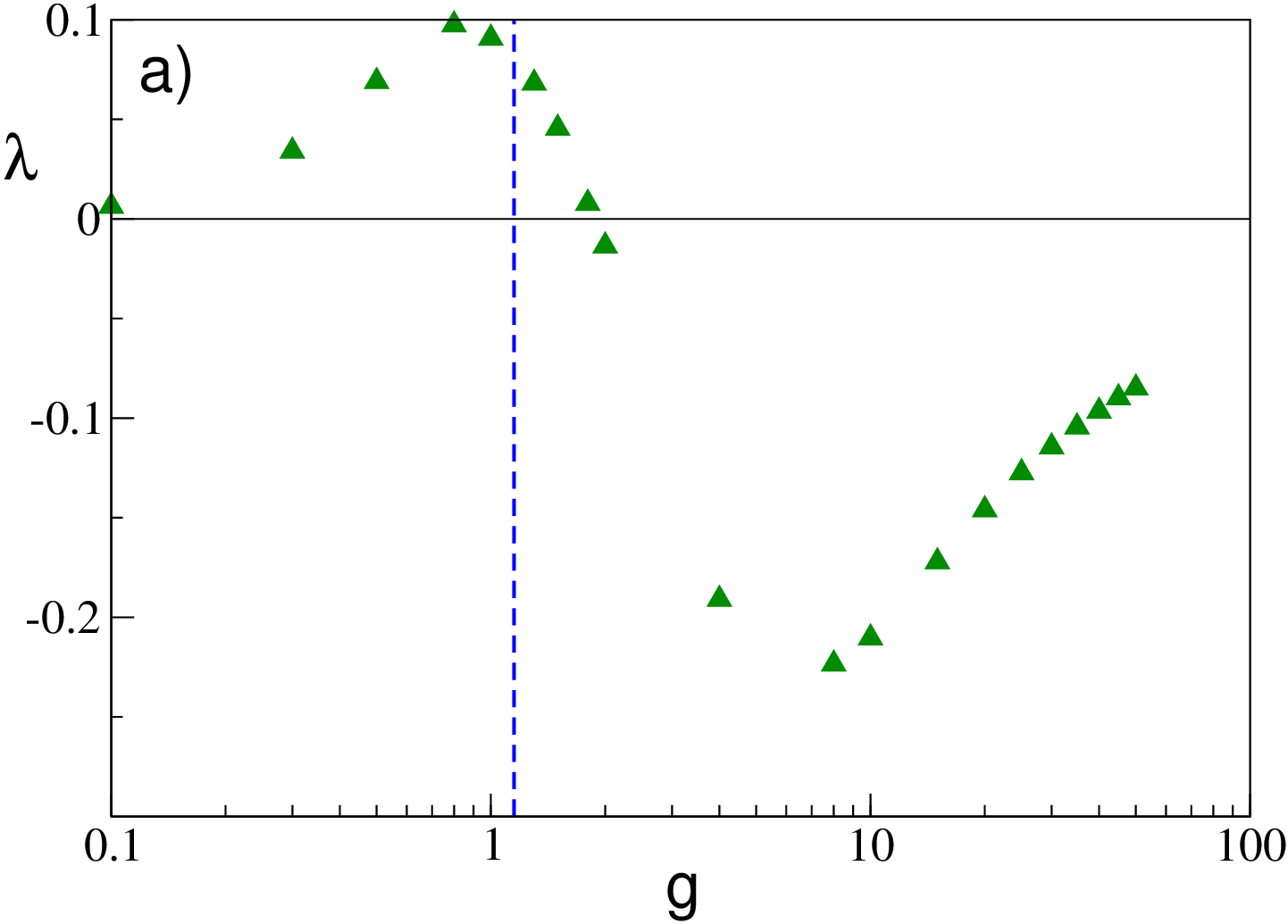}
\includegraphics[width=0.32\linewidth]{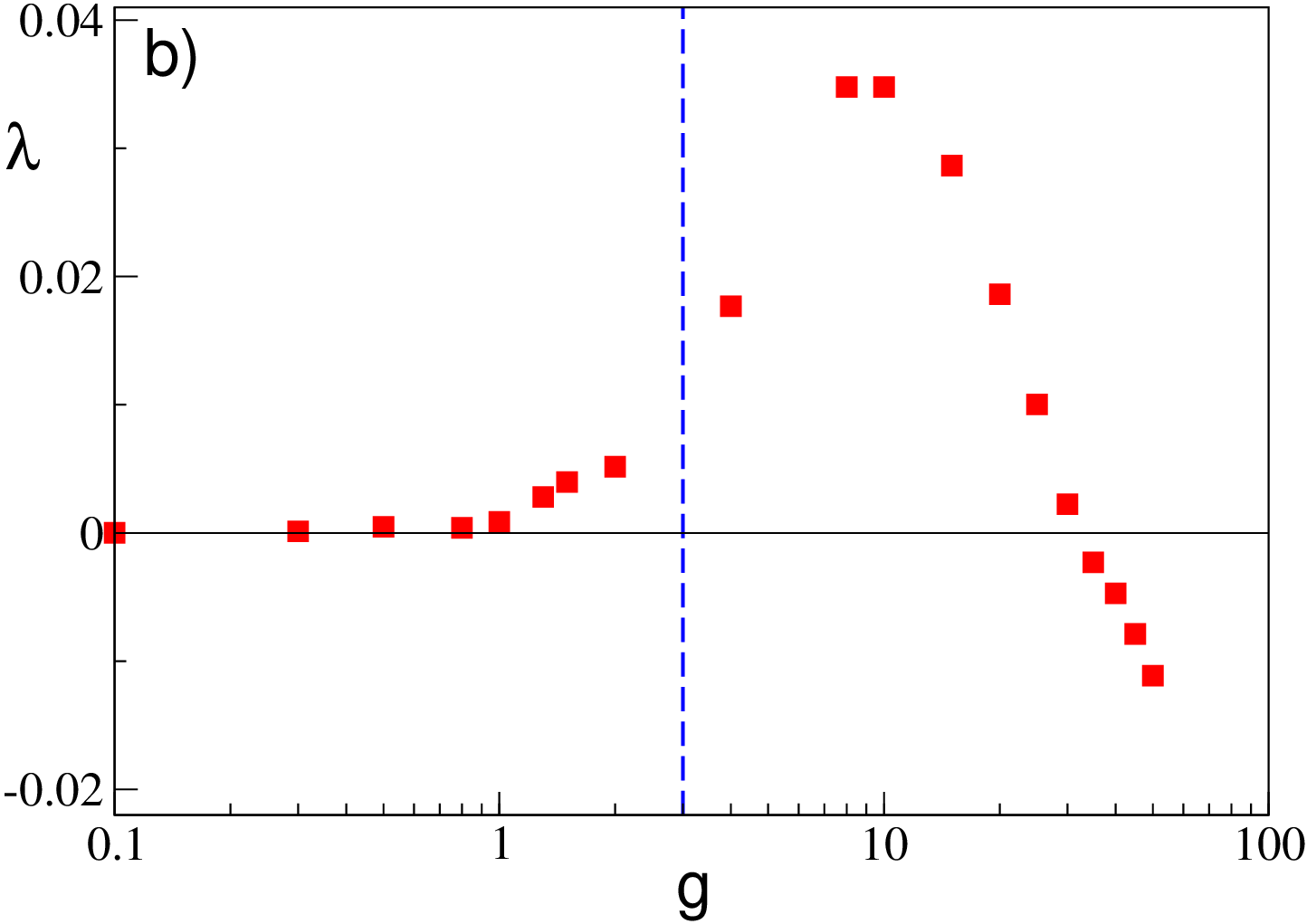}
\includegraphics[width=0.32 \linewidth]{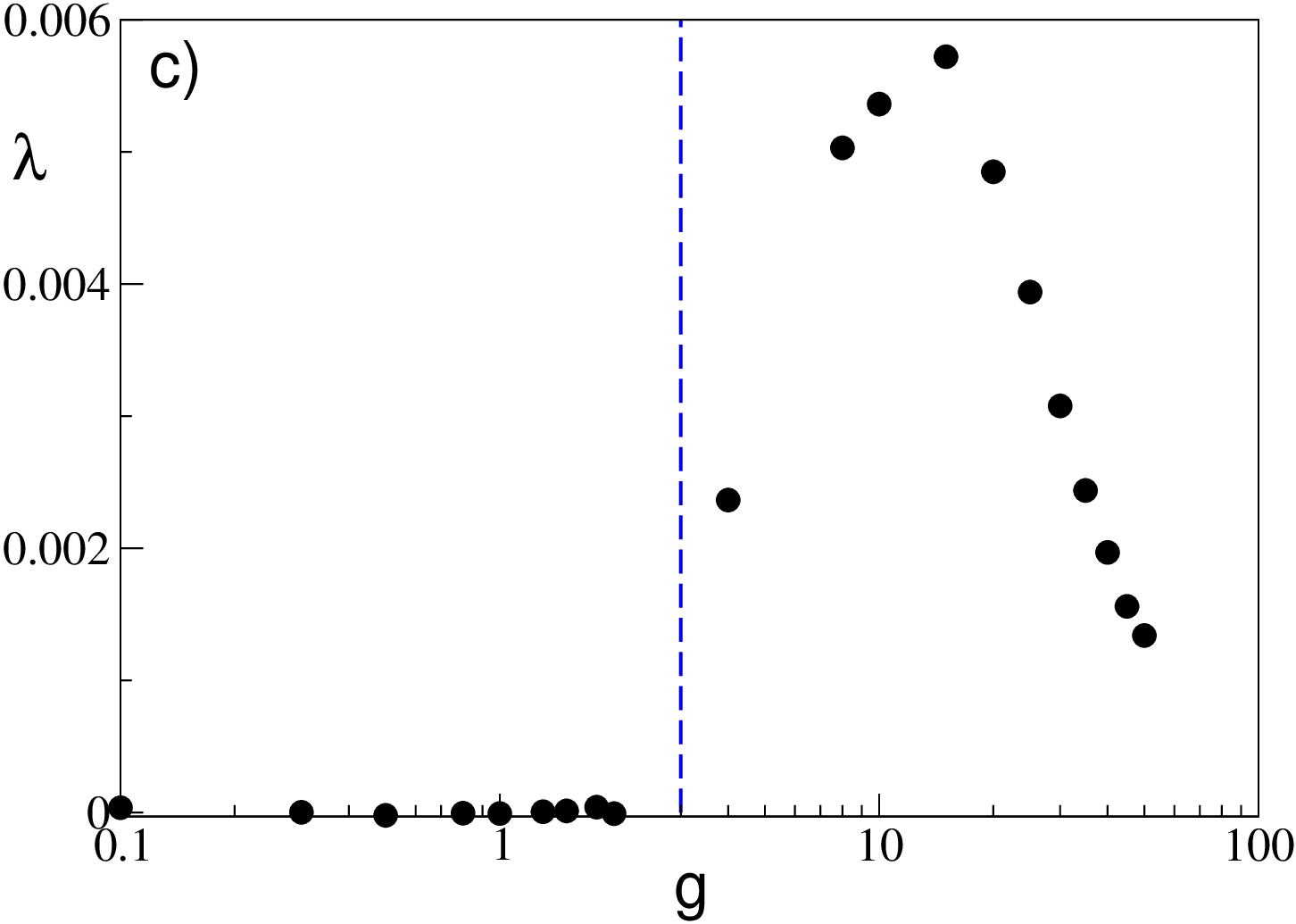}
\caption{Maximal Lyapunov exponent $\lambda$  versus $g$ for a network of $\alpha$-pulse 
coupled neurons, for $\tau_\alpha = 0.125$ (a), $\tau_\alpha=2$ (b) and $\tau_{\alpha} = 10$ (c).
The blue dashed vertical line  denote the  $g_m$ value.
All the reported data  were calculated for a system size $N = 400$ and $K = 20$ and for 
simulation times  $5 \times 10^4 \le t_S \le 7 \times 10^5$ ensuring a good convergence of $\lambda$ to its asymptotic value. The other parameters are as in Fig.~\ref{fig:alfa_Nstar_vs_Time}.} 
\label{fig:alfa_maxly}
\end{figure*}

As a final point, to better understand the dynamical origin of the measured fluctuations in this
deterministic model,  we have estimated the maximal Lyapunov exponent $\lambda$.
As expected from previous analysis, 
for non-instantaneous synapses we can observe the emergence of regular chaos in purely inhibitory networks
~\cite{Timme2009ChaosBalance,Zillmer2009LongTrans,angulo2015Striatum}.
In particular, for sufficiently fast synapses, we typically note a transition
from a chaotic state at low coupling to a linearly stable regime (with $\lambda < 0$) at large
synaptic strengths, as shown in Fig.~\ref{fig:alfa_maxly} (a) for $\tau_\alpha = 0.125$.
Despite the fact that current fluctuations are monotonically increasing with the synaptic strength.
Therefore, fluctuations are due to chaos, at small coupling,  while at larger
$g$ they are due to finite amplitude instabilities, as expected for stable
chaotic systems~\cite{angulo2014}. However, the passage from positive to negative values of 
the maximal Lyapunov exponent is not related to the transition occurring at $g_m$ from a 
mean-driven to a fluctuation-driven dynamics in the network. 

For slow synapses,  $\lambda$ is essentially zero at small coupling in
the frozen phase characterized  by tonic spiking of the neurons, 
while it becomes positive by approaching $g_m$.
For larger synaptic strengths $\lambda$, after reaching a maximal value,
decreases and it can become eventually negative at $g >> g_m$,
as reported in  Fig.~\ref{fig:alfa_maxly} (b-c).
Only for extremely slow synapses,  as shown in Fig.~\ref{fig:alfa_maxly} (c)
for $\tau_\alpha =10$, the chaos onset seems to coincide with the transition occurring at $g_m$.
These findings are consistent with recent results concerning the emergence of 
asynchronous rate chaos in homogeneous inhibitory LIF networks with 
deterministic~\cite{harish2015} and stochastic~\cite{kadmon2015}  evolution.
However, a detailed analysis of this aspect goes beyond the scope of the present
 paper.

%
%\begin{figure}
%\centering
%\includegraphics[width=0.9\linewidth]{Figures/averageISI.eps}
%\caption{Average ISI  ${\bar T}_{ISI}$ (black circles) as a function of the
%synaptic coupling $g$. Red squares denote the average ISI within the
%bursting phases, while the blue diamonds represents the average inter-burst
%interval.  The magenta horizontal dashed line is the value of the pulse duration $\tau_P$, while
%the green dotted vertical line denotes $g_m=3$.
%Parameters of the simulation $\tau_{\alpha} = 10$, $N = 400$, $K = 20$, 
%the data are averaged over a time interval $t_s=1 \times 10^5$ after discarding a transient
%correpsonding to the emission of $10^6$ spikes.} 
%\label{fig:aveISI}
%\end{figure}

\section{Discussion}
\label{sec:discussion}

In this paper we have shown that the effect reported in \cite{ponzi2013optimal,angulo2015Striatum}
is observable whenever two sources of quenched disorder are present in the network: namely,
a random distribution of the neural properties and a random topology. 
In particular, we have shown that neuron's death due to synaptic inhibition is
observable only for heterogeneous distributions of the neural excitabilities.
Furthermore, in a globally coupled network
the less excitable neurons are 
silenced for increasing synaptic strength until only one or few neurons remain active.
This scenario corresponds to the winner-takes-all mechanism via
lateral inhibition, which has been often invoked in neuroscience 
to explain several brain functions~\cite{yuille1989}.
WTA mechanisms have been proposed to model hippocampal 
CA1 activity~\cite{coultrip1992cortical}, as well as 
to be at the basis of visual velocity estimate~\cite{grzywacz1990}, 
and to be essential for controlling visual attention~\cite{itti2001}.

However, most brain circuits are characterized by  
sparse connectivity~\cite{laughlin2003communication,plenz2003,bullmore2009},
in these networks we have shown that an increase in inhibition
can lead from a phase dominated by neuronal death to a regime
where neuronal rebirths take place. Therefore  
the growth of inhibition can have the counter-intuitive effect to
activate silent neurons due to the enhancement of current fluctuations.
The reported transition is characterized by a passage from
a regime dominated by the almost tonic activity of a group
of neurons, to a phase where sub-threshold 
fluctuations are at the origin of the irregular firing of large part
of the neurons in the network.
For instantaneous synapses, the average first and second
moment of the firing distributions have been obtained 
together with the fraction of active neurons
within  a mean-field approach, where the neuronal
rebirth is interpreted as an activation process
driven by synaptic shot noise~\cite{richardson2010firing}.

For a finite synaptic time smaller than the characteristic membrane time constant, the
scenario is similar to the one observed for instantaneous
synapses. However, the transition from mean-driven to fluctuation-driven dynamics
becomes dramatic for sufficiently slow synapses. In this situation one
observes for low synaptic strength a frozen phase, where the 
synaptic filtering washes out the current fluctuations leading to
an extremely regular dynamics controlled only by a WTA mechanism.
As soon as the inhibition is sufficiently strong to lead
the active neurons below threshold, the neuronal activity
becomes extremely irregular exhibiting long silent phases
interrupted by bursting events. The origin of these
bursting periods can be understood in terms of the emergence
of correlations in the current fluctuations induced by the slow synaptic timescale,
as explained in~\cite{moreno2004}.

In our model, the random dilution of the network connectivity is a fundamental ingredient 
to generate current fluctuations, whose intensity is controlled by the average
network in-degree $K$. A natural question is whether the reported scenario will be
still observable in the thermodynamic limit. On the basis of previous studies
we can affirm that this depends on how $K$ scales with the system size~\cite{golomb2001,Tattini2011Coherent,Luccioli2012PRL}.
In particular, if  $K$ stays finite for  $N \to \infty$ the transition will still be observable. 
For $K$ diverging with $N$, the fluctuations become negligible for 
sufficiently large system sizes, impeding neuronal rebirths and
the dynamics will be controlled only by the WTA mechanism.

An additional source of randomness present in the network is related to the 
variability in the number of active pre-synaptic neurons. 
In our mean-field approach we have
assumed that each neuron is subject to $n_A  K$ spike trains, 
however this is true only on average. The number of 
active pre-synaptic neurons is a random variable
binomially distributed with average $n_A  K$
and variance $n_A(1-n_A)K$. Future developments of the
theoretical approach here reported should include also
such variability in the modeling of the network dynamics~\cite{BrunelHakim1999}.

\begin{table*}[t]
\begin{tabular}{||  c | c | c | c | c | c ||} \toprule
\hfill & & & & & \\
$\tau_\alpha/\tau_m$ & $\tau_m$ (msec) \hfill & Spike Rate (Hz) & 
 $\overline{CV}$  &  Burst Duration (msec) &   Spike Rate within Bursts  (Hz) \\ % Column names row
\hfill & & & & & \\
\hline
\hfill & & & & & \\
$2$ & \hfill $10$  \hfill & 4-6   & $\simeq 1.8$  &  $100 \pm 40$ &  $ 42 \pm 2$ \\
 & \hfill $20$  \hfill & 2-3   & $\simeq 1.8$  &  $200 \pm 80$ &  $ 21 \pm 1$ \\
\hfill & & & & & \\
\hline
\hfill & & & & & \\
$10$ & \hfill $10$ \hfill &  4-6  & $\simeq 4.2$ & $400 \pm 150$ &  $41\pm2$ \\
 & \hfill $20$ \hfill &  2-3  & $\simeq 4.2$ & $800 \pm 300$ &  $20\pm1$ \\
\hfill  & & & & & \\
\hline
\hline
\hfill & & & & & \\
& Experimental data & 2-3 & $\simeq 1.5-3$ & $500-1100$ &  $31 \pm 15$ \\
\hfill & & & & & \\
\botrule
\end{tabular}
\caption{Comparison between the results obtained for slow $\alpha$- synapses 
and experimental data for MSNs.The numerical data refer to results obtained in the bursting phase, namely for synaptic strenght $g$ in the range  $[10:50]$, for simulation times $t_S =1 \times 10^5$, $N=400$ and $K=20$.
The experimental data refer to MSNs population in striatum for free behaving wild type mice,
data taken from \cite{miller2008dysregulated}.}
\label{tab:MSN}
\end{table*}

As a further aspect, we show that the considered model is not chaotic for instantaneous synapses,
in such a case we observe irregular asynchronous states due to stable chaos~\cite{politi2010stable}.
The system can become truly chaotic only for finite synaptic times~\cite{Timme2009ChaosBalance,angulo2014}.
However, we report clear indications that for synapses faster than the membrane time constant $\tau_m$  the passage from mean-driven to fluctuation-driven dynamics is not related to the onset of chaos.  Only for extremely slow synapses we 
have numerical evidences that the appearance of the bursting regime could be related to a passage
from a zero Lyapunov exponent to a positive one, in agreement with the results reported in~\cite{kadmon2015,harish2015} for homogeneous inhibitory networks. These preliminary  indications demand 
for future, more detailed investigations of  deterministic spiking networks  in order to relate fluctuation-driven regime and chaos onsets. 
Moreover, we expect that it will be hard to distinguish whether the 
erratic current fluctuations are due to regular chaos or stable chaos on the basis
of the analysis of the network activity, as also pointed out in~\cite{Timme2009ChaosBalance}.

For what concerns the biological relevance of the presented model, 
we can attempt a comparison with experimental data obtained for MSNs in the 
striatum. This population of neurons is fully inhibitory with sparse lateral connections
 (connection probability $\simeq 10-20\%$~\cite{tunstall2002inhibitory,taverna2004sparseStriatum}), 
unidirectional and relatively weak~\cite{tepper2004gabaergic}.
Furthermore, for MSNs within the same collateral network
the axonal propagation delays are quite small $\simeq 1-2$ ms~\cite{taverna2008}
and they can be safely neglected.
The  dynamics of these neurons in behaving mice, reveals a low average firing rate
with irregular firing activity (bursting) with associated large coefficient of variation~\cite{miller2008dysregulated}.
As we have shown, these features can be reproduced by sparse networks of LIF neurons
with sufficiently slow synapses at $g > g_m$ and $\tau_\alpha >  \tau_m$.
For  values of the membrane time constant which are comparable to the ones
measured  for MSNs~\cite{plenz1998up,planert2013membrane} (namely, $\tau_m \simeq 10-20$ msec),
the model is able to capture even quantitatively some of the  main aspects of the MSNs dynamics,
as shown in Table \ref{tab:MSN}. 
We obtain a reasonable agreement with the experiments for sufficiently slow synapses, where the 
interaction among MSNs is mainly mediated by  GABA$_A$ receptors, which are 
characterized by IPSP durations  of the order of $\simeq 5 -20$ ms~\cite{tunstall2002inhibitory,koos2004comparison}. However,
apart the burst duration, which is definitely shorter, 
all the other aspects of the MSN dynamics can be already captured for
$\tau_\alpha = 2 \tau_m$  (with $\tau_m =10$ ms)  as shown in Table \ref{tab:MSN}.
Therefore, we can safely affirm, as also suggested in~\cite{ponzi2013optimal}, that the fluctuation 
driven regime emerging at $g>g_m$ is the most appropriate in order to reproduce the dynamical evolution of 
this population of neurons. 

 Other inhibitory populations are present in the basal ganglia. In particular
two coexisting inhibitory populations, {\it arkypallidal} (Arkys) and {\it prototypical} 
(Protos) neurons, have been recently discovered in the external globus pallidus~\cite{mallet2012dichotomous}.
These populations have distinct physiological and dynamical characteristics and
have been shown to be fundamental for action suppression during the performance
of behavioural tasks in rodents~\cite{mallet2016arkypallidal}.
Protos are characterized by a high firing rate $\simeq 47$ Hz and a not too large coefficient of
variation (namely, $CV \simeq 0.58$) both in awake and slow wave sleep (SWS) states,
while Arkys have a clear bursting dynamics with $CV \simeq 1.9$~\cite{mallet2016arkypallidal,dodson2015distinct}.
Furthermore, the firing rate of Arkys is definitely larger in the awake state (namely, $\simeq 9$ Hz)
with respect to the SWS state, where the firing rates are $\simeq 3-5$ Hz~\cite{mallet2016arkypallidal}.
  
On the basis of our results, on the one hand Protos can be modeled as LIF neurons with 
reasonable fast synapses in a mean driven regime, namely with
a synaptic coupling $g < g_m$. On the other hand, Arkys should be characterized
by IPSP with definitely longer duration and they should be in a fluctuation
driven phase as suggested from the results reported in Fig.~\ref{fig:alfa_sparse_nu_nStar}.
Since, as shown in Fig.~\ref{fig:alfa_sparse_nu_nStar} (a),
the firing rate of inhibitory neurons decrease by increasing the synaptic strenght $g$
we expect that the passage from awake to slow wave sleep should be characterized by a reinforcement 
of Arkys synapses. Our conjectures about Arkys and Protos synaptic properties based on their
dynamical behaviours ask for for experimental verification,  which we hope will happen shortly.

Besides the straightforward applicability of our findings to networks of 
pulse-coupled oscillators~\cite{mirollo1990synchronization}, it has been recently shown that LIF networks 
with instantaneous and non-instantaneous synapses can be transformed into
the Kuramoto-Daido model~\cite{politi2015,daido1996,kuramotobook}.
Therefore, we expect that our findings should extend to phase oscillator 
arrays with repulsive coupling~\cite{tsimring2005}. 
This will allow for a wider applicability of our results, due to the 
relevance of  limit-cycle oscillators not only for modeling  biological
systems~\cite{winfree2001}, but also for the many scientific and
technological applications~\cite{strogatz2001,dorfler2014,pikovsky2015,rodrigues2016}.

\acknowledgments
Some preliminary analysis on the instantaneous synapses has been performed in collaboration with A. Imparato, the complete results will be reported elsewhere~\cite{alb2017}. We thank for useful discussions J. Berke, B. Lindner, G. Mato, G. Giacomelli, S. Gupta,  A. Politi,  MJE Richardson,  R. Schmidt, M.Timme. This work has been partially supported by the European Commission under the program ``Marie Curie Network for Initial Training", through the project N. 289146, ``Neural Engineering Transformative Technologies (NETT)" (D.A.-G., S.O., and A.T), by the A$^\ast$MIDEX grant (No. ANR-11-IDEX-0001-02) funded by the French Government ``program Investissements d'Avenir'' (D.A.-G. and A.T.), and by ``Departamento Adminsitrativo de Ciencia Tecnologia e Innovacion - Colciencias" through the program ``Doctorados en el exterior - 2013" (D.A.-G.). This work has been completed at Max Planck Institute for the Physics of Complex Systems in Dresden (Germany) as part of 
the activity of the Advanced Study Group 2016 entitled "From Microscopic to Collective Dynamics in Neural Circuits".

\section*{APPENDIX A: Event Driven Maps}

By following \cite{Zillmer2007,Olmi2014linear} the ordinary differential
equations \eqref{eq:generic_model} and \eqref{eq:recurrent_current} describing
the evolution of the membrane potential of the neurons can be rewritten 
exactly as discrete time maps connecting successive firing events 
occurring in the network. In the following we will report explicitly
such {\it event driven maps} for the case of instantaneous and $\alpha$ synapses.

For instantaneous PSPs, the event-driven map for neuron $i$ takes the following expression:
\begin{equation}
\label{eq:v_delta_map}
v_i(n+1) = v_i(n) \text{e}^{-T_\delta} + I_i(1 - \text{e}^{-T_\delta}) 
- \frac{g}{K}C_{mi} \; ,
\end{equation}
where the sequence of firing times $\{t_n\}$ in the network 
is denoted by the integer indices $\{n\}$, $m$ is the index of the neuron firing at time $t_{n+1}$ and 
$T_\delta \equiv t_{n+1}-t_n$ is the inter-spike interval associated with two successive neuronal firing. 
This latter quantity is calculated from the following expression:
\begin{equation}
\label{eq:T_delta}
T_\delta  = \log \left[ \frac{I_m - v_m}{I_m - 1}\right] \; .
\end{equation}

For $\alpha$-pulses, the evolution of the synaptic current $E_i(t)$,
stimulating the $i$-th neuron can be  expressed in terms of a second order 
differential equation, namely 
\begin{equation}
\label{eq:E_secondOrder}
\ddot E_i(t) + 2 \alpha \dot E_i(t)+\alpha^2 E_i(t) = \frac{\alpha^2}{K}  \sum_{j \ne i} \sum_{n|t_n < t} C_{ij} \delta(t-t_n) \qquad .
\end{equation} 

Eq. \eqref{eq:E_secondOrder} can be rewritten as two first order
differential equations  by introducing the auxiliary variable 
$Q \equiv \dot E_i -\alpha E_i $, namely 
\begin{equation}
\label{eq:E_and_P}
\dot{E_i} = Q_i - \alpha E_i, \quad \dot{Q_i}=-\alpha Q_i + \frac{\alpha^2}{K}   
\sum_{n|t_n < t} C_{ij} \delta(t-t_n) 
\end{equation}

Finally, the equations \eqref{eq:generic_model} and \eqref{eq:E_and_P} can be exactly integrated from the time 
$t_n$, just after the deliver of the
$n$-th pulse, to time $t_{n+1}$ corresponding to the emission
of the $(n+1)$-th spike, to obtain the following event driven map:
\begin{subequations}
\label{eq:diluted_map}
\begin{align}
\label{qq}
Q_i(n+1)&=Q_i(n) {\rm e}^{-\alpha T_{\alpha}}+\frac{\alpha^2}{K} C_{mi} 
\\
\label{eq:E_map}
E_i(n+1)&=E_i(n) {\rm e}^{-\alpha T_{\alpha}}+Q_i(n) T_{\alpha}
{\rm e}^{-\alpha T_{\alpha}} \\
\label{V_map}
v_{i}(n+1)&=v_i(n){\rm e}^{-T_{\alpha}}+I_i(1-{\rm e}^{-T_{\alpha}})-g H_i(n) \, ,
\end{align}
\end{subequations}

In this case, the inter-spike interval $T_{\alpha} \equiv t_{n+1} - t_{n}$ should be estimated
by solving self-consistently the following expression 
\begin{equation}
\label{eq:tau_implicit}
T_\alpha = \ln\left[\frac{I_m-v_m(n)}{I_m-g H_m(n)-1}\right] \ ,
\end{equation}
where the explicit expression for $H_i(n)$ appearing in equations (\ref{V_map}) and (\ref{eq:tau_implicit}) is
\begin{eqnarray}
\label{eq:H_i}
\nonumber H_i(n) &=& \frac{{\rm e}^{-T_\alpha} - {\rm e}^{-\alpha T_\alpha}}{\alpha-1} \left(E_i(n)+\frac{Q_i(n)}{\alpha-1} \right) \\
	&-&\frac{T_\alpha {\rm e}^{-\alpha T_\alpha}}{\alpha-1} Q_i(n) \, .
\end{eqnarray}

The model so far introduced contains only adimensional units,
however, the evolution equation for the membrane potential \eqref{eq:generic_model}
can be easily re-expressed in terms of dimensional variables as follows
\begin{equation}
\label{eq:dim}
\tau_m \dot{V}_{i}(\tilde t)={\tilde I}_i - {V}_{j}(\tilde t) - \tau_m {\tilde g} {\tilde E}_i( \tilde t) \quad i=1,\cdots, N \quad ;
\end{equation}
where we have chosen $\tau_m = 10$ ms as the membrane time constant, ${\tilde I_i}$ represents 
the neural excitability and the external stimulations.
Furthermore, ${\tilde t} = t \cdot \tau_m$,
the field ${\tilde E}_i = E_i / \tau_m$ has the dimensionality of a frequency 
and ${\tilde g}$ of a voltage. The currents $\{{\tilde I}_i\}$ have also the dimensionality
of a voltage, since they include the membrane resistance.

For the other parameters/variables the transformation to physical units is simply given by
\begin{eqnarray}
{V}_{i} &=& {V}_r + ({V}_{th} - {V}_{r}) v_i\\
{\tilde I}_i &=& {V}_r + ({V}_{th} - {V}_{r}) I_i\\
{\tilde g} &=& ({V}_{th} - {V}_{r}) g 
\end{eqnarray}
where ${V}_{r}= -60$ mV and ${V}_{th}=-50$ mV are realistic values of the membrane 
reset and threshold potential. The isolated $i$-th LIF neuron is supra-threshold
whenever ${\tilde I}_i > {V}_{th}$.

\section*{APPENDIX B: Average firing rate \\ for instantaneous synapses}

In this Appendix, by following the approach in \cite{richardson2010firing}
we  derive the average firing rate of a supra-threshold LIF neuron subject 
to inhibitory synaptic shot noise of constant amplitude $G$, namely
\begin{equation}
\label{eq:shot_model}
\dot{v}(t) = I - v(t) - G \sum_{\{t_k\}} \delta(t-t_k) \qquad ;
\end{equation}
where $I > 1$.The post-synaptic pulses reaching the neuron 
are instantaneous and their arrival times 
are Poisson-distributed and characterized by a rate $R$. In order to find the firing rate response of the LIF
neuron we introduce the probability density $P(v)$ and the flux $J(v)$ associated
to the membrane potentials, these satisfy the continuity equation:

\begin{equation}
\label{eq:ContinuityEqaution}
\frac{\partial P}{\partial t} + \frac{\partial J}{\partial v}  =  
\rho(t) [\delta (v-v_{r}) - \delta (v-\theta)] \quad ;
\end{equation}
where $\rho(t)$ is the instantaneous firing rate of the neuron. The flux can be decomposed
in an average drift term plus the inhibitory part, namely
\begin{eqnarray}
\label{eq:FluxTot}
J & = & (I-v) P + J_{inh}  \\
\label{eq:FluxInh_der}
\frac{\partial J_{inh}(v,t)}{\partial v} & = & R  [P(v,t) - P(v-G,t)] \quad ;
\end{eqnarray}

The set of equations \eqref{eq:ContinuityEqaution} to \eqref{eq:FluxInh_der} is 
complemented by the boundary conditions:

$$
J(\theta,t) = \rho(t) \qquad J_{inh}(\theta,t) = 0 \qquad P(\theta,t) = 0 \; ;
$$
and by the requirement that membrane potential distribution should be normalized, i.e
$$\int_{-\infty}^{\theta} P(v,t) dv = 1 \;.$$

By introducing bilateral Laplace transforms $\tilde f(s) = \int_{-\infty}^{\infty} 
dv {\rm e}^{sv} f(v)$ and by performing some algebra along the lines described
in \cite{richardson2010firing} it is possible to derive the analytic
expression for the average firing rate
\begin{equation}
\label{eq:nu0}
\frac{1}{\nu_0} =  \int_0^{\infty} \frac{ds}{s}\frac{{\rm e}^{s \theta}- {\rm e}^{s v_r}}
{\tilde Z_0(s)} .
\end{equation}
where ${\tilde Z_0(s)}$ is the Laplace transform of the sub-threshold probability density. 
Namely, it reads as
\begin{equation}
\label{eq:Z0}
\tilde Z_0(s) = E \left[s^{-R} {\rm e}^{sI + R {\cal E}(Gs)} \right] \qquad ;
\end{equation}
where ${\cal E}(y) = - \int_{-y}^{\infty} dt {\rm e}^{-t}/t $ is the exponential integral, 
$E= {\rm e}^{-R ( \Gamma +\ln G )}$
is the normalization constant ensuring that the distribution $Z_0(v)$ is properly
normalized, and $\Gamma$ is the Euler-Mascheroni constant.

\begin{figure}
\centering
\includegraphics[width=\linewidth]{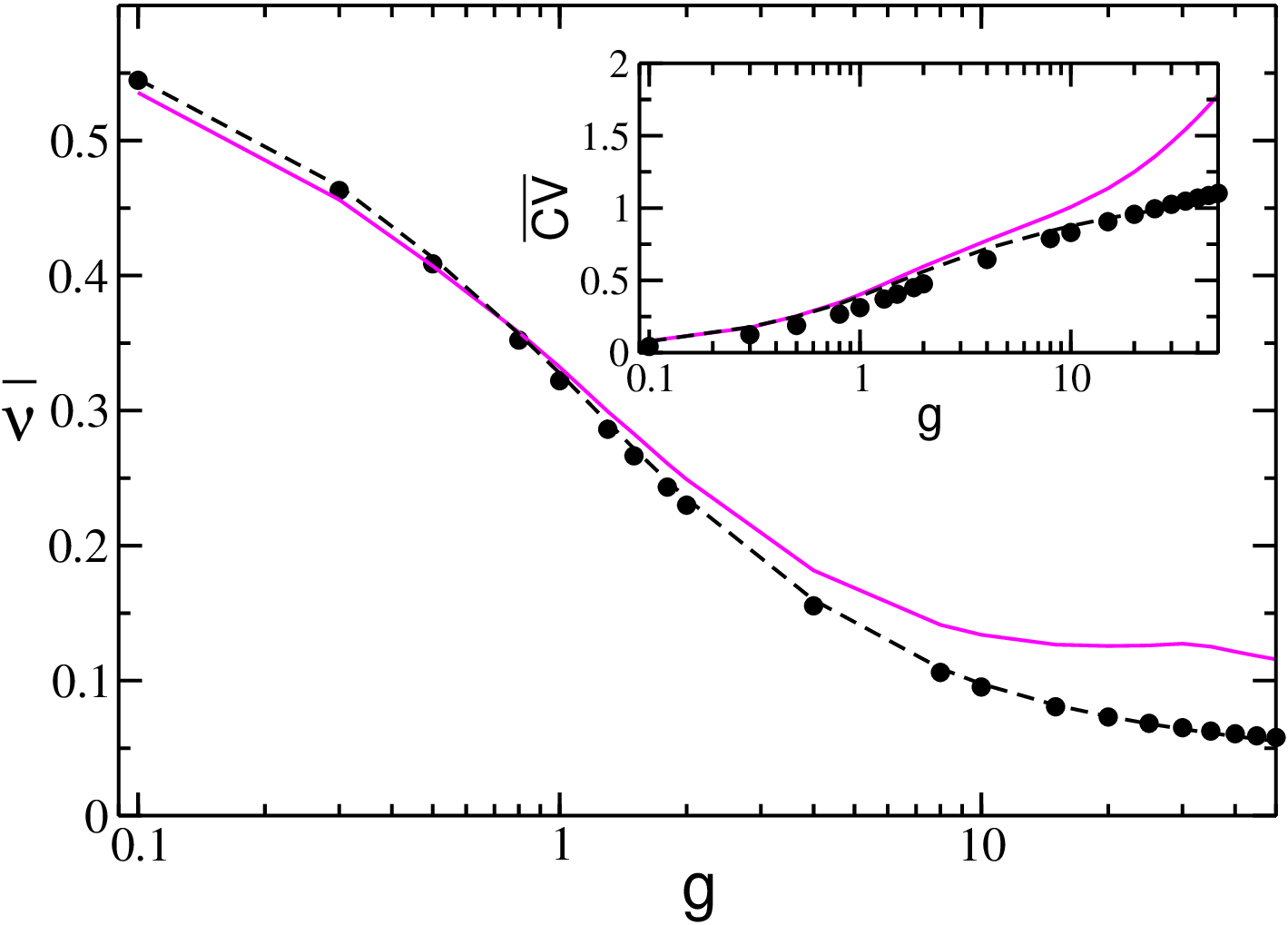}
\caption{Average network's frequency $\bar \nu$ 
as a function of the synaptic strength $g$ for 
instantaneous synapses and uniform distributions 
$P(I)$ with support $[l_1, l_2] = [1.0 , 1.5]$.
Inset: average coefficient of variation $\overline{CV}$ versus $g$.
Filled symbols refer to numerical simulation for $N = 400$ and $K=20$, 
dashed lines to the corresponding analytic solutions reported in Appendices B and C
and the solid (magenta) lines to the diffusion approximation. 
The data have been averaged over a time interval $t_S=1\times 10^6$ after discarding a transient of $10^6$ spikes.
}
\label{fig:shot_diffusion}
\end{figure}

In order to validate the method here outlined to obtain the average firing frequency
$\bar \nu$, we compare the theoretical
estimates given by \eqref{eq:nu_sparse} 
with numerical data obtained for sparse networks with in-degree $K$ and instantaneous inhibitory synapses.
The agreement is quite remarkable as shown in Fig.~\ref{fig:shot_diffusion}.
In the same figure the solid magenta line refers to the results obtained
by employing the diffusion 
approximation~\cite{ricciardi2013diffusion,tuckwell2005,BrunelHakim1999,Brunel2000Sparse}:
clear discrepancies are evident already for $g \geq 1$. 
In particular, for the diffusion approximation 
and the evaluation of \eqref{eq:nu_sparse} we assume that each neuron receives
a Poissonian spike train with an arrival rate given by $R=n_A \bar \nu K$.
Furthermore, it should be stressed
that in this case we limit to test the quality of the approach described in this Appendix 
versus the diffusion approximation, therefore $n_A$, required to estimate
$\bar \nu$, is obtained from the
simulation and not derived self-consistently as done in Sect. IV.

\section*{APPENDIX C: Coefficient of variation \\ for instantaneous synapses}

In order to derive the coefficient of variation for the shot noise case 
it is necessary to obtain the first two moments of the 
first-passage-time density $q(t)$. By following the same
approach as in Appendix B, the time-dependent continuity equation 
with initial condition $P(v,0) = v_r$ is written as
\begin{widetext}
\begin{equation}
\label{eq:ContinuityEq_nonztaz}
\frac{\partial P}{\partial t} + \frac{\partial J}{\partial v}  =  
\rho(t) [\delta (v-v_{r}) - \delta (v-\theta)] + \delta (t)\delta (v-v_r) \;.
\end{equation}
\end{widetext}
As suggested in \cite{richardson2010firing}, Eq.~\eqref{eq:ContinuityEq_nonztaz} can
be solved by performing a Fourier transform in time and a bilateral Laplace
transform in membrane potential. This allows to obtain the Fourier
transform of the spike-triggered rate, namely
\begin{equation}
\hat{\rho}(\omega) =  \frac{\displaystyle\int_0^{\infty} ds \enskip s^{\mathrm{i}\omega}A^\prime (s)}
{\displaystyle \int_0^{\infty} ds \enskip s^{\mathrm{i}\omega} \left[B^\prime(s) - A^\prime (s)\right]
} \quad ;
\label{rhow}
\end{equation}
where $A(s) = {\rm e}^{s v_r}/\tilde Z_0(s)$ and
$B(s) = {\rm e}^{s \theta}/\tilde Z_0(s)$. The Fourier transform
of the first-passage-time density is $\hat{q}(\omega) 
= \frac{\hat{\rho}(\omega)}{1+\hat{\rho}(\omega)}$ and the first
and second moment of the distribution are given by
\begin{eqnarray}
\label{eq:fOmega}
\frac{\partial \hat{q}}{\partial \omega} |_{\omega = 0} = - \mathrm{i} \langle t \rangle \\
\frac{\partial^2 \hat{q}}{\partial \omega^2} |_{\omega = 0} = - \langle t^2 \rangle
\end{eqnarray}
The integrals appearing in Eq.~\eqref{rhow} cannot be exactly solved, 
therefore we have expanded it to the second order obtaining
\begin{equation}
\hat{\rho}(\omega) \simeq \frac{n_0 + n_1 \omega + n_2 \omega^2}{d_0 + d_1 \omega + d_2 \omega ^2} \quad;
\end{equation}
where $n_0 = -1$, $d_0=0$, $d_1 = -\mathrm{i} n_0/\nu_0$ and
\begin{eqnarray}
\label{eq:ns_ds}
n_1 & = & \mathrm{i} \int_0^{\infty} \log s A^\prime(s) ds\\
d_2 & = & \int_0^{\infty} \frac{\log s}{s}[B(s)-A(s)] ds \qquad .
\end{eqnarray}

From the expression \eqref{rhow} we can finally obtain
the first and second moment of $q(t)$, namely

\begin{equation}
\langle t \rangle = \frac{1}{\nu_0} \quad,\quad
\langle t^2 \rangle = 2 [d_1^2 + d_2  + d_1 n_1] \quad .
\end{equation}

Once these quantities are known the coefficient of variation can be easily
estimated for each neuron with excitability $I$.

The results obtained for the average coefficient of variation 
$\overline{CV}$ for a sparse network are compared with numerical data
and with the diffusion approximation in the inset of 
Fig.~\ref{fig:shot_diffusion}. It is evident that the approximation here derived
is definitely more accurate than the diffusion approximation for synaptic
strengths larger than $g \simeq 1$.

\section*{APPENDIX D: Average Firing Rate \\ for Slow Synapses}

In Section \ref{sec:alpha} we have examined the average activity of the network for
non instantaneous IPSPs with $\alpha$-function profiles. In presence of synaptic filtering,
whenever the synaptic time constant
is larger than the membrane time constant one can apply the so-called adiabatic approach 
to derive the firing rate $\nu_0$ of a single neuron, as described in~ \cite{moreno2004,moreno2010response}. 

In this approximation, the output firing rate $\nu_0$ of the single neuron driven by a slowly varying 
stochastic input current $z$ with an arbitrary distribution $P(z)$ is given by
\begin{equation}
\label{eq:nu_adiabatic}
\nu_0 \simeq \int dz P(z)\nu(z)
\end{equation}
where $\nu(z)$ is the input to rate transfer function of the neuron under a stationary input which for the LIF neuron is simply:
\begin{equation}
\label{eq:inputRate_TF}
\nu(z) = \left[ \ln \left( \frac{z - v_r}{z - \theta}\right)  \right]^{-1}\quad .
\end{equation}

The  synaptic filtering induces temporal correlations in the input current $z$, which
can be written as:
\begin{equation}
\langle (z(t) -{\mu})(z(t') -{\mu}) \rangle = \frac{\sigma ^2}{2\tau_s}\exp \left[-\frac{|t-t'|}{\tau_s}\right]
\quad ;
\end{equation}
here $\tau_s$ is the synaptic correlation time. In the case of $\alpha$-pulses,
where the rise and decay synaptic times coincide, we can assume that the correlation 
time is given by $\tau_s = \tau_P = 2 \tau_{\alpha}$.

Analogously to the diffusion approximation~\cite{Ricciardi1971,BrunelHakim1999,Brunel2000Sparse}, 
the input currents are approximated as a Gaussian noise with 
mean $\mu$ and variance $\sigma_z^2 = \sigma / 2\tau_s$.
In our network model, a single neuron receives an average current $\mu$
given by Eq. \eqref{eq:eff_input_GC} with a standard deviation $\sigma$ given by
Eq. \eqref{eq:sigma}. In particular, the fraction of active neurons $n_A$ entering in the expressions
of $\mu$ and $\sigma$ is in this case obtained by the numerical simulations.

Therefore,  the single neuron output firing rate reads as
\begin{eqnarray}
\label{eq:nu_adiabatic_appendix_NotSelf}
\nu_0(I) & = & \int \frac{dz}{\sqrt{2\pi} \sigma_z} {\rm e}^{-\frac{(z-\mu(I))^2}{2\sigma_z^2}} 
\left[ \ln\left( \frac{z - v_r}{z - \theta} \right) \right]^{-1} \; ;
\end{eqnarray}
where $I$ is the neuronal excitability.

The average firing rate of the LIF neurons in the network, characterized by 
an excitability distribution $P(I)$, can be estimated as
%%\begin{widetext}
\begin{equation}
\label{eq:nu_adiabatic_appendix_Self}
\bar{\nu}  =  \int_{\{ I_A \}} dI P(I) \int_{\theta} \frac{dz}{\sqrt{2\pi} \sigma_z} {\rm e}^{-\frac{(z-\mu(I))^2}{2\sigma_z^2}} 
\left[\ln\left( \frac{z - v_r}{z - \theta} \right)\right]^{-1}
\end{equation}
%%\end{widetext}
where we impose the self-consistent condition that the 
average output frequency is equal to the average input one.

%\bibliographystyle{plain}
%\bibliography{//home/torcini/Dropbox/ANGULO/BibliographyComplete/Bibliography_report}
%%\bibliography{//home/dangulog/Dropbox/ANGULO/BibliographyComplete/Bibliography_report}

\end{document}